\documentclass[12pt,preprint]{aastex} 
\usepackage{natbib}
\usepackage{morefloats}
 \usepackage{times} 
\usepackage{amsmath}

 \shortauthors{Lionello et al.}
\shorttitle{Time Delays Expected from Impulsive Heating}

\def\bhat{\bf\skew{-5}\hat b} 

\begin{document}


\title{CAN LARGE TIME DELAYS OBSERVED IN LIGHT CURVES OF CORONAL LOOPS BE
EXPLAINED BY IMPULSIVE HEATING?} 
\author{Roberto Lionello,}
\affil{Predictive Science, Inc., 9990 Mesa Rim Rd., Ste. 170,
San Diego, CA 92121-2910} \email{lionel@predsci.com} \and
\author{Caroline E.\ Alexander, Amy R.\ Winebarger,} \affil{NASA
Marshall Space Flight Center, ZP 13, Huntsville, AL 35805}
\email{\{caroline.e.alexander,amy.r.winebarger\}@nasa.gov} \and
\author{Jon A.\ Linker, Zoran Miki\'c,} \affil{Predictive
Science, Inc., 9990 Mesa Rim Rd., Ste. 170, San Diego, CA
92121-3933} \email { \{linkerj,mikicz\}@predsci.com }


\begin{abstract} 
The light curves of solar coronal loops often peak first in channels associated with higher temperatures and then in those associated with lower. The time delays between the different narrowband EUV channels have been measured for many individual loops and recently for every pixel of an active region observation. Time delays between channels for an active region exhibit a wide range of values, with maxima $>$ 5,000\,s. These large time delays make up 3-26\% (depending on the channel pair) of the pixels where a significant, positive time delay is measured. It has been suggested that time delays can be explained by impulsive heating. In this paper, we investigate whether the largest observed time delays can be explained by this hypothesis by simulating a series of coronal loops with different heating rates, loop lengths, abundances, and geometries to determine the range of expected time delays between a set of four EUV channels. We find that impulsive heating cannot address the largest time delays observed in two of the channel pairs and that the majority of the large time delays can only be explained by long, expanding loops with photospheric abundances. Additional observations may rule out these simulations as an explanation for the long time delays. We suggest that either the time delays found in this manner may not be representative of real loop evolution, or that the impulsive heating and cooling scenario may be too simple to explain the observations and other heating scenarios must be explored.

\end{abstract} 

\section{INTRODUCTION} 

Finding the mechanism that is primarily responsible for heating the solar corona has been a continuous challenge to generations of solar physicists. Much is still not understood about the fundamental processes that heat active region coronal loops, which are closed magnetic structures that appear bright in X-ray and EUV images.  The free energy of the magnetic field and/or energy in magnetohydrodynamic waves is most likely converted into heat, but it is not known how and where. 

An important, accurate, and relatively easy measurement is the timescale for the evolution of coronal loops. With the development of narrowband EUV imagers, such as the Atmospheric Imaging Assembly (AIA; \citealt{2011SoPh..tmp..115L}) on the {\it Solar Dynamics Observatory (SDO)}, it is possible to track the loops as they cool through different EUV channels.  The timing of the intensity evolution of coronal loops provides information on the cooling of the plasma. There have been several studies of individual loops that find what appears to be a general cooling trend, as the loop intensity peaks first in channels that see higher temperature plasma and later in those that see lower temperature plasma
\citep{2003ApJ...593.1164W,2006ApJ...643.1245U,2011ApJ...733...59M}.

Recently, \citet{2012ApJ...753...35V} measured the most likely time delay, or time lag, at every pixel in a data cube (instead of individual loop structures) by performing cross-correlation analysis to evaluate the temporal delays between different EUV channels on AIA in each pixel in a 12-hour (or 2-hour) active region observation. Applying this technique to AIA observations in different channels, they produced a map of time lags and concluded that the majority of the structures in the active region, including regions where no discernible loop was present, showed evidence of this cooling trend.  In many of the channel pairs, such as those including the 335, 211, 193, and 171\,\AA\, channels, they found a wide range of time lags.  The maximum time lag in each of these channel pairs exceeds 5,000\,s. 

Most previous studies of individual loops involved only two channels from the {\it Transition Region and Coronal Explorer (TRACE)}.  The 195 and 171\,\AA\ channels on {\it TRACE} have similar wavelength and temperature responses to the 193 and 171\,\AA\ channels of AIA.  In a study of 8 loops, \cite{2011ApJ...733...59M} found a time delay between the two {\it TRACE} channels between 94\,s and 650\,s.  In a study of 5 loops, \cite{2006ApJ...643.1245U} found time delays between the two {\it TRACE} channels of 5 - 251\,s.  \cite{2003ApJ...587..439W} studied the evolution of 5 loops in the {\it TRACE} 195 and 171\,\AA\ channels.  Four of the loops had time delays of 150 - 1500\,s, while the longest loop had a time delay of 11,000\,s.  This loop had an estimated loop length of 287 - 424\,Mm.   Hence, in the previous studies of individual loops, only 1 loop out of 18 ($\sim 5$\%) had a time delay between {\it TRACE} 195 and 171\,\AA\ longer than 5,000\,s.  However, this could have been a selection bias in previous studies or the difficulty in obtaining long, interrupted data sets by {\it TRACE}.

Additional studies incorporated higher temperature plasma observations to measure time delays. \cite{2005ApJ...626..543W} used Yohkoh/SXT and TRACE 171\,\AA\ and measured time delays of 1-3 hours for 5 loops where longer loops had longer delay times. \cite{2009ApJ...695..642U} used EIS to study core loops and measured the length of time each loop was observed in various lines. Further investigations looked at cooling loops \citep[see e.g.,]{2006A&A...449.1177R, 2001ApJ...556..896S, 2011ApJ...738..146S,2002ApJ...571..999W,2008ApJ...677.1395W,1999JGR...104.9753D,2008A&A...481L..53T,2011ApJ...732...81A} and evaluated the temperature evolution using a variety of methods,  but did not place specific limits on the observed time delay between peaks in EUV emission. See \cite{2010LRSP....7....5R} for a comprehensive review.

In \citet{2013ApJ...771..115V}, they demonstrated that the persistent time lags in the diffuse corona could be a consequence of so-called ``nanoflare storms,'' a phrase coined by \cite{2006SoPh..234...41K}.  In this heating scenario, an observed loop is a bundle of unresolved ``strands,'' a fundamental structure where the density and temperature are uniform across the structure.  These strands are impulsively heated \citep[e.g.,][]{1995ApJ...439.1034C, 1997ApJ...478..799C,2002ApJ...579L..41W, 2003ApJ...593.1174W} at different times. If the length of the storm, i.e., the interval between the first and last heating event on all strands, is relatively short compared to the cooling time of the plasma, the loop would then appear in channels associated with progressively lower temperatures, as the plasma in the strands cools and drains down. If the length of the storm is longer than the cooling time of the plasma, the structures may have steadier emission.  \citet{2013ApJ...771..115V} simulated an arcade of coronal strands along the line-of-sight with lengths from 60-200\,Mm.  They heated these strands impulsively with varying magnitudes and durations,  which means that they continuously added newly heated strands to the line-of-sight arcade. Then they calculated the expected light curves, summed along the line-of-sight, and calculated the time lags.  They found time lags of 100s of seconds between the channel pairs.   \citet{2013ApJ...771..115V}  focused on how multiple heating events occurring along the line-of-sight in the so-called diffuse corona (where no discernible loop was present) would predict measurable time lags. However, they did not include a complete parameter space study to determine if
the full range of observed time lags could be explained by nanoflare storms.

Although \citet{2013ApJ...771..115V} focused exclusively on impulsive 
heating to explain their observations, 
there may be other possible explanations.  In particular,
one potential scenario that might yield long time delays
  was explored in a recent series of papers \citep{2013ApJ...773...94M,2013ApJ...773..134L,2014ApJ...795..138W}.  These authors argued that loops could be undergoing thermal non-equilibrium \citep{1982A&A...108L...1K,1983A&A...123..216M, 1991ApJ...378..372A,1999ApJ...512..985A,2001ApJ...553L..85K, 2003ApJ...593.1187K, 2003A&A...411..605M,2004A&A...424..289M, 2006ApJ...637..531K}, which is caused by highly stratified heating concentrated at the footpoints. Under some conditions, there are no steady solutions for this type of heating.  Instead, cold condensations can be periodically formed in the corona and then slide down along the field lines into the photosphere \citep[which may be observed as ``coronal rain'', e.g.][]{2001SoPh..198..325S}. Although thermal non-equilibrium was dismissed by \citet{2010ApJ...714.1239K} for allegedly giving unrealistic solutions, \citet{2013ApJ...773..134L} showed that these solutions can be perfectly compatible with observations, provided the geometrical properties of the loops are taken into proper account in the model \citep{2013ApJ...773...94M}.
A preliminary analysis of non-thermal equilibrium solutions (Winebarger et al., in preparation) shows that it can produce the large time lags found by \citet{2013ApJ...771..115V}.

In this paper, we have simulated a series of loops to investigate: 1) how different parameters affect the expected time lag between the channel pairs, and 2) under what conditions (if any) it is possible to achieve time lags longer than 5,000\,s between these channel pairs.  The parameters we have investigated are  1) the magnitude of the heating event, 2) 
the loop length, 3) the abundances of the plasma, and 4) whether the cross-sectional area of the loop expands or is constant.  
We have also considered the duration of the heating event and the inclination of the loop, but have found that the time lags are not greatly influenced by those parameters.
To complete this analysis, we have used one-dimensional, time-dependent, 
hydrodynamic (HD) models.  For each solution, we have  calculated
 the expected light curves in the AIA 335, 211, 193, and 171\,\AA\ channels 
and obtained the time lags of the emission peaks in the different channel pairs. In this study, we find that most loops do not show
temporal delays of several thousands of seconds between channels, such as observed by \citet{2012ApJ...753...35V}.   
Long loops, loops that expand, and loops that have photospheric abundances have the longest time delays.  These results may have several implications.  The time lags measured
by \citet{2012ApJ...753...35V} may not be representative of individual evolving loops, and instead may be indicative of other time scales in the active region evolution.
Additionally, a simple impulsive heating and cooling model may be too simple to explain the active region structures, and other heating scenarios need to be investigated.   
It further indicates that the time delay between when a structure appears in two channels may be a powerful diagnostic to differentiate between different heating scenarios.


\section{OBSERVATIONS}

\begin{figure}
\includegraphics[width=0.24\textwidth]{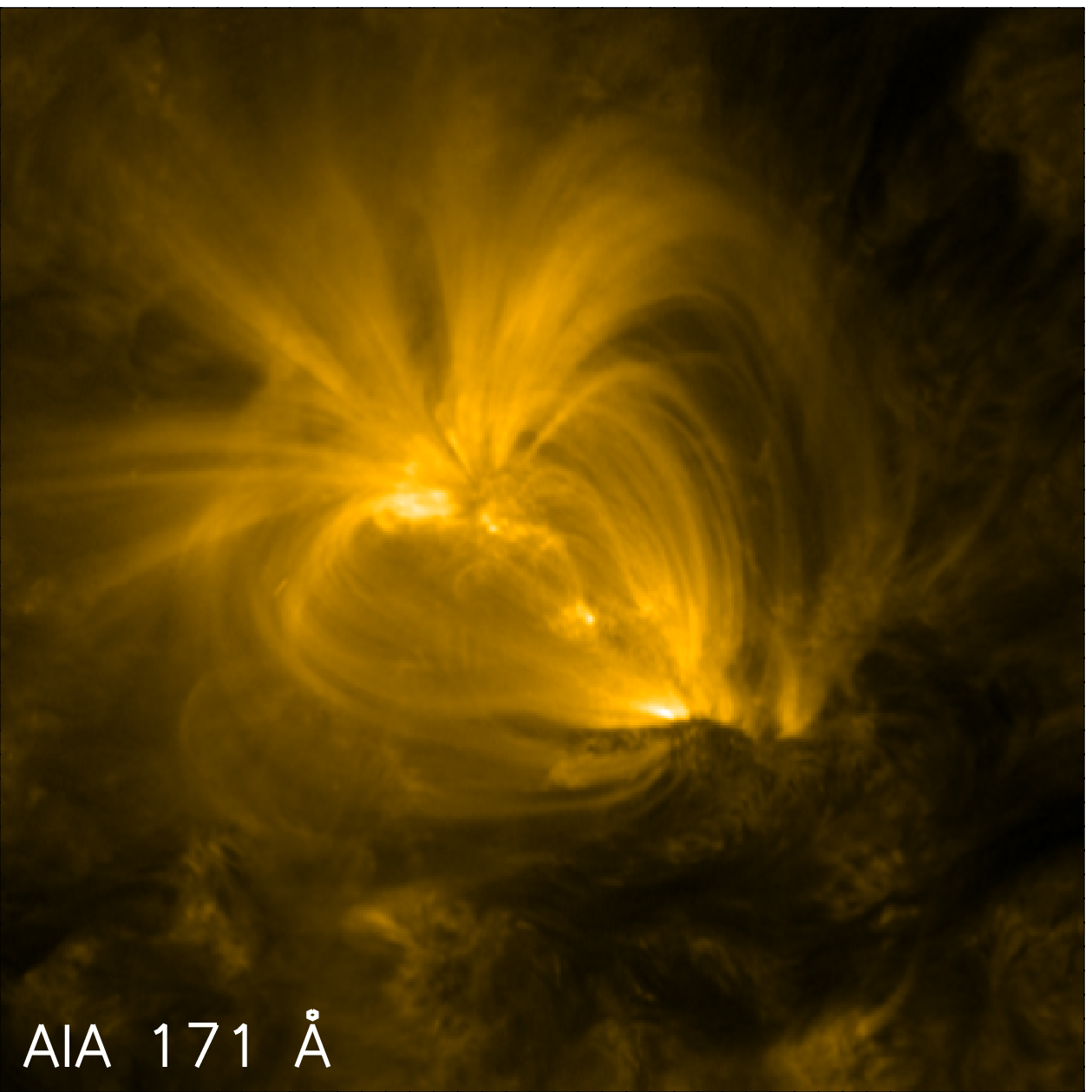}
\includegraphics[width=0.24\textwidth]{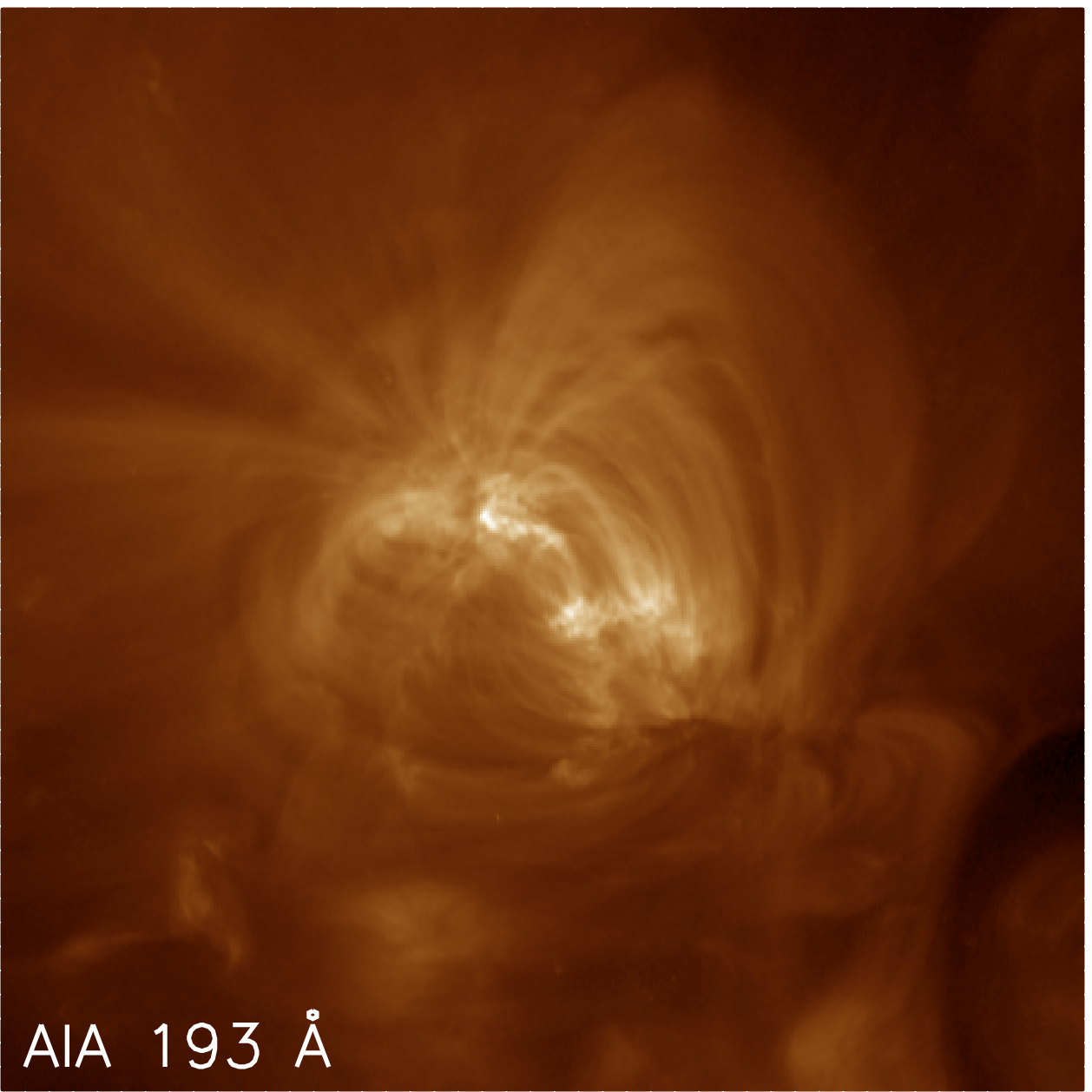}
\includegraphics[width=0.24\textwidth]{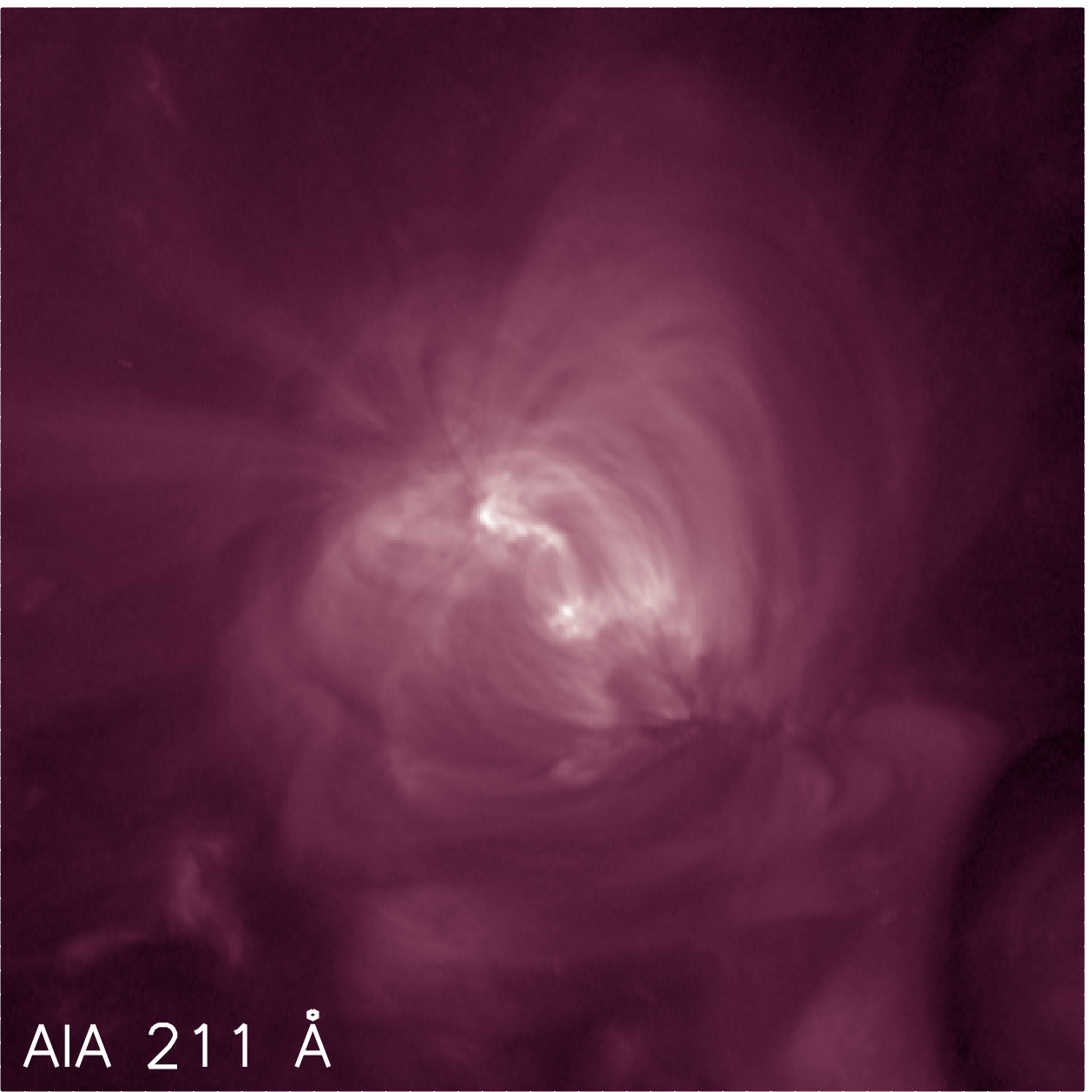}
\includegraphics[width=0.24\textwidth]{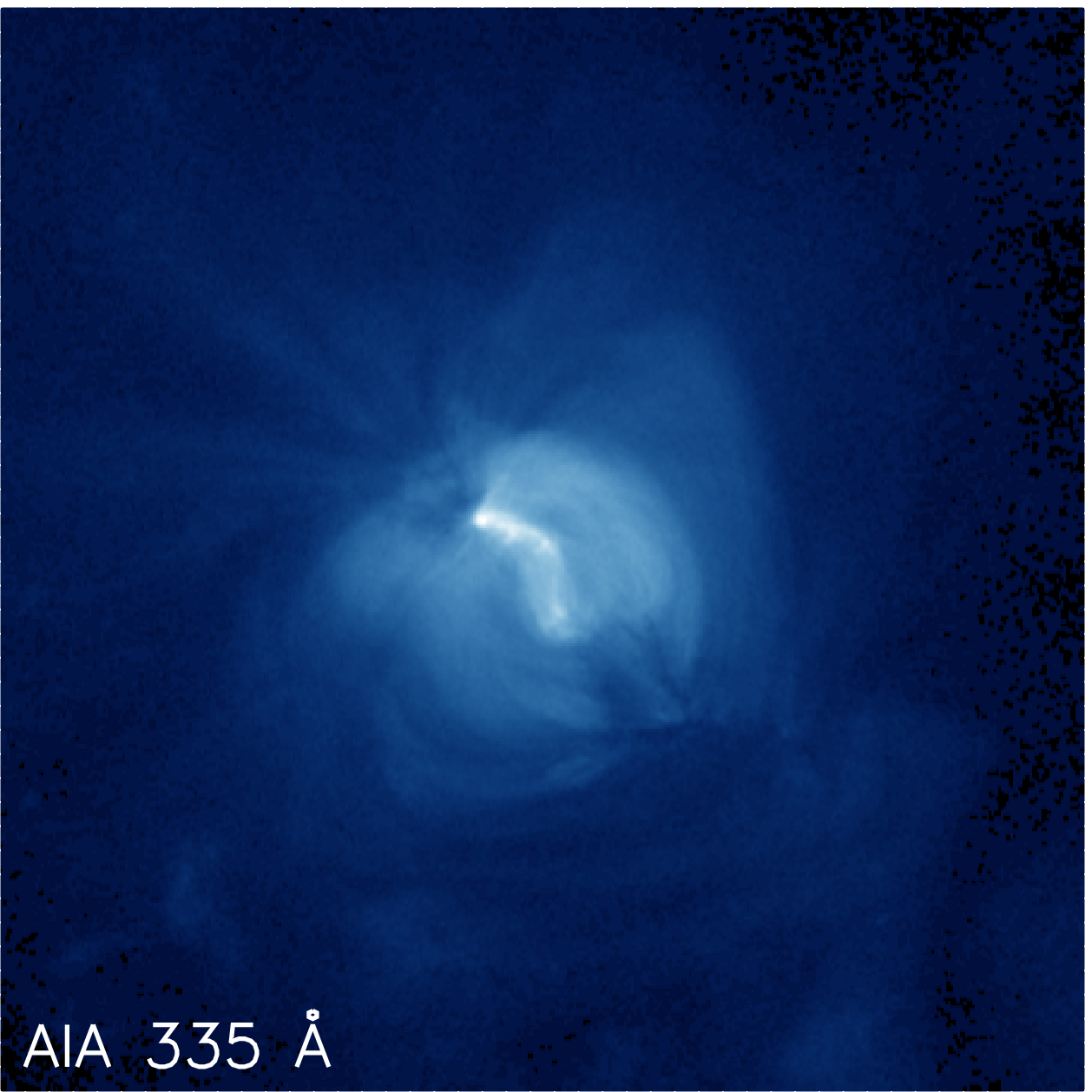}
\caption{Active Region 11082 observed in the AIA 171, 193, 211, and 335\,\AA\ channels on 19 June 2010 00:00. The square root of the intensity is shown.} 
\label{fig:aia_images} 
\end{figure}

\citet{2012ApJ...753...35V} examined Active Region 11082 observed on 19 June 2010.  This active region is shown in Figure~\ref{fig:aia_images} in four of the AIA channels.  They performed time lag analysis of this active region, calculating the time lag with the highest correlation coefficient between each channel pair at each pixel in the data cube.  In this paper, we focus on the time lags pairs for the four channels shown in Figure~\ref{fig:aia_images}.  The response functions for these channels is given in the top two panels of Figure~\ref{fig_tresp_abun}. Note that the 211, 193, and 171\,\AA\ channel have a narrow, single-peaked temperature response, while the temperature response function of 335\,\AA\ channel is broader with multiple peaks.  The width of the high temperature component of the 335\,\AA\ response function is roughly three times the width of the 193 and 211\,\AA\ temperature response functions and six times the width of the 171\,\AA\ temperature response function.  The temperature at the high temperature peak of the response functions of these channels and the difference in temperature for each channel pair is given in Table~\ref{tab:combs}.  The time lag maps of the remaining two channels (94 and 131\,\AA) used by \citet{2012ApJ...753...35V} were difficult to evaluate due to the bimodal response function and we do not consider them in this analysis.  

\begin{figure}
\begin{center}
\includegraphics[width=0.32\textwidth]{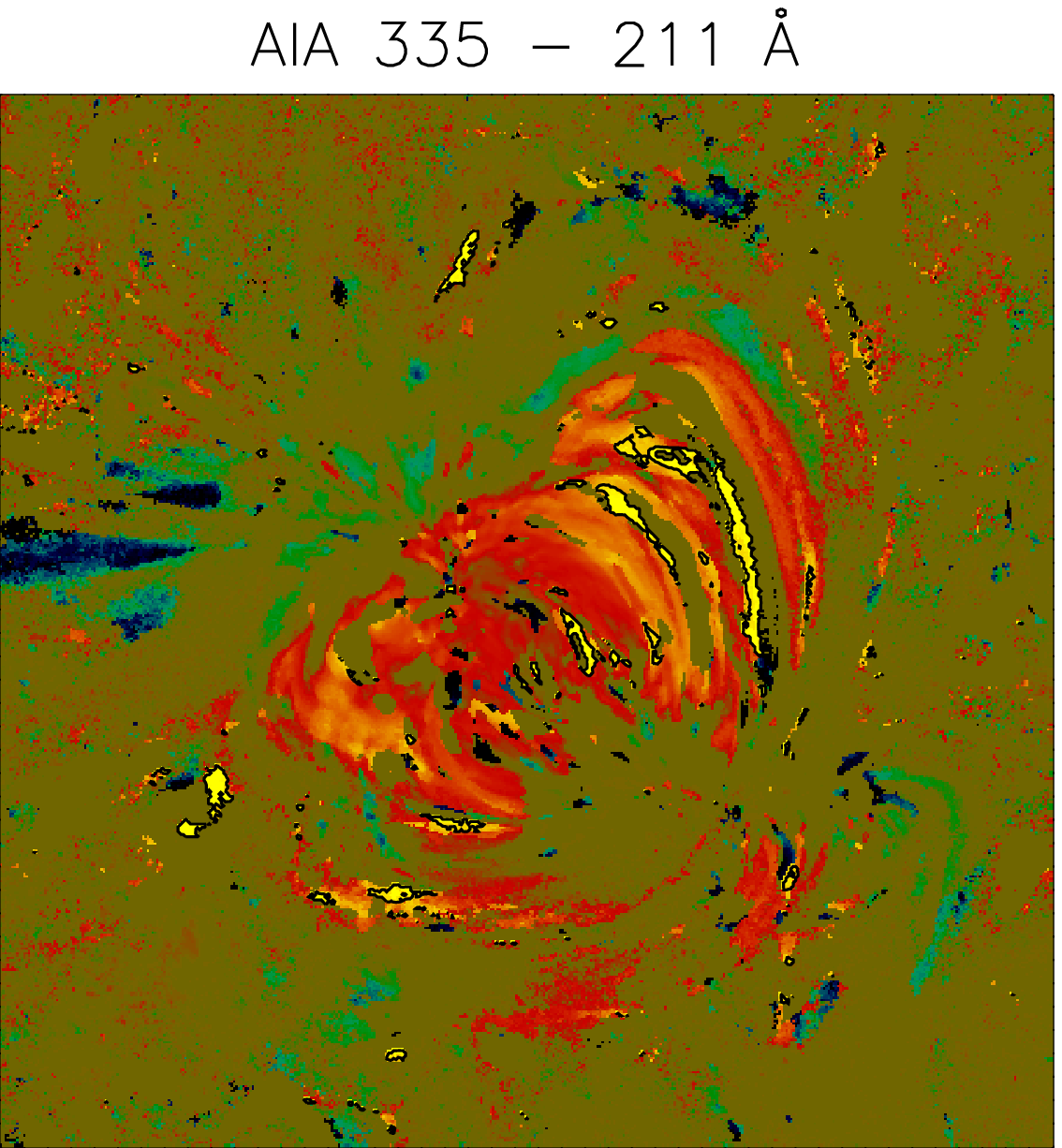}
\includegraphics[width=0.32\textwidth]{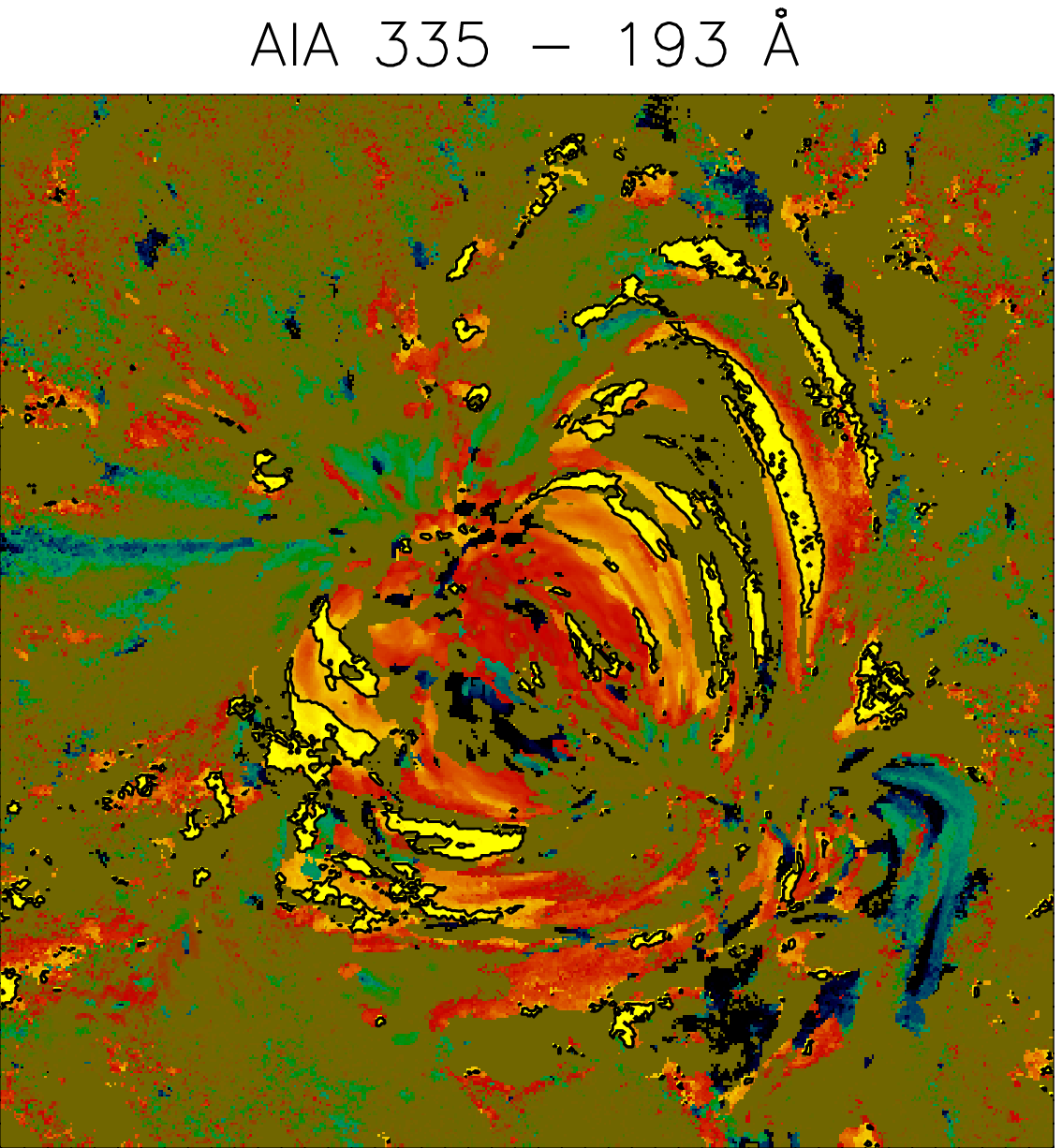}
\includegraphics[width=0.32\textwidth]{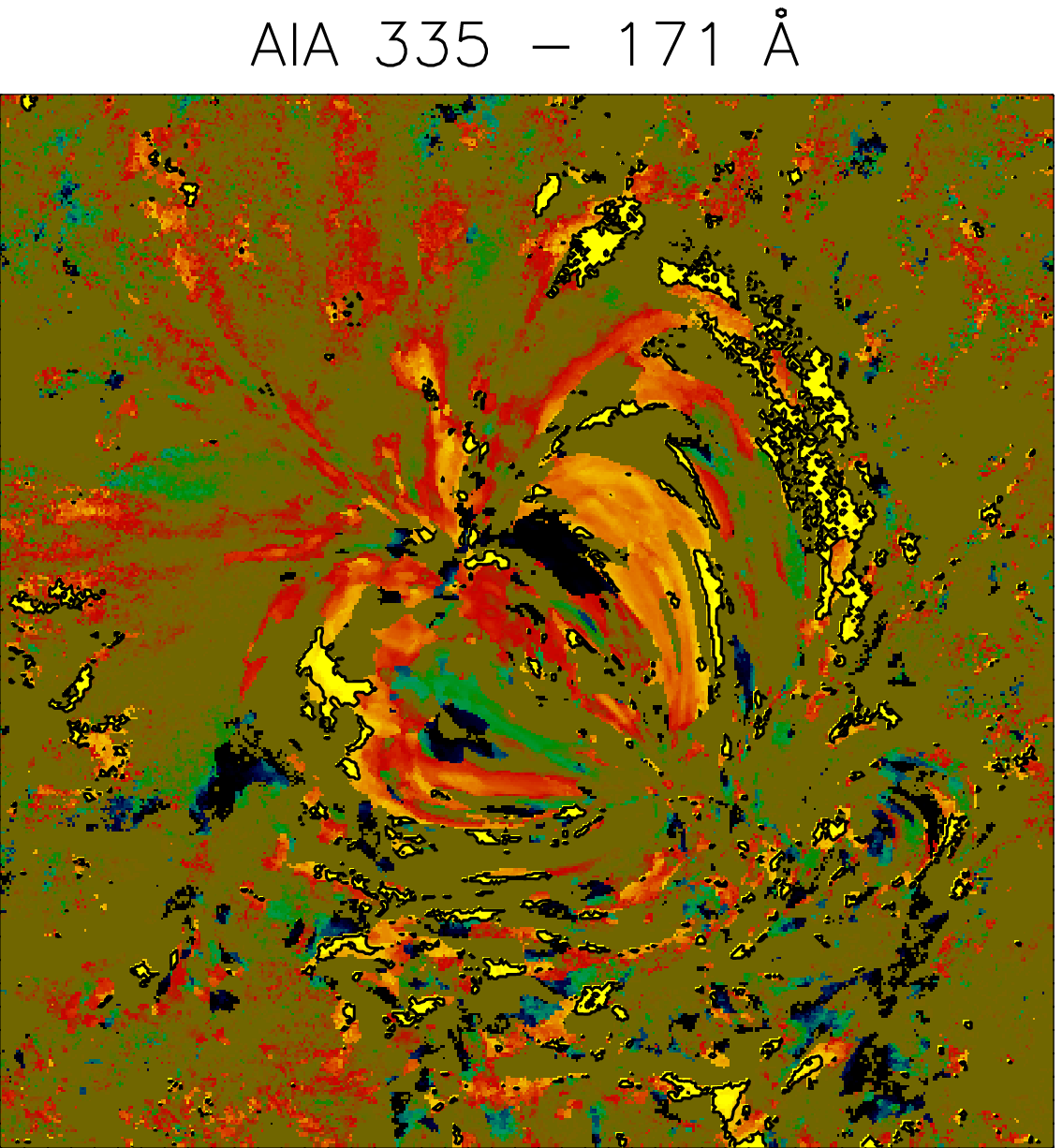}\\

\includegraphics[width=0.32\textwidth]{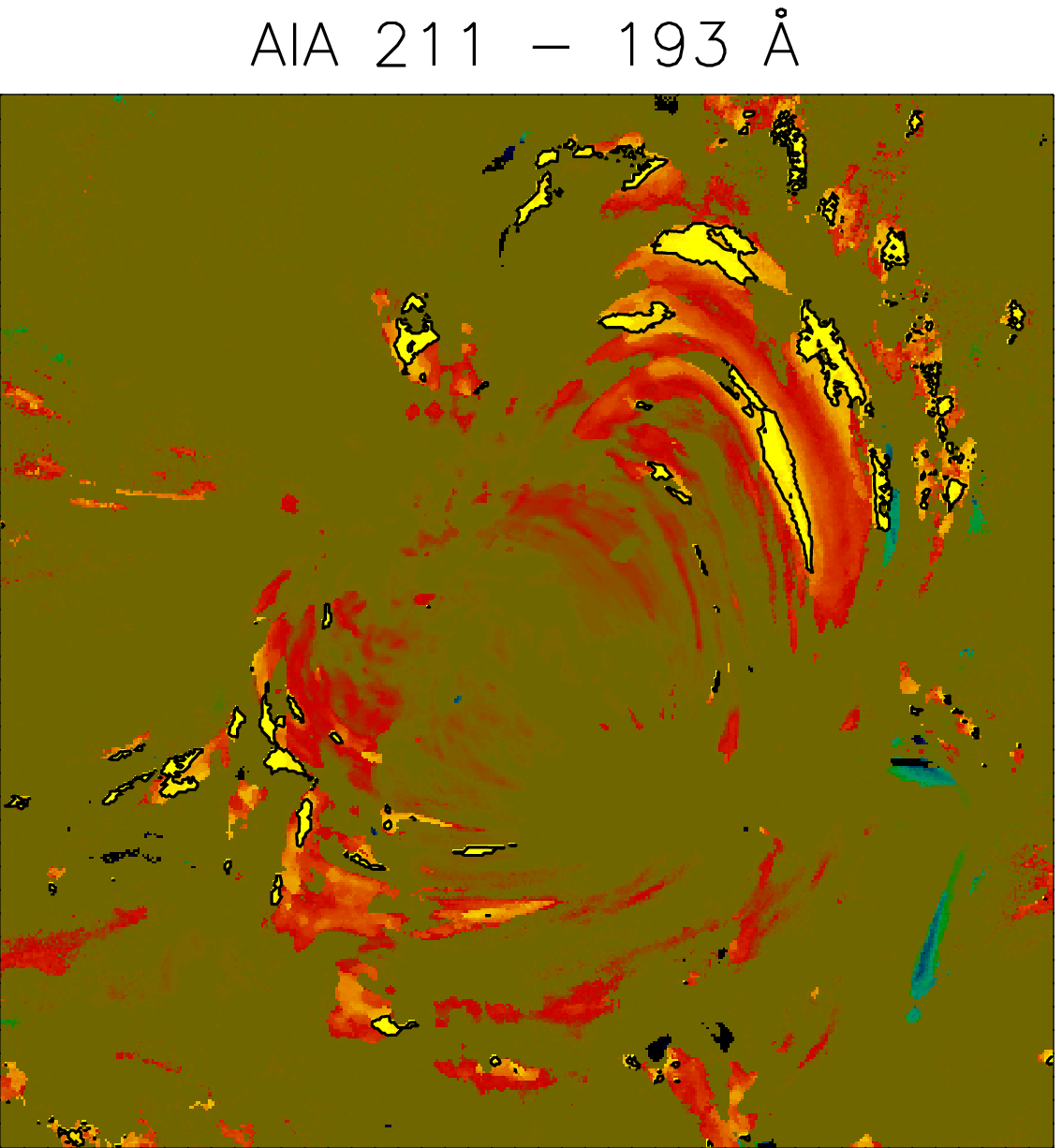}
\includegraphics[width=0.32\textwidth]{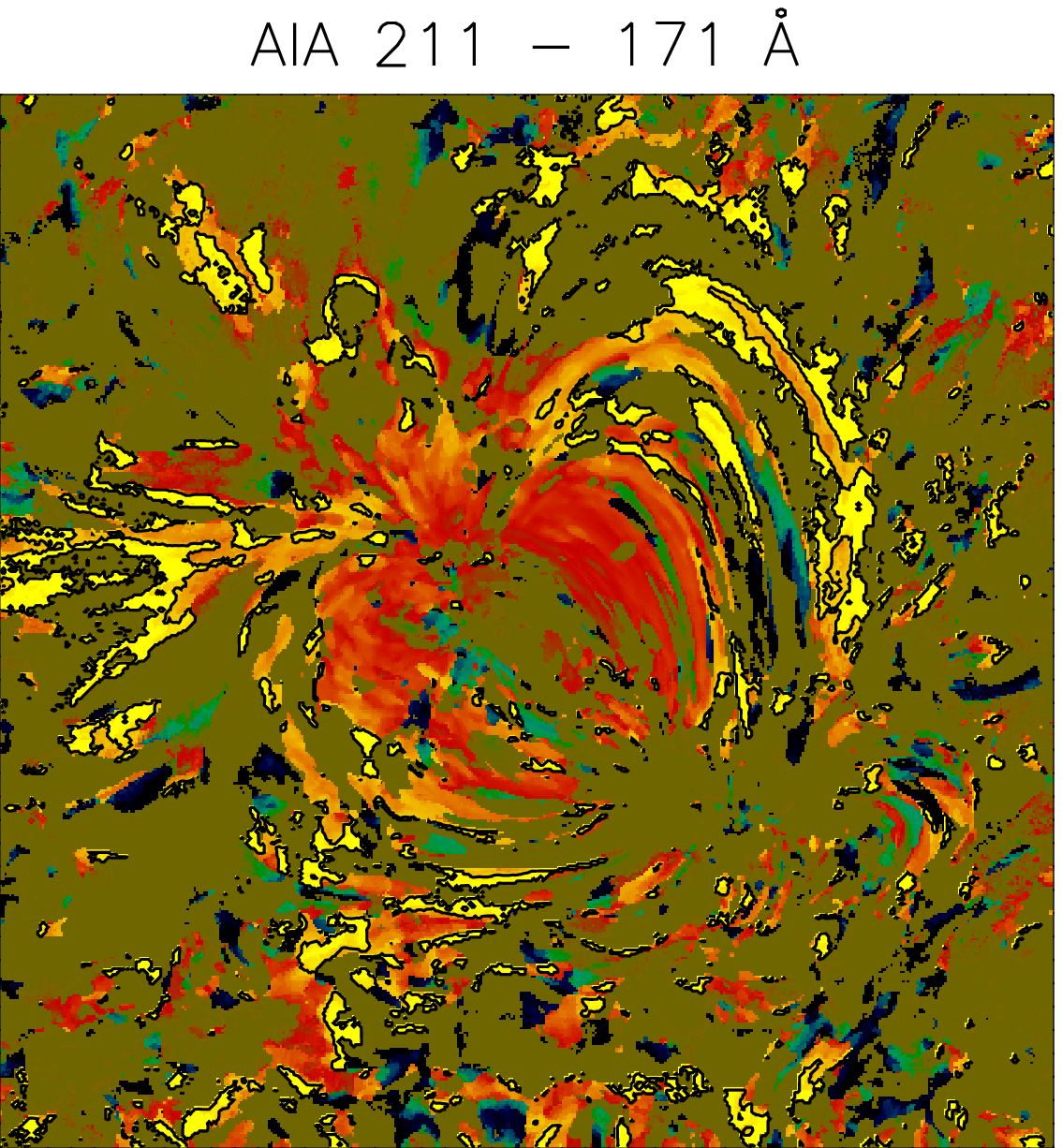}
\includegraphics[width=0.32\textwidth]{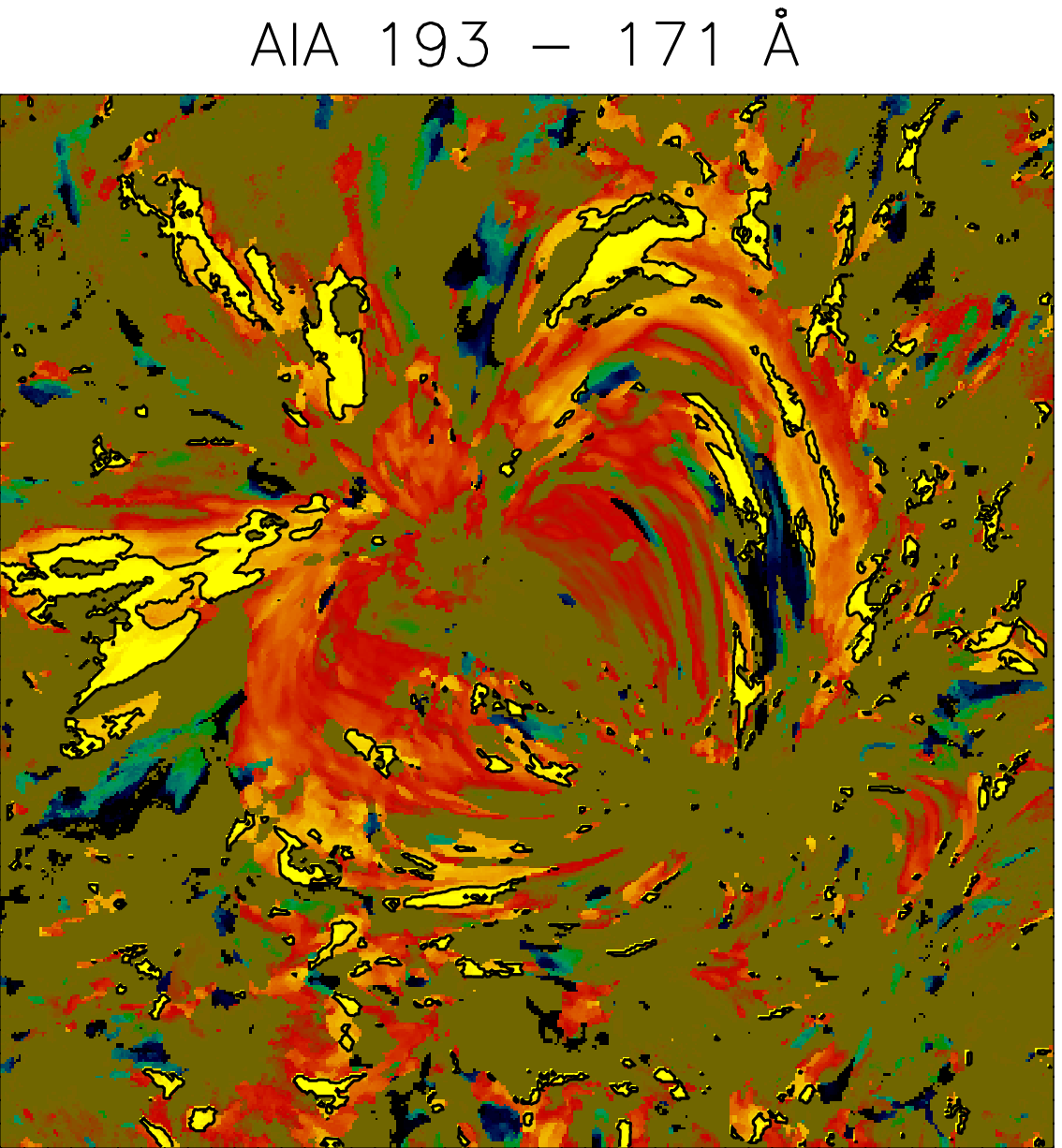}\\
\includegraphics[width=0.5\textwidth]{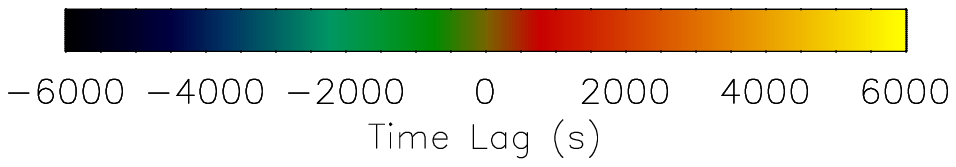}
\caption{Time lag maps from active Region 11082 from a 12 hour window of data 19 June 2010 00:00 - 12:00.  If the maximum correlation coefficient is less than 0.2 or if the time lag ``saturates'' (i.e., the most probable time lag is $\pm$ 7,200), we set the time lag to 0. } 
\label{fig:tl_map} 
\end{center}
\end{figure}

We calculate the time lag maps for all the channel pairs using an aligned data cube from 19 June 2010 00:00 - 12:00 UT following the same analysis method of \citet{2012ApJ...753...35V}.  We allow there to be a maximum of a 2-hour ($\pm$ 7,200\,s) time lag.   These time lag maps are shown in Figure~\ref{fig:tl_map}.   These maps are analogous to Figure 5 in  \citet{2012ApJ...753...35V} with two exceptions.  If the maximum correlation coefficient was less than 0.2, we set the time lag to 0, i.e., the pixel would show as olive green in the images.  (Note that this is similar to the process described in the Appendix of  \citet{2012ApJ...753...35V} and that we use the same limit of correlation coefficient.) Also, if the time lag ``saturated", which is to say,  the most likely time lag was either 7,200\,s or -7,200\,s, we set the time lag to 0.   Finally, we add contours on each map at the level of 5,000\,s.  Each of these channel pairs show pixels with time lags larger than 5,000\,s.  These pixels appear to trace out loops embedded in and at the periphery of the active region.  

To quantify the significance of these long time lags, we first calculate the percentage of pixels that show a positive time lag with a correlation coefficient higher than 0.2.  The percentage of positive pixels are given in Table~\ref{tab:combs}.  Roughly one-third of the pixels in each map show evidence of cooling.  We then calculate the percentage of these pixels where the time lag is greater than 5,000\,s.  These values are given in the last column of Table~\ref{tab:combs} and range from $\sim 3$\% (in the 335-211\,\AA\ channel pair) to $\sim 26$\% (in the 211-171\,\AA\ channel pair.) 

\begin{deluxetable}{cccc|cc}
\tablecaption{Channel pairs examined for time lag calculations}
\tabletypesize{\scriptsize}
\tablewidth{0pt}
\tablehead{
\colhead{Channel } & \colhead{Temp. at } & \colhead{Temp. at } & \colhead{Difference}&\colhead{Percentage} & \colhead{Percentage}\\
\colhead{Pair} & \colhead{Peak Response} & \colhead{Peak Response} & \colhead{ [MK]} & \colhead{Pixels $>$ 0\,s} &  \colhead{Pos. Pixels $>$ 5,000\,s}\\
\colhead{[Ch. 1 - Ch. 2]}& \colhead{Channel 1 [MK]} & \colhead{Channel 2 [MK]}
} 
\startdata
335 \AA\ - 211 \AA\  & 2.5$^{*}$ & 2.0 & 0.5$^{*}$ & 35.2\% & 3.4\% \\
335 \AA\ - 193 \AA\ & 2.5$^{*}$ & 1.5 & 1.0$^{*}$ & 35.1\% & 14.4\% \\
335 \AA\ - 171 \AA\  & 2.5$^{*}$ & 1.0 & 1.5$^{*}$ & 38.3\% & 11.9\% \\
211 \AA\ - 193 \AA\  & 2.0 & 1.5 & 0.5 & 31.9\% & 7.2\% \\
211 \AA\ - 171 \AA\  & 2.0 & 1.0 & 1.0 & 35.8\% & 26.3\% \\
193 \AA\ - 171 \AA\ & 1.5 & 1.0 & 0.5 & 43.5\% & 15.9\%\\
\enddata
\tablecomments{Characteristic temperatures of the pairs of EUV channels used in the time lag analysis. The difference in temperature between each combination is also shown to indicate which pairings should show the largest time lag between emission peaks. $^{*}$The 335\,\AA\ channel has a broader response curve than the other channels.}
\label{tab:combs}
\end{deluxetable}

\begin{figure}
\begin{center}
\includegraphics[width=0.29\textwidth]{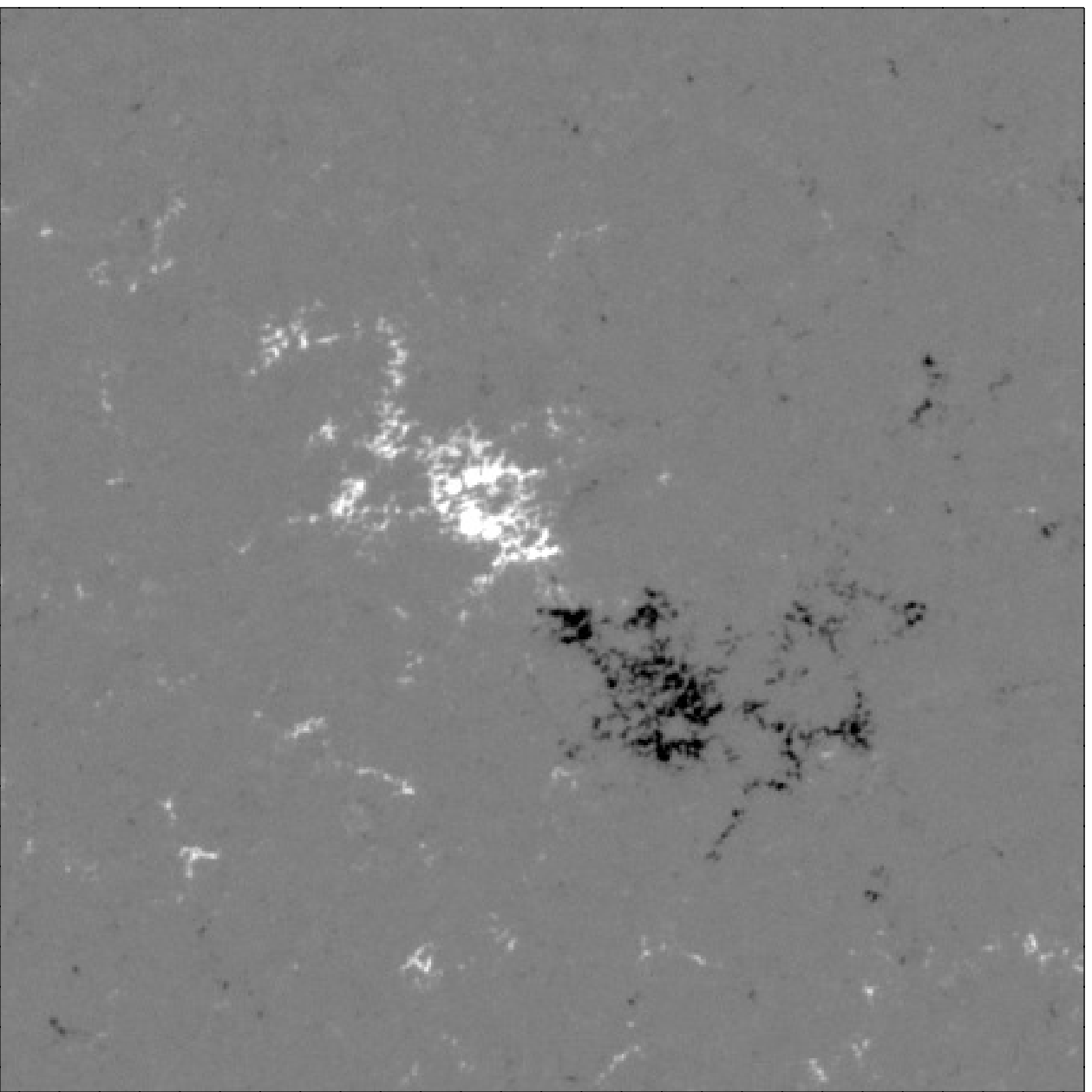}
\includegraphics[width=0.29\textwidth]{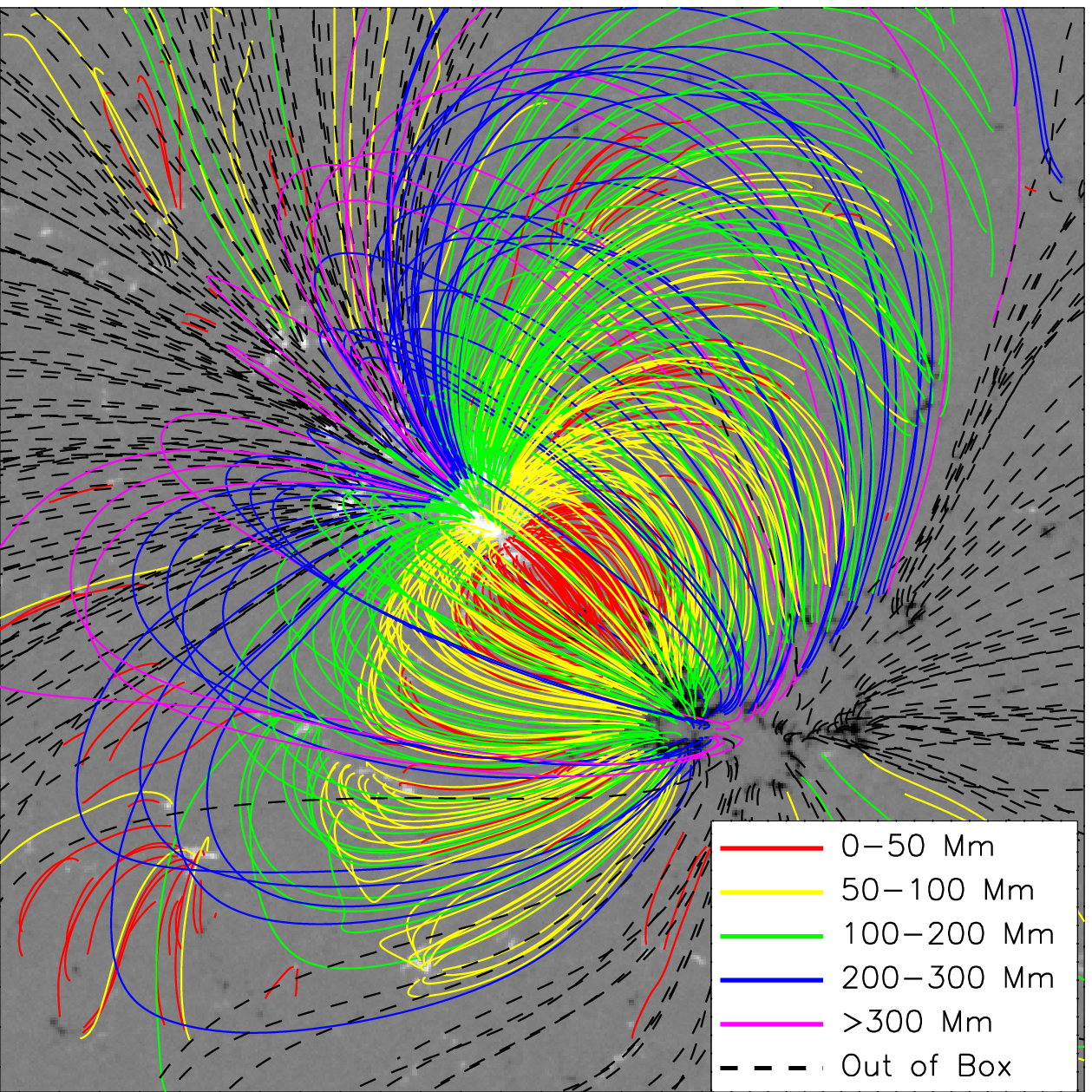}
\includegraphics[width=0.4\textwidth]{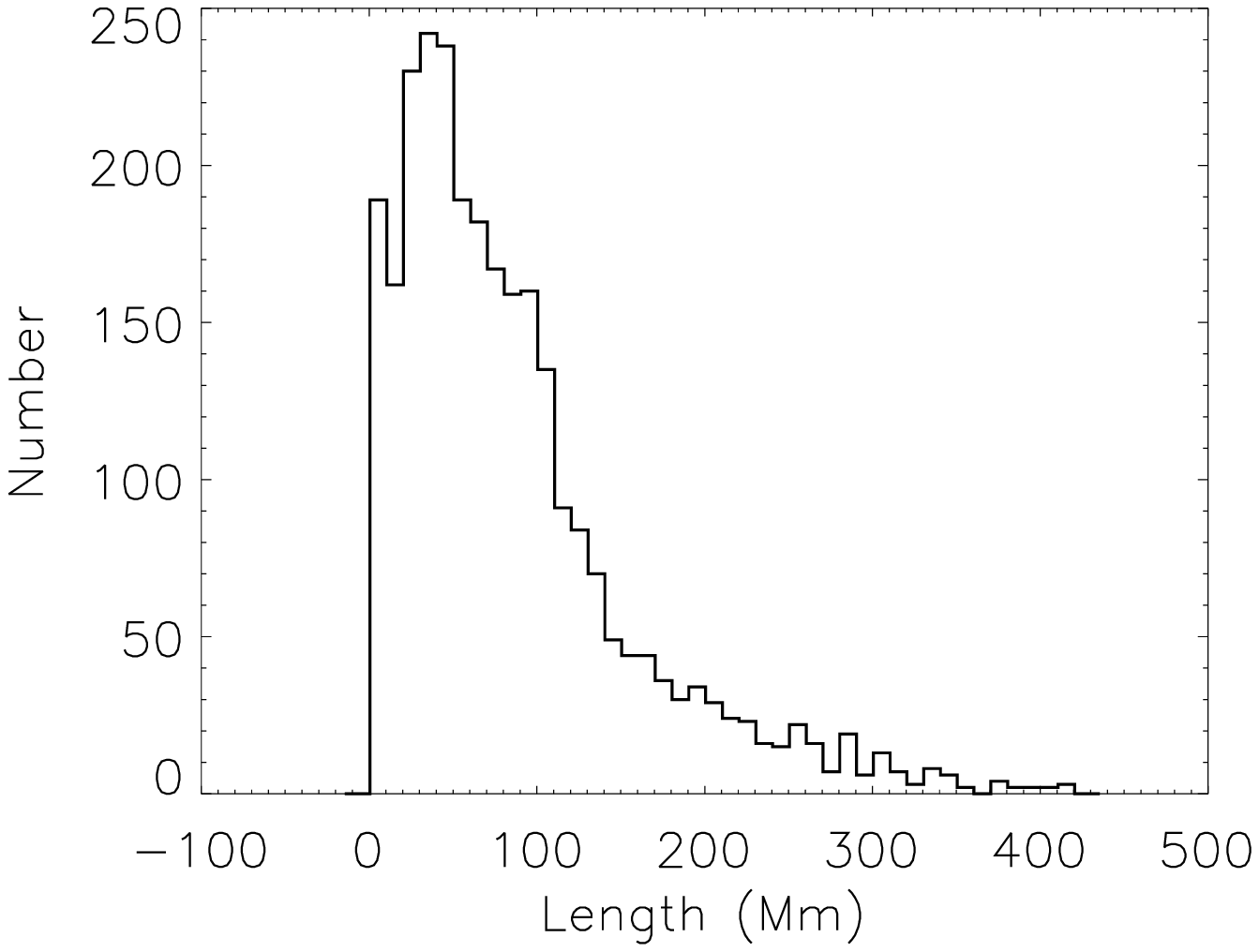}
\caption{Line-of-sight magnetic field measured at 19 June 2010 06:00 in Active Region 11802 (left) and with example field lines drawn from field strength 50-500\,G (middle).  A histogram of the loop lengths in this active region (right).} 
\label{fig:mag} 
\end{center}
\end{figure}

How the coronal plasma cools is highly dependent on the loop length. Therefore we approximate the loop lengths in this active region using a potential field extrapolation of the photospheric field. First, we start with the full Sun photospheric magnetic field measurements from the Solar Dynamic Observatory's Helioseismic and Magnetic Imager (HMI; \citealt{2012SoPh..275..207S}) observed at 19 June 2010 06:00 (approximately half-way through the AIA time series.)  The left panel of Figure~\ref{fig:mag} shows the HMI line-of-sight magnetogram of this active region.  We extract the magnetic field around the active region so that the pixels are approximately square.  After correcting the magnetic field, we numerically solve
the equations $\mathbf{\nabla \times B} = 0 $ and $\mathbf{\nabla \cdot B} = 0$ to determine the
magnetic field vectors in the volume above the photospheric
field \citep[e.g.,][]{1989ApJS...69..323G}. We then trace field lines from pixels where magnetic field strength is greater than 50\,G and less than 500\,G.  The upper limit is defined to avoid sunspots which are known to be faint in EUV and X-ray emission \citep[e.g.,][]{1997soco.book.....G}. A subset of these field lines is  shown in the middle panel of Figure~\ref{fig:mag}; the lines are color-coded to indicate their length.  Field lines that leave the computational box are shown with black, dashed lines.  A histogram of the lengths of these field lines that terminate in the computational domain is also given in Figure~\ref{fig:mag}.  The majority of field lines are less than 100 Mm, while longest field lines are $\sim$ 400 Mm.  

A comparison of the time lag maps (Figure~\ref{fig:tl_map}) and magnetic field extrapolation (Figure~\ref{fig:mag}) reveals that some of the large time lags originate from ``closed'' structures in the active region core (i.e., with loop lengths less than 400 Mm).   Additionally, these large time lags can be seen in the ``open'' structures, or fans, at the edge of the active region in the 211-171\,\AA\ and 193-171\,\AA\ channel pairs.  It is difficult to assess what percentage of the large time lag structures are closed and open in these two channel pairs, but we estimate roughly half the pixels are in the open structures.  These structures are not modeled as part of this work.

\section{DESCRIPTION OF SIMULATIONS}

To calculate the evolution of these loops we rely on two different hydrodynamic codes, the Naval Research Laboratory's Solar Flux Tube Model (NRLSOLFTM; \citealt{1982ApJ...255..783M}) and the Predictive Science One-dimensional Hydrodynamic Model (PSOHM; \citealt{2013ApJ...773...94M}).   NRLSOLFTM will be used to address the dependence of time lag on the heating magnitude, heating duration, loop length,  and abundances.  PSOHM will be used to investigate how area expansion impacts the time lag, which is not possible with the NRLSOLFTM.  We set up both models with many of the same parameters and initial conditions.  

Both models solve the following 1D
hydrodynamic equations along the length of a loop:

\begin{equation}
\frac{\partial \rho}{\partial t}
+\frac{1}{A}\frac{\partial}{\partial s}\left(A\rho v\right)=0, 
\label{eq-mass}
\end{equation}
\begin{equation}
\rho\left(\frac{\partial v}{\partial t}+v\frac{\partial v}{\partial s}\right)= 
 -\frac{\partial p}{\partial s}+\rho g_s 
 +\frac{1}{A}\frac{\partial}{\partial s}\left(A\nu\rho
\frac{\partial v}{\partial s}\right),
\label{eq-mom}
\end{equation}
\begin{multline}
\frac{\partial T}{\partial t}
+\frac{1}{A}\frac{\partial}{\partial s}\left(ATv\right)=
-(\gamma-2)\frac{T}{A}\frac{\partial}{\partial s}\left(Av\right)+ \\
\frac{(\gamma-1)T}{p}
\left[\frac{1}{A}\frac{\partial}{\partial s}\left(A
\kappa_\parallel\frac{\partial T}{\partial s}\right)
-n_en_p{Q(T)}+{H}\right],
\label{eq-energy}
\end{multline}
where $s$ is the length along the loop, $T$, $p$, and $v$ are
the plasma temperature, pressure, and velocity, $n_e$ and $n_p$
are the electron and proton number density (assumed to be
equal), and $\gamma=5/3$ is the
ratio of specific heats. The mass density for the        plasma
is $\rho=\frac{1+4f_\mathrm{He}}{1+2f_\mathrm{He}}m_pn_e$, where $m_p$ is the proton mass and $f_\mathrm{He}$ is the fraction of helium to hydrogen particles.
The plasma pressure is $p=\frac{2+3f_\mathrm{He}}{1+2f_\mathrm{He}}kn_eT$; $k$ is Boltzmann's constant.
The gravitational acceleration projected
along the loop is $g_s=\bhat\cdot{\bf g}$, where $\bhat$ is the
unit vector along the magnetic field ${\bf B}$. The Spitzer thermal conductivity coefficient is   $\kappa_\parallel(T)=C_0
T^{5/2}\,\text{[erg/cm/s/K]}$.


$H(s,t)$ is the coronal heating function.  In this investigation, the heating function will be
\begin{equation}
H(s,t) = H_0 + H_{\rm imp}g(t)
\end{equation}
where $H_0$ is a small background heating, applied uniformly and constantly, $H_{\rm imp}$ is the magnitude of the
impulsive heating that is also uniformly applied along the loop over a short time.  The time dependence of the impulsive heating event, $g(t)$, is a triangular pulse
defined as 
\begin{eqnarray}
g(t) = t/\delta  \hspace{0.5in} 0&<t\le&\delta \\
g(t) = (2\delta - t)/\delta \hspace{0.5in}\delta &< t\le&2\delta
\end{eqnarray}
where $\delta$ is the total duration of the impulsive heating.

In the PSOHM, a uniform
kinematic viscosity $\nu$ is used to damp out waves, and is made
very small, corresponding to a diffusion time $L^2/\nu \sim
2{,}000\,\text{hours}$; this term is absent in NRLSOLFTM. 
When considering solutions with loop expansion, the loop
cross-sectional area, $A(s)$, is inversely proportional to the
magnitude of the magnetic field $B(s)$ measured along the loop.

The small values of $\kappa_\parallel$ at
low temperatures produce very steep gradients in temperature in
the transition region that are difficult to resolve in numerical
calculations. In NRLSOLFTM, a fine mesh is centered on the steepest temperature
gradient.   The location of this fine mesh is re-evaluated at each time step.  
In  PSOHM, in order to minimize the mesh resolution
requirements, we have developed special treatment of the thermal
conduction and radiation loss function to artificially broaden
the transition region at low temperatures
\citep{2009ApJ...690..902L} without significantly affecting the
coronal solution \citep{2013ApJ...773...94M}. 

\begin{figure}
\begin{center}
\includegraphics[width=8cm]{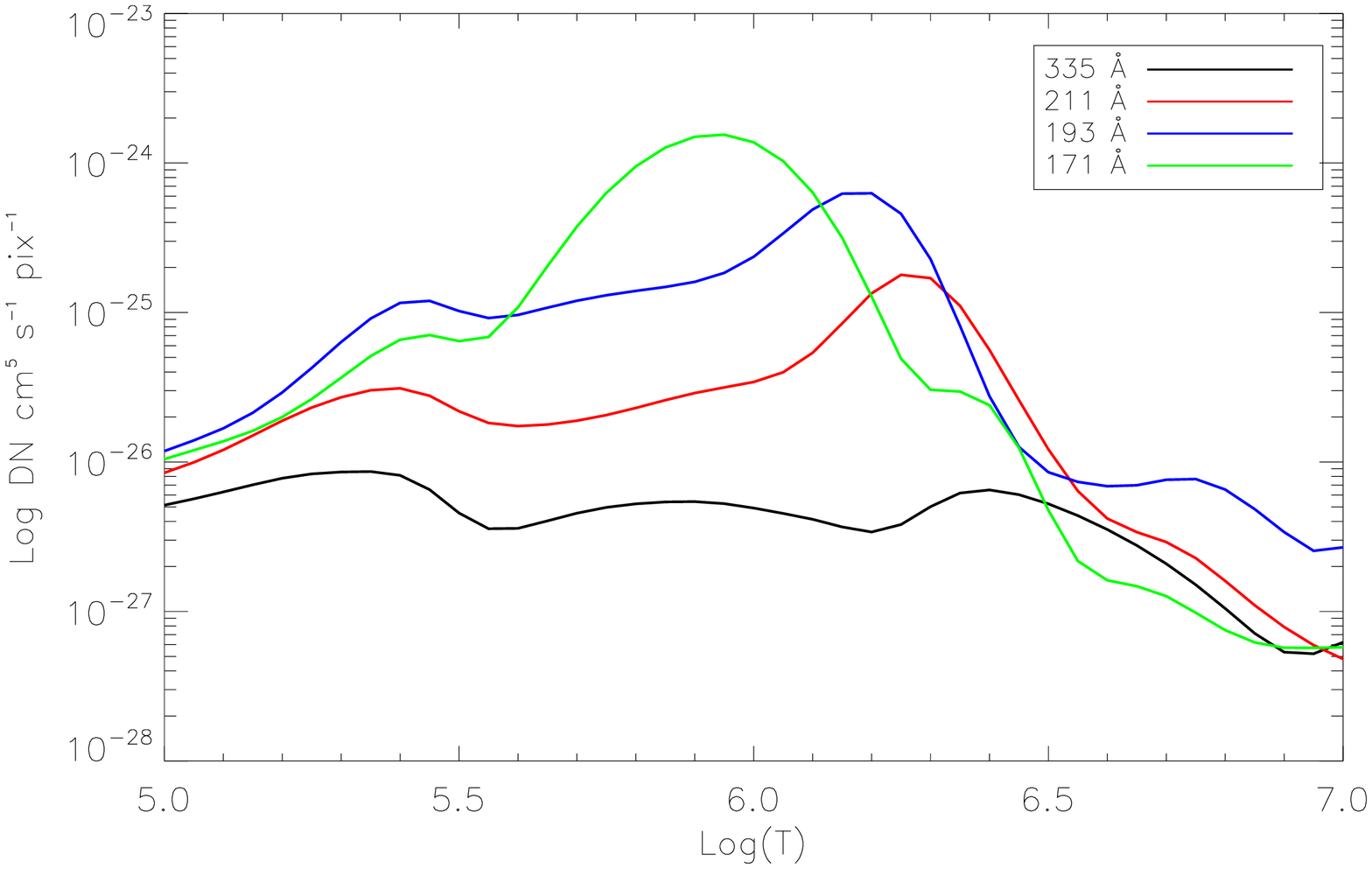}
\includegraphics[width=8cm]{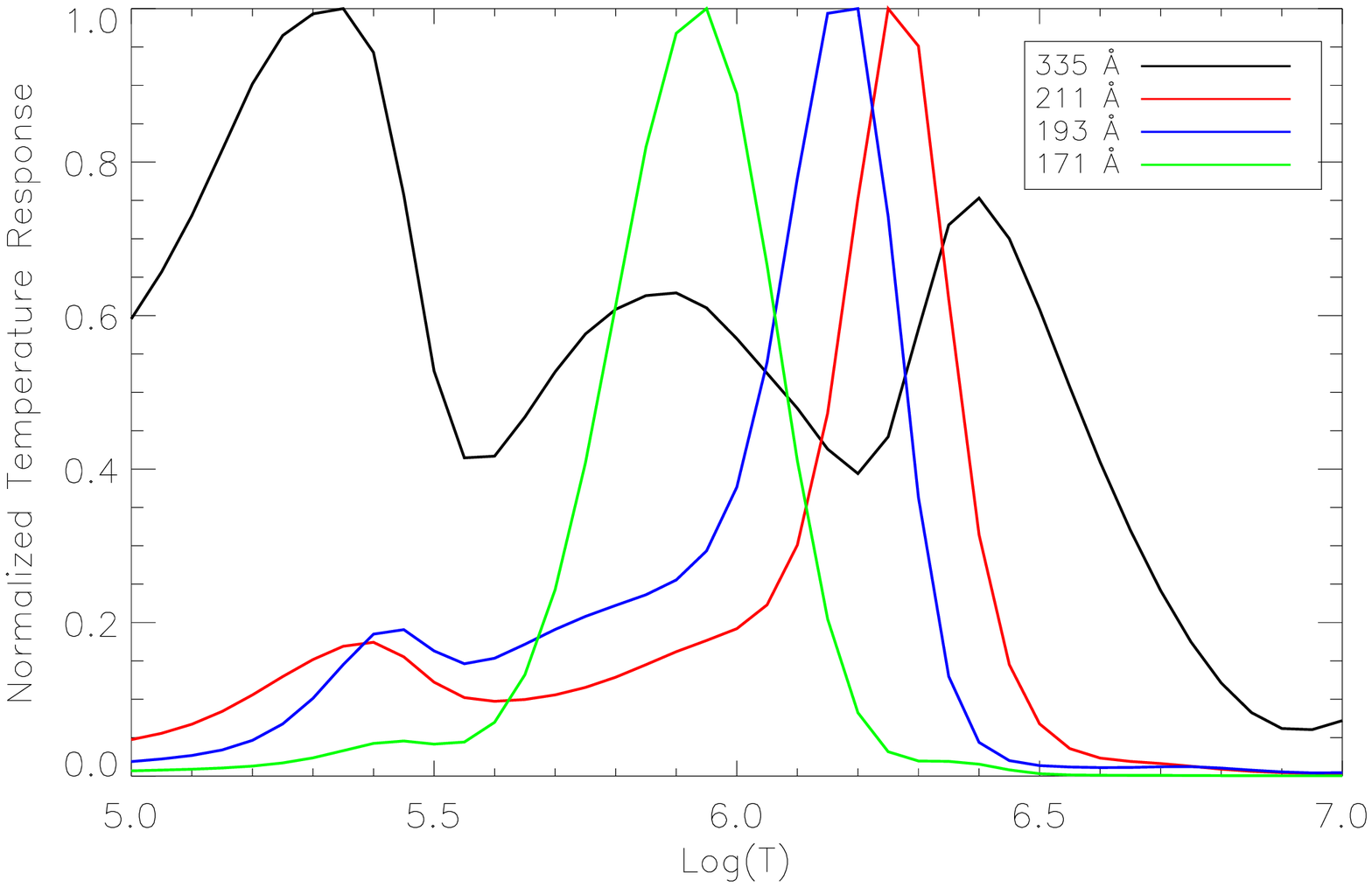}
\includegraphics[width=8cm]{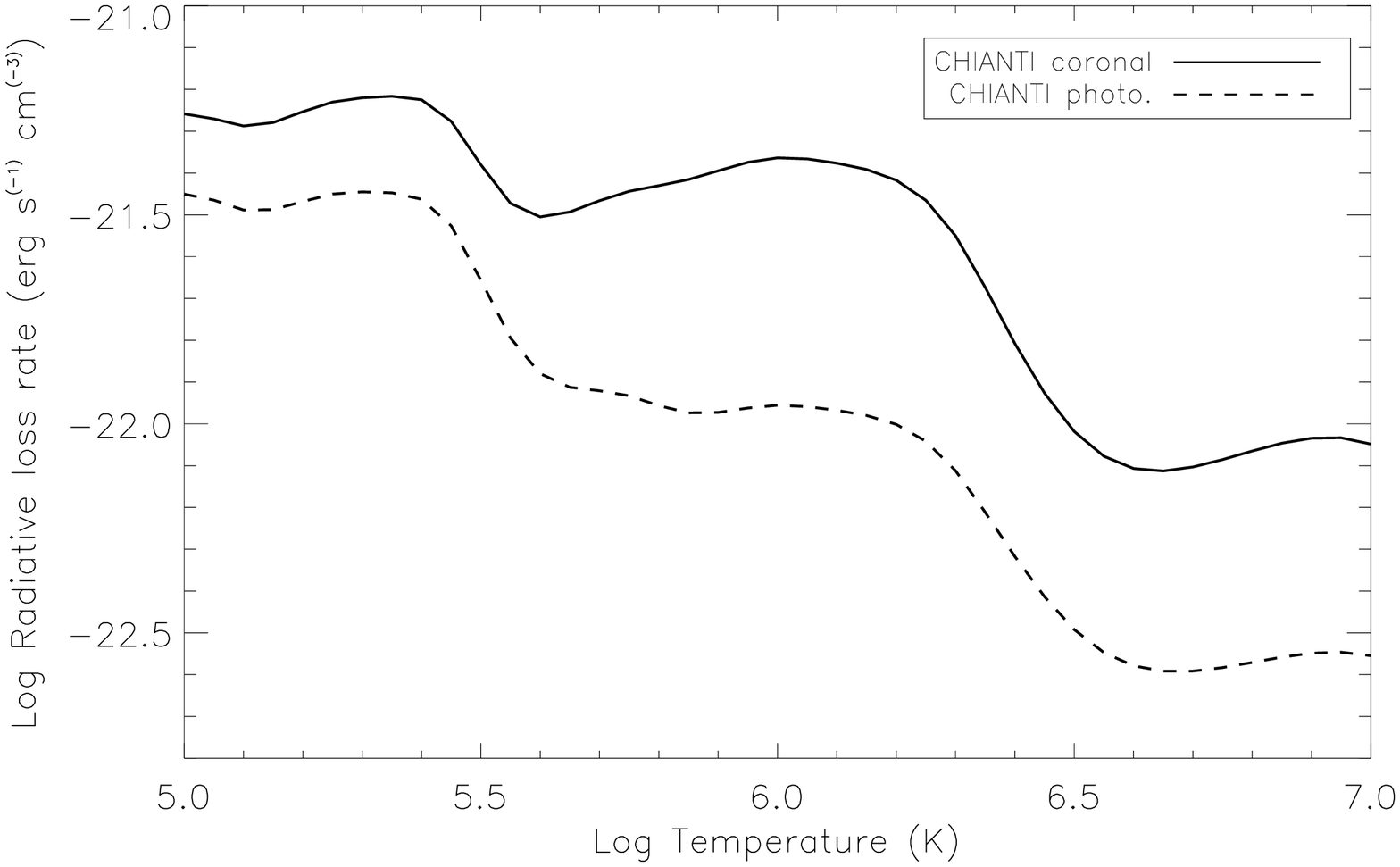}
\caption{Top: The temperature response curves of the four AIA channels used in this analysis plotted on a logarithmic scale and again with the responses normalized in order to highlight the breadth of the 335\,\AA\ response. Bottom: radiative
loss functions calculated using CHIANTI 7.1 with coronal or photospheric abundances.} 
\label{fig_tresp_abun} 
\end{center}
\end{figure}

In the above equations, $Q(T)$ represents the radiative loss function.  
We consider two
versions of the radiative loss function derived from the CHIANTI (version~7.1) 
atomic database
\citep{1997A&AS..125..149D,2013ApJ...763...86L}, one with coronal abundances
\citep[\texttt{sun\_coronal\_ext.abund},][]{1992ApJS...81..387F,2002ApJS..139..281L,1998SSRv...85..161G}
the other with photospheric abundances 
\citep[\texttt{sun\_photospheric\_2007\_grevesse.abund},][]{2007SSRv..130..105G},
both
using the CHIANTI collisional ionization equilibrium model \citep{2009A&A...498..915D}.
 These two radiative loss functions are shown in the lower panel of Figure \ref{fig_tresp_abun}.

All the loops will be semi-circular and perpendicular to the solar surface.   
For both models, we start the loops in a cool, tenuous steady state atmosphere with a transition region temperature.  Both models include a dense, cool chromosphere as a mass well for the solution.  Radiation is turned off at chromospheric temperatures.  

Finally, we have run a test case to compare the output of the two codes.  With identical loop length, heating functions, geometries, and radiative loss functions, we find near identical ($\pm$5\%) time lags in 
all channel pairs, even with different initial conditions, boundary conditions, and other simulation parameters such as $f_\mathrm{He}$ and $C_0$.

\section{AN EXAMPLE SIMULATION}

\begin{figure}
\begin{center}\includegraphics[width=0.48\textwidth]{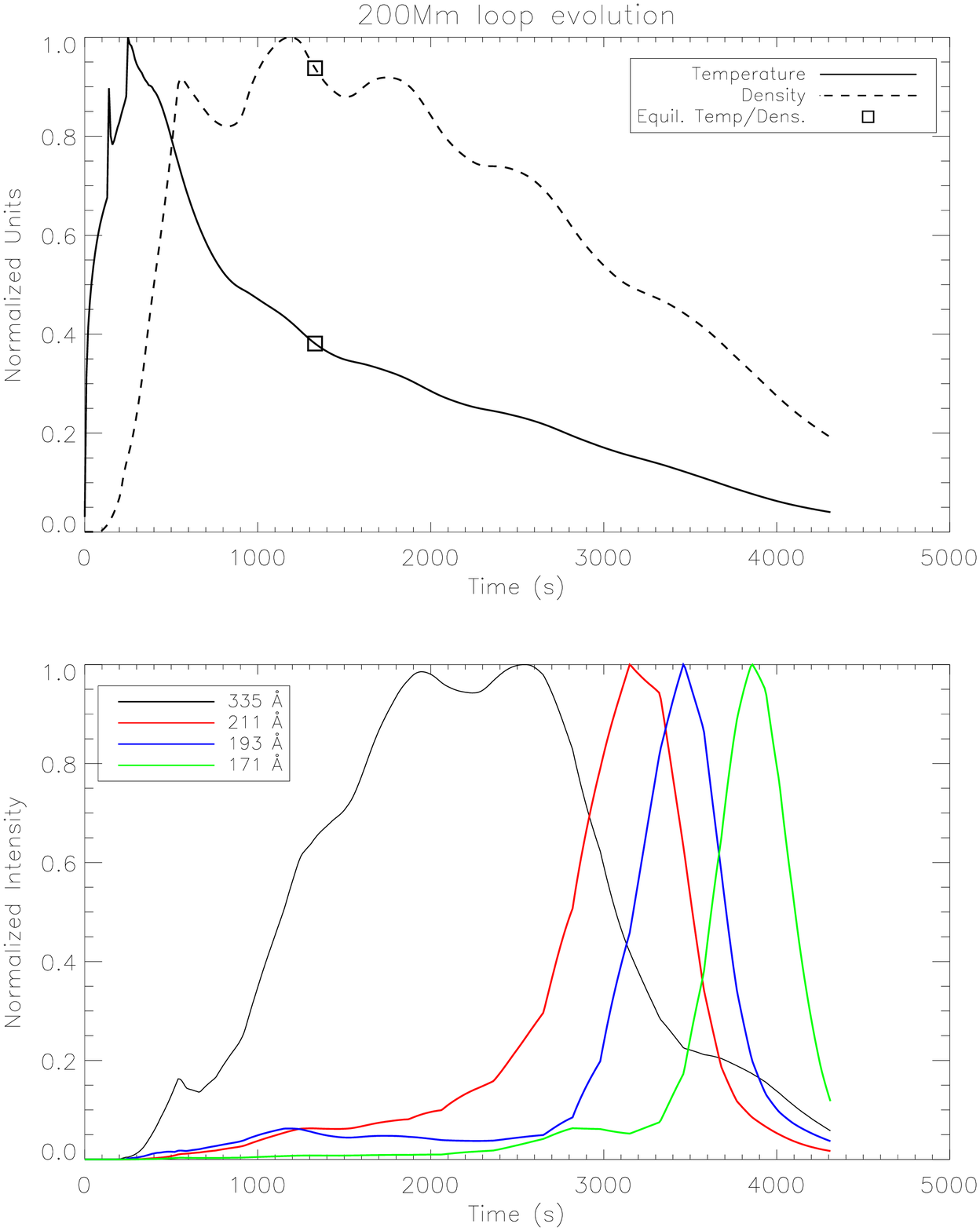}\\
\includegraphics[width=.48\textwidth]{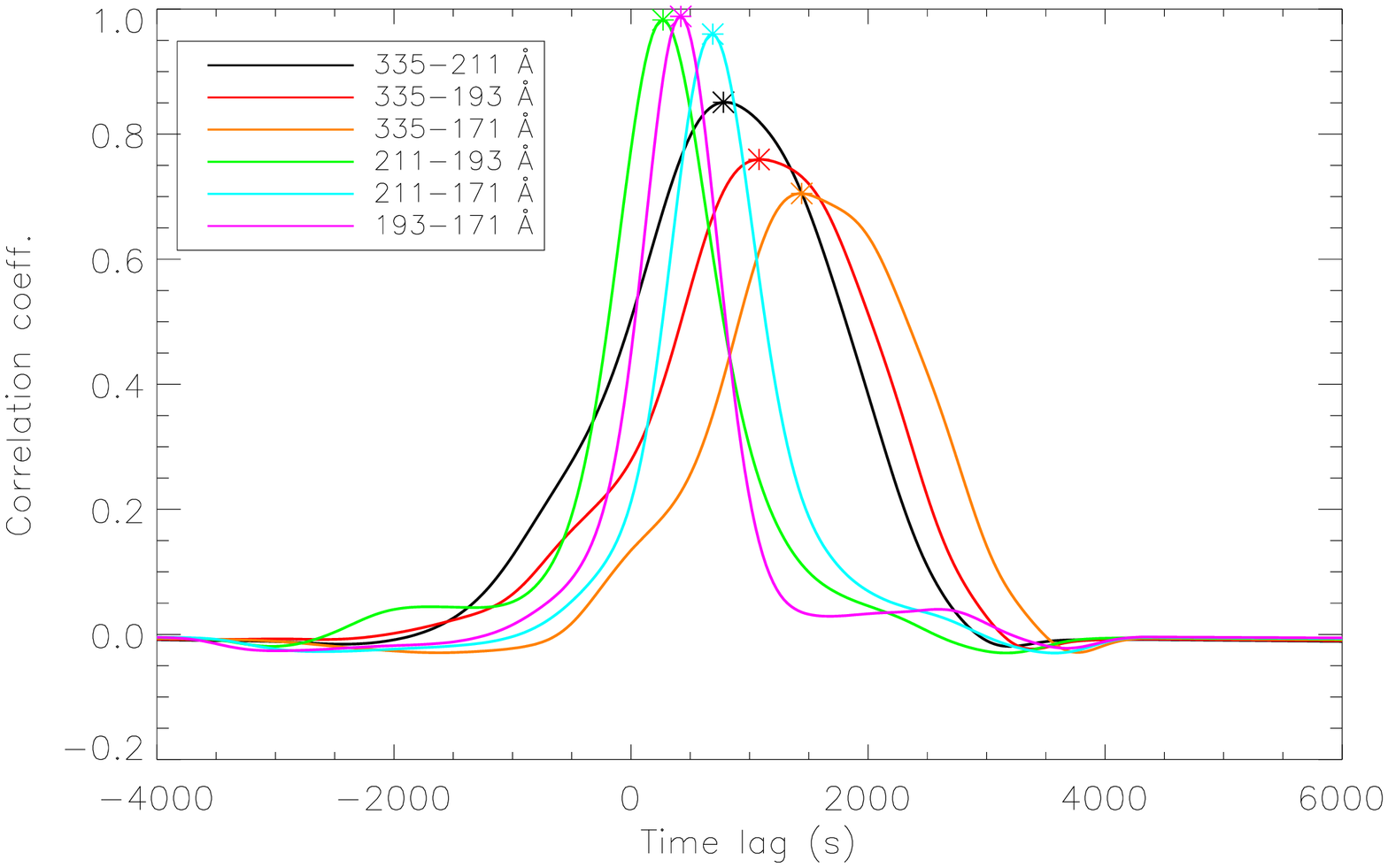}
\caption{Top: The average apex temperature and density evolution of the example loop. The temperature is normalized by 12.9 MK and the density is normalized by 2.5$\times 10^9$ cm$^{-3}$.  The equilibrium temperature and density are shown with a square symbol.  Middle: Normalized light curves in the AIA 335, 211, 193, and 171\,\AA\ channels. Bottom: The correlation coefficient as a function of time lag for the six channel pairs.  The asterisk marks the time lag at the peak of each curve.} 
\label{fig:tempdenslc}
\end{center} 
\end{figure}

In this section, we demonstrate the analysis with an example simulation.  We use a 200 Mm loop with constant cross section.  The magnitude of the impulsive heating event is 0.03 ergs cm$^{-3}$ s$^{-1}$ with the heating duration of
500\,s.  We use the coronal radiative loss function.   We calculate the resulting temperature and density evolution of the loop using NRLSOLFTM.   We then average the density and temperature over the upper 50\% of the coronal part of the loop.  The evolution of this average apex temperature and density is shown in the top panel of Figure~\ref{fig:tempdenslc}.

Often, the maximum temperature a simulation reaches is used to characterize the simulation.  Of course, because the density is so low during this initial, high temperature phase, the likelihood of detecting its maximum temperature is small.  Additionally, the duration of the heating significantly impacts the maximum temperature; the same amount of total energy deposited in the loop over a shorter duration would result in a higher maximum temperature, while if it were deposited over a longer duration, it would result in a lower maximum temperature \citep{2004ApJ...610L.129W}.  For these reasons, we prefer to use the so-called ``equilibrium temperature'' to characterize the solution.  The equilibrium temperature is the temperature at which the density and temperature in the loop are the same as expected for hydrostatic equilibrium.  It is also roughly the time when the cooling changes from being dominated by conduction to dominated by radiation.  Finally, simulations with the same total energy deposited will share the same equilibrium temperature, as long as the heating duration is ``short,'' e.g., less than the sound crossing time of the loop.  See \cite{2004ApJ...610L.129W} for a complete description of equilibrium temperature.  The equilibrium temperature and density are shown with a square in Figure~\ref{fig:tempdenslc}.  The equilibrium temperature for this simulation is 4.9 MK.

Using these temperatures and densities, we calculate the expected light curve in the AIA 335, 211, 193, and 171\,\AA\ channels with
\begin{equation}
I_{\rm ch}(t) = n(t)^2 R_{\rm ch}(T(t)) ds
\end{equation}
where $R_{\rm ch}$ is the response function of a channel (see Figure~\ref{fig_tresp_abun}) and $ds$ is the depth of the loop.   Previous studies have found that loops are 1-2 Mm wide \citep{2005ApJ...634L.193A, 2008ApJ...680.1477A}.  We use 1\,Mm as the depth in this study.  The calculated light curves  are shown in the middle panel of Figure~\ref{fig:tempdenslc}.  

It is interesting to note that the light curves for the 211, 193, and 171\,\AA\ channels have fairly narrow peaks,  which are reached after the equilibrium temperature occurs, in the phase during which the loop is dominated by radiative cooling.  The 335\,\AA\ light curve, however, has a broad peak, with the light curve rising before the equilibrium temperature is reached.  Because of this, the rise of the 335\,\AA\ light curve may depend on heating duration.  We investigate and discuss this further in Section~\ref{sec:hm}.

Finally, we use the IDL C\_CORRELATE.PRO function to determine the time lag with the highest correlation coefficient between the different channel pairs.  As in \citet{2012ApJ...753...35V}, we investigate time lags up to $\pm$ 2 hours.  The correlation coefficient as a function of time lag is shown in the bottom panel of Figure~\ref{fig:tempdenslc}.  The asterisk highlights the time lag where the correlation coefficient is maximum.   Note that the correlation coefficient functions for the channel pairs that include 335\,\AA\ channel tend to be broader than the other channel pairs. The time lags associated with the peak of the correlation coefficients  are given in Table~\ref{tab:example}.   These time lags range between 270 - 1,440\,s.  We complete identical analysis for each loop simulated in this study.  

\begin{deluxetable}{rcc}
\tablecaption{Time lags calculated for example simulation}
\tabletypesize{\scriptsize}
\tablewidth{0pt}
\tablehead{
\colhead{Channel Pair} & \colhead{Time lag (s)} & \colhead{Correlation Coefficient}}
\startdata
335-211\,\AA\ &      780 & 0.85 \\
335-193\,\AA &     1,080 &  0.76\\
335-171\,\AA\ &      1,440 &  0.71 \\
211-193\,\AA\  &      270 &  0.98 \\
211-171\,\AA\ &      690 &  0.96\\
193-171\,\AA\ &      420 &  0.99\\
\enddata
\label{tab:example}
\end{deluxetable}

\section{RESULTS}

In the following sections, we investigate the effect varying different parameters has on the time lags.  

\subsection{Heating Magnitude}
\label{sec:hm}

In this section, we investigate how the time lag in the channel pairs depends on the magnitude of the heating.   To investigate this parameter, we use a 200\,Mm loop with constant cross section and coronal radiative loss function. We calculate the solutions with NRLSOLFTM.  The duration of the heating is 500\,s for all simulations.  We vary the magnitude of the heating, $H_{\rm imp}$, from 0.002 to 0.1 ergs cm$^{-3}$ s$^{-1}$.  

\begin{figure}
\begin{center}
\includegraphics[width=0.48\textwidth]{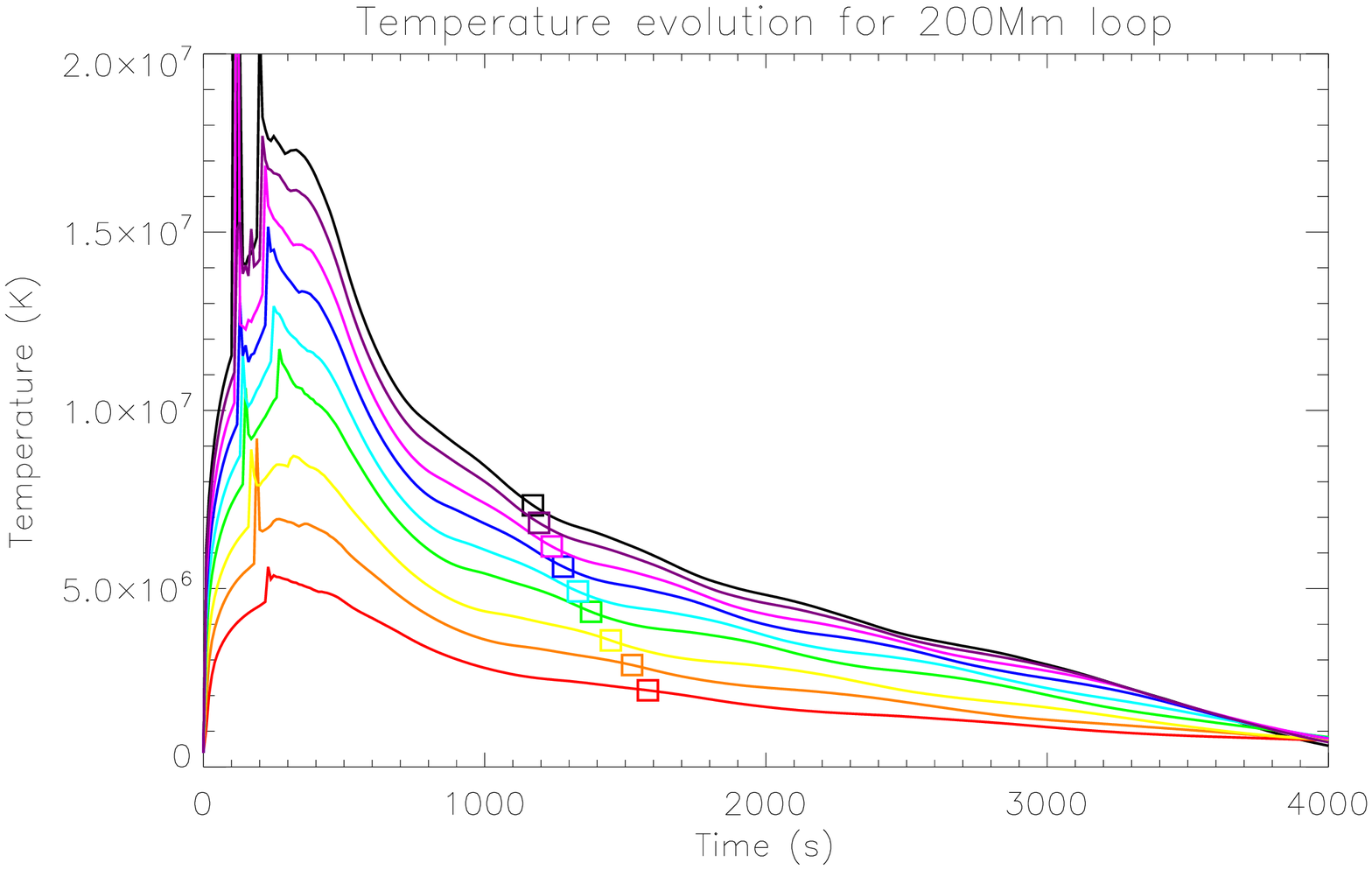}
\includegraphics[width=0.48\textwidth]{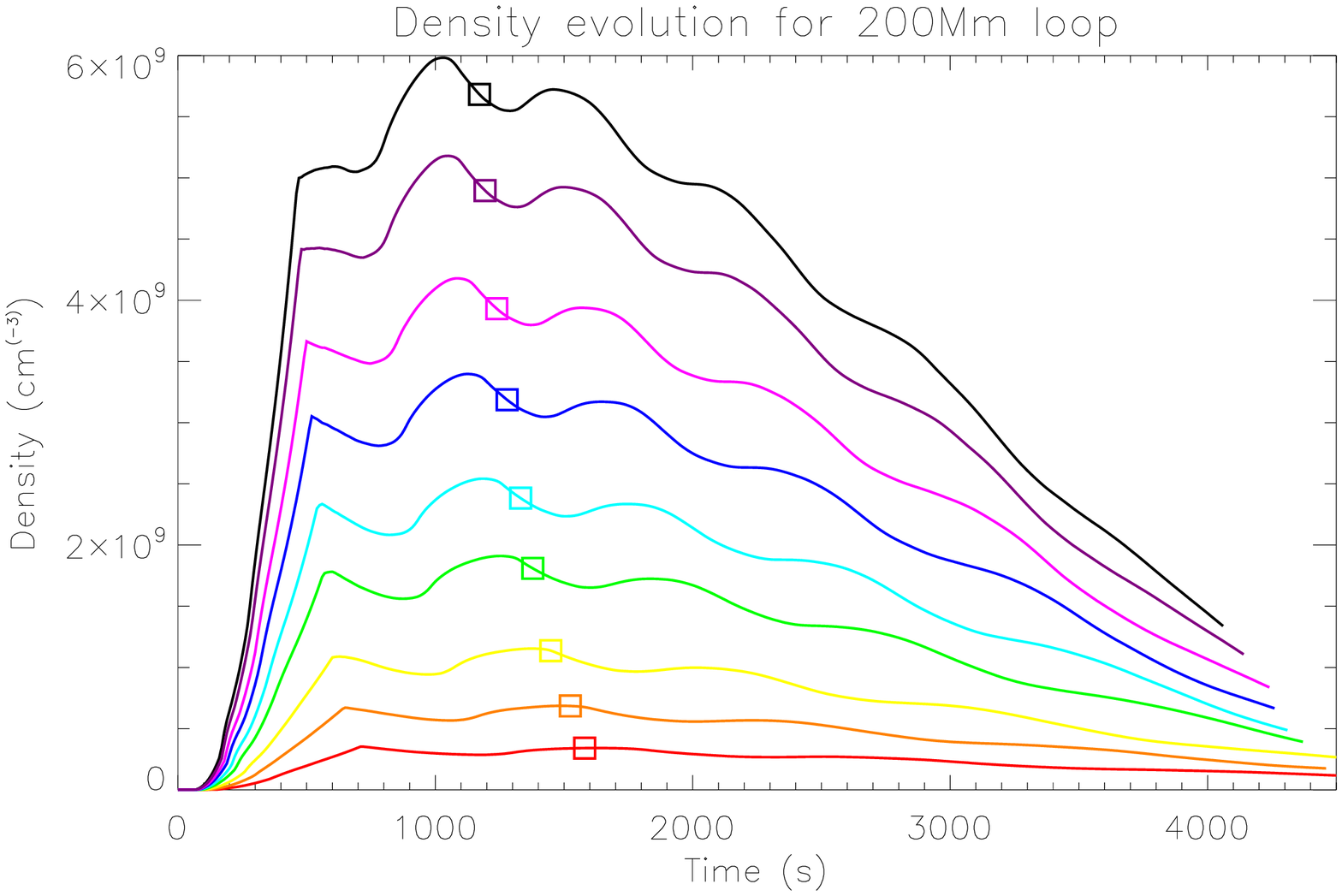}
\caption{The average apex temperature (left) and density evolution of a series of simulations for 200\,Mm loops. The square symbols represent the equilibrium temperature and density reached in each case.} 
\label{fig:energy_tn}
\end{center} 
\end{figure}

Figure~\ref{fig:energy_tn} shows the evolution of the average apex temperature and density for each of these solutions.    We calculate the equilibrium temperature and density for each simulation, these are shown as squares on Figure~\ref{fig:energy_tn}.  The higher equilibrium temperatures correspond to the higher heating rates.

\begin{figure}
\begin{center}
\includegraphics[width=0.32\textwidth]{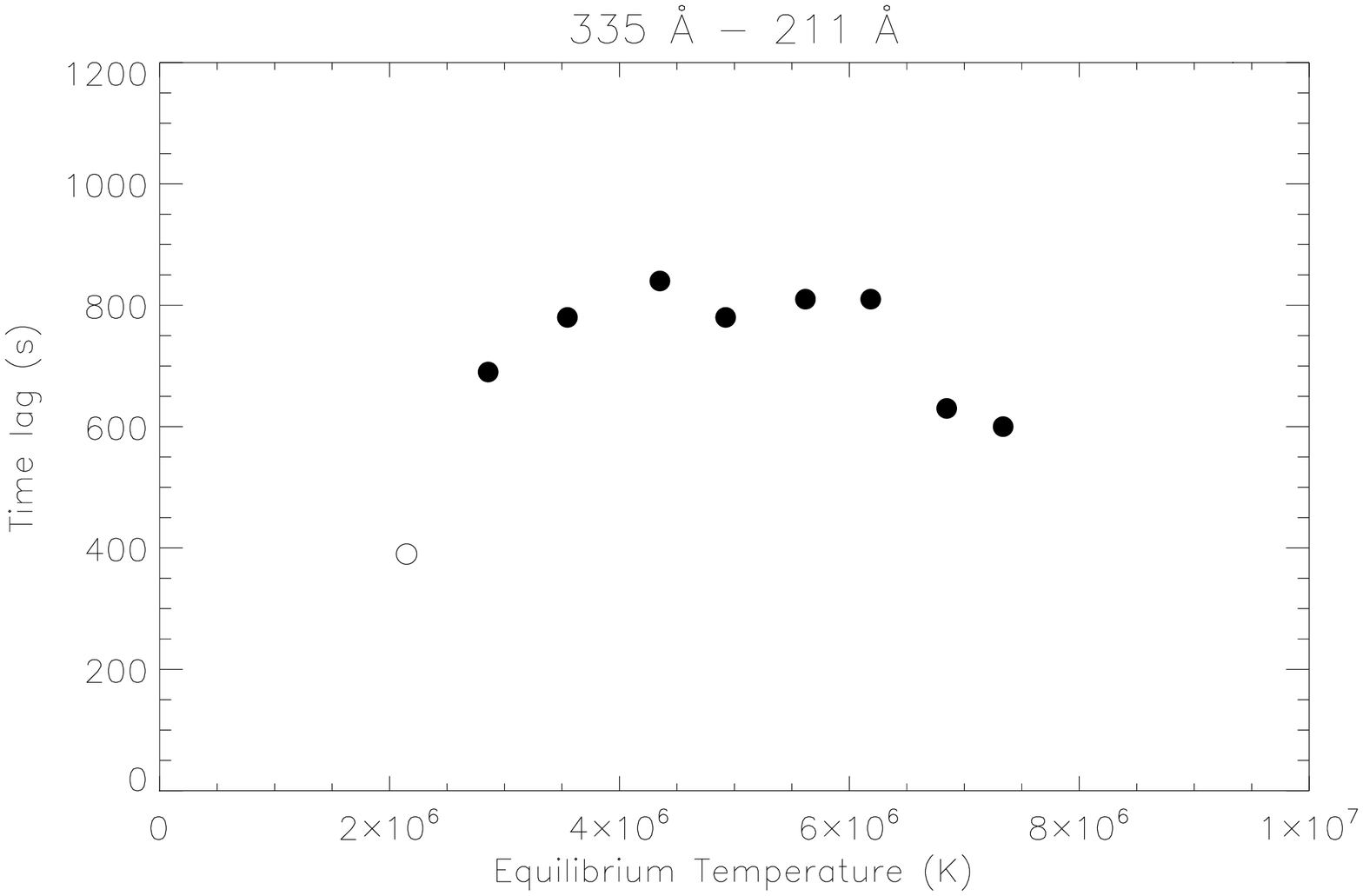}
\includegraphics[width=0.32\textwidth]{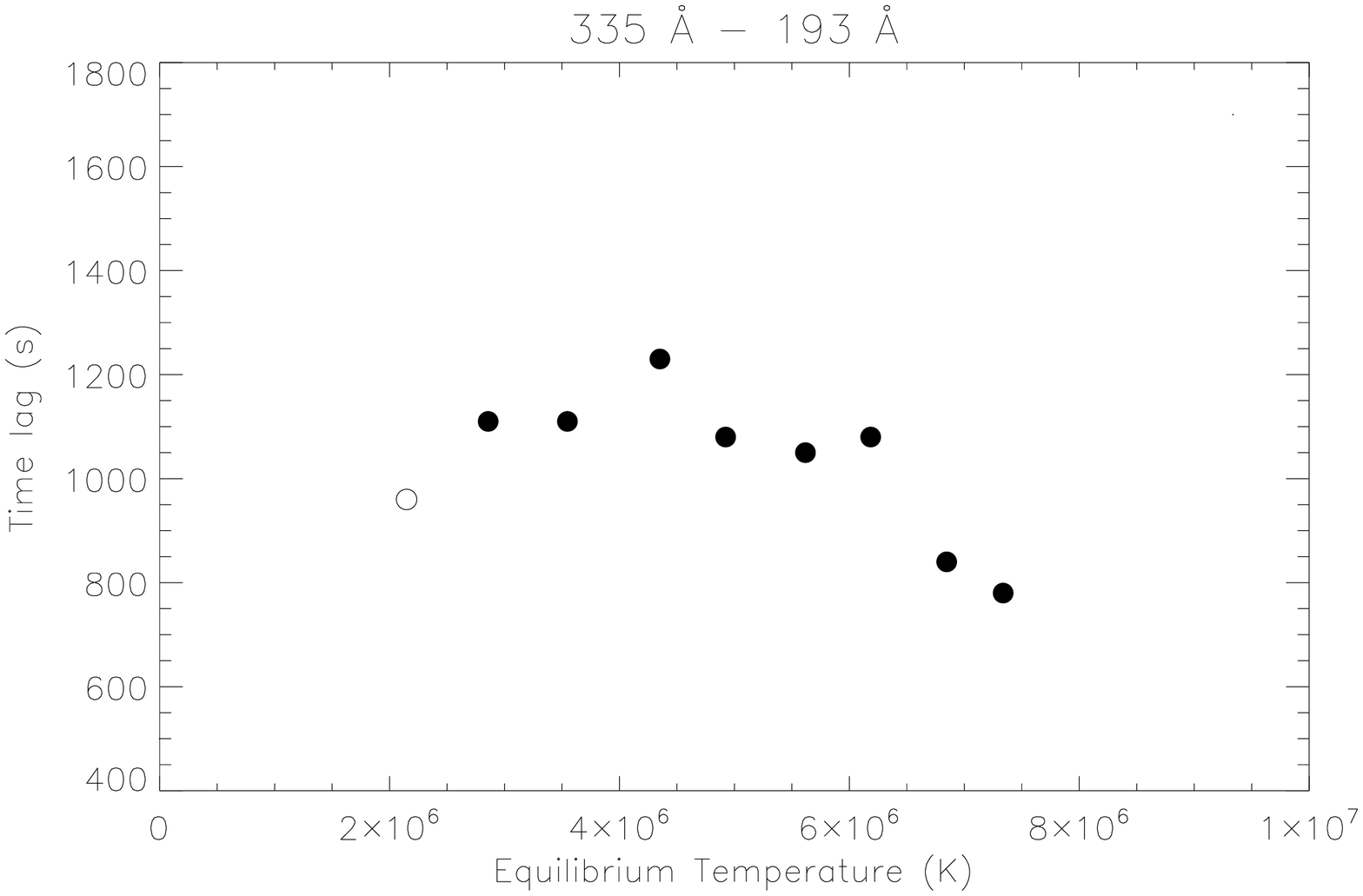}
\includegraphics[width=0.32\textwidth]{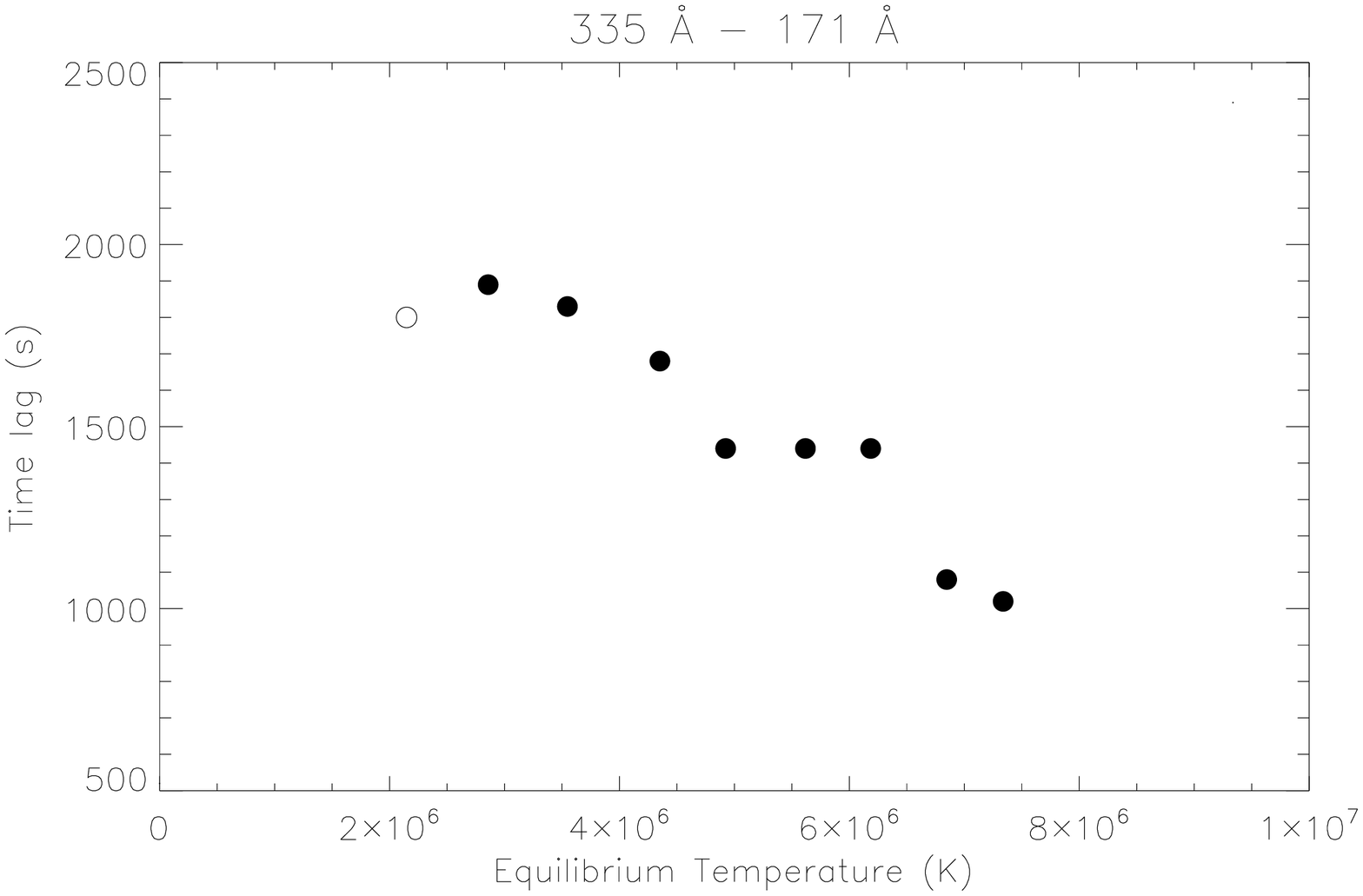}\\
\includegraphics[width=0.32\textwidth]{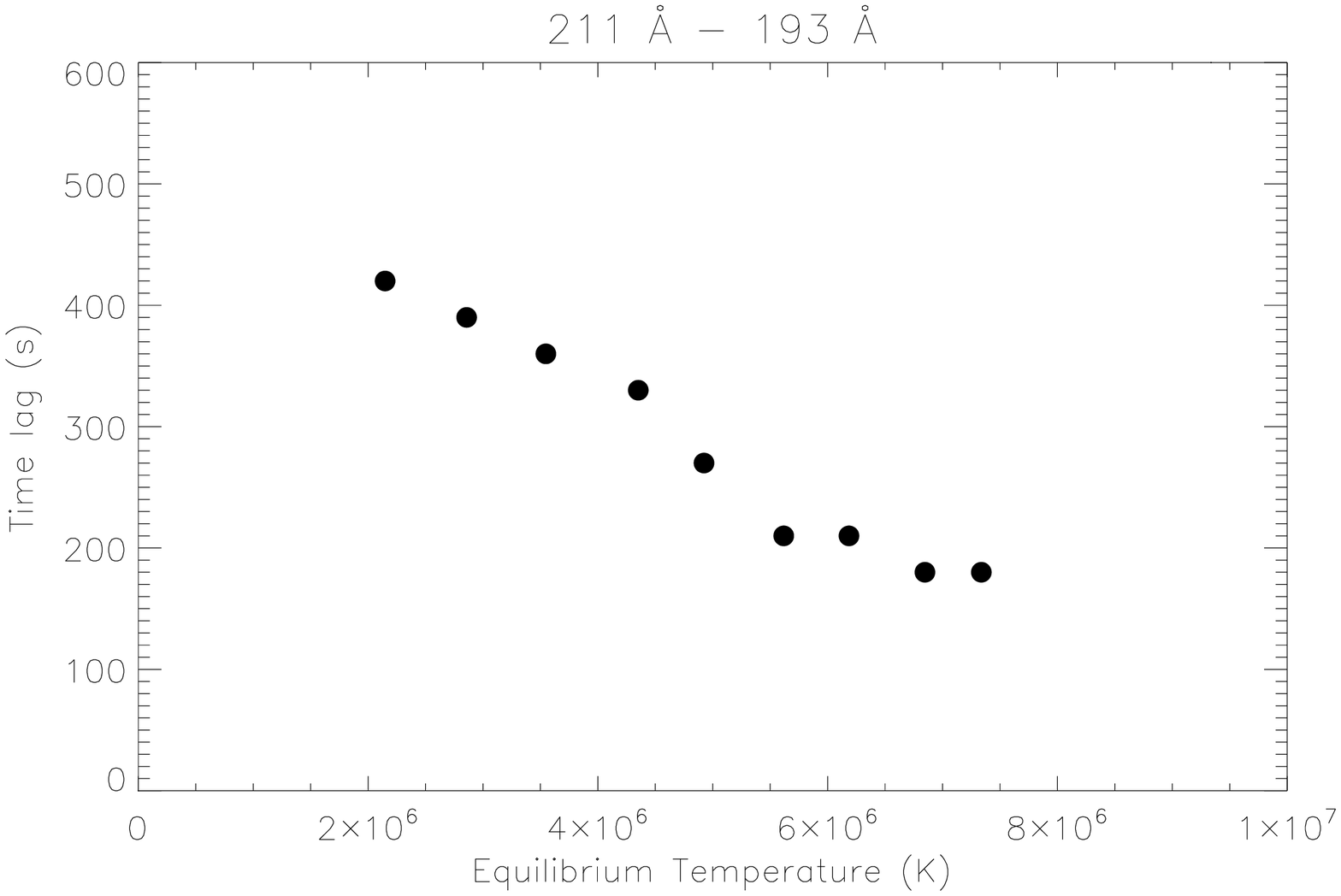}
\includegraphics[width=0.32\textwidth]{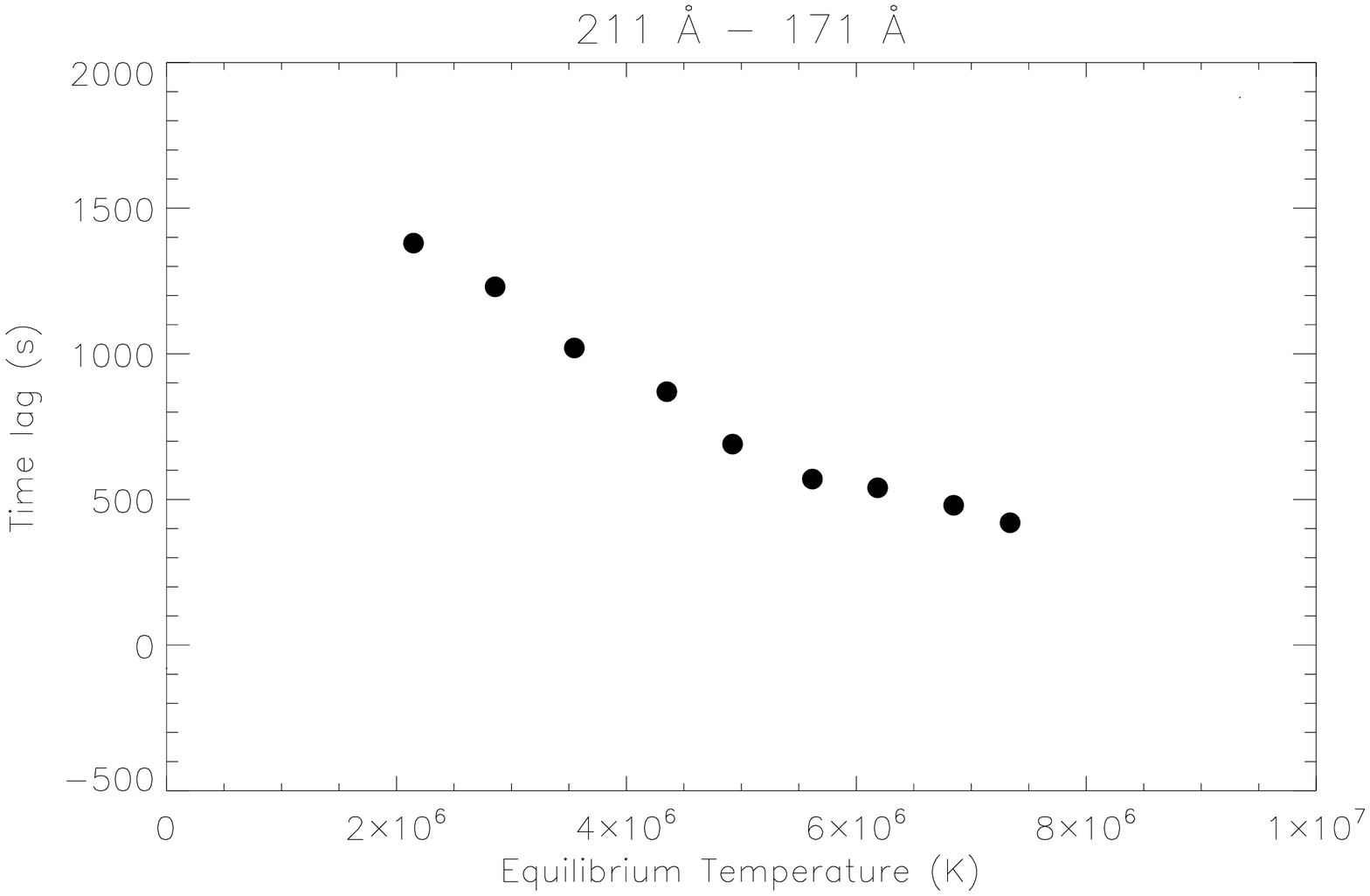}
\includegraphics[width=0.32\textwidth]{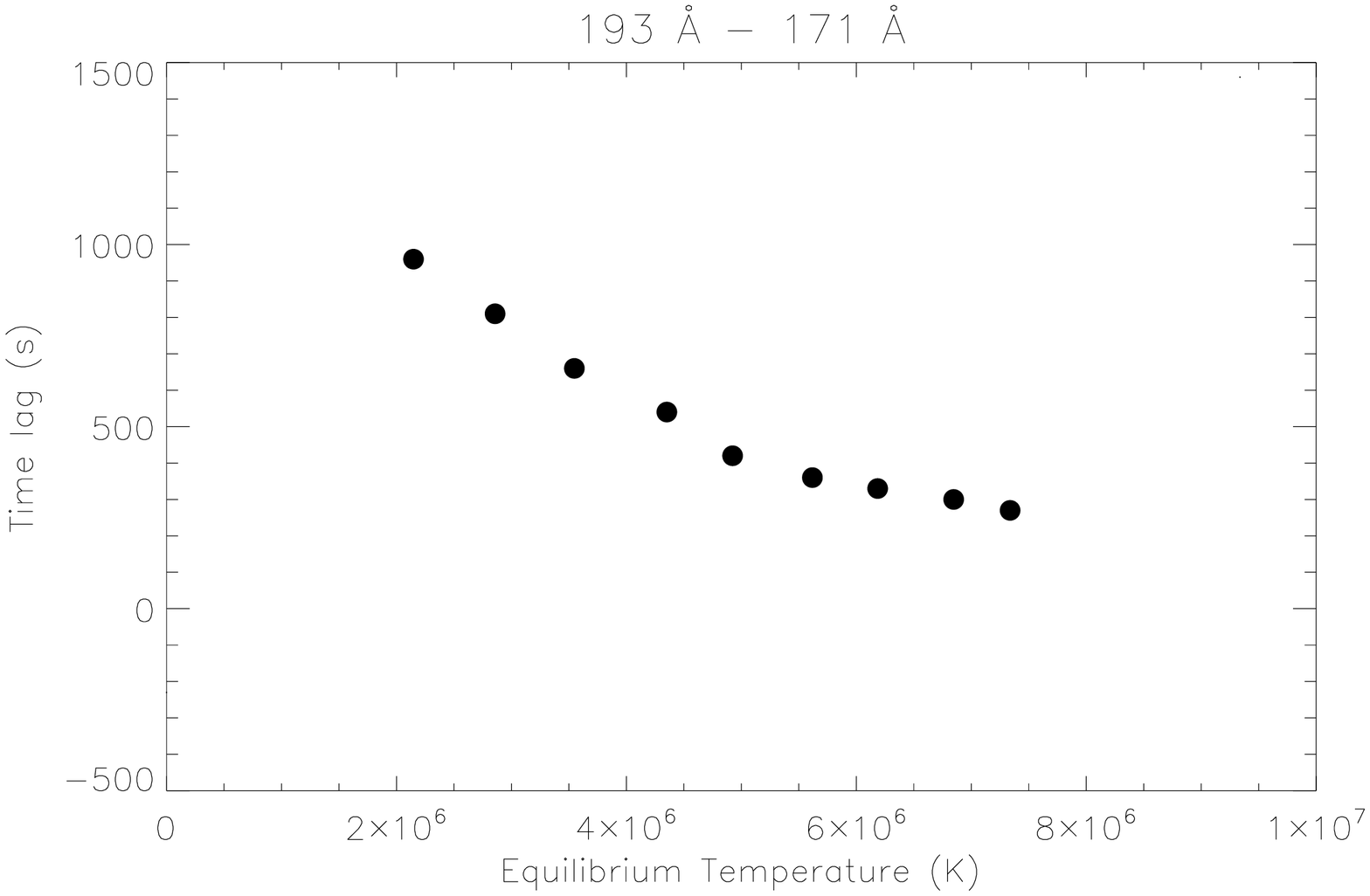}
\caption{The time lag in the channel pairs as a function of the equilibrium temperature of the solution.  The empty symbols indicate the intensity in the one or both channels was less than 0.3 DN s$^{-1}$.} 
\label{fig:energy_tl}
\end{center} 
\end{figure}

For each of these solutions, we calculate AIA 335, 211, 193, and 171\,\AA\ light curves. We then calculate the time lag in the channel pairs.  The time lag for each channel pair as a function of the equilibrium temperature of the solution is shown in Figure~\ref{fig:energy_tl}.   

Another concept to consider in completing this analysis is whether or not a loop (and associated time lag) is ``observable.''  When the magnitude of the heating is small, the loop may never be hot or dense enough to generate measurable intensity in the channels we are considering, even though we can calculate a time lag.  We suggest that for a loop to be observable, it must generate at least 1 DN in a single exposure.  Given that AIA exposure times in all channels are typically $\sim$ 3\,s, we use an open circle to denote loops where the intensity in one or both of the channel pairs is expected to be less than 0.3 DN s$^{-1}$.

The magnitudes of the time lags in this series of simulations are all relatively short, less than 2,000\,s, in all channel pairs.  In the 211, 193, and 171\,\AA\ channel pairs, we find that longer time lags are associated with loops with lower heating magnitudes or equilibrium temperatures and shorter time lags are associated with higher heating rates or equilibrium temperatures.   This is expected, as the larger the magnitude of heating, the higher the resulting density in the loop.  This higher density then leads to more rapid radiative cooling and shorter time lags.  In the 335\,\AA\ channel pairs, this trend is absent.  This is due, in part, to the fact that 335\,\AA\ has a broader response function, broader light curve, and that 335\,\AA\ intensity is rising early in the simulation during the conduction-dominated phase.   Additionally, for solutions where the equilibrium temperature is less than the temperature at the peak of the 335\,\AA\ response function, the emission in the 335\,\AA\ channel no longer peaks when the temperature of the plasma is at 2.5\,MK so the time lag does not represent the cooling time between the two temperatures given in Table~\ref{tab:combs}.  { This is the case for all channels where the plasma temperature does not fully reach the peak temperature of the response, but it is especially problematic for the 335\,\AA\  channel due to its larger width. This makes the time lags in the 335\,\AA\  channel difficult to interpret.}

We have repeated this series of simulations with two additional parameter changes.  First we consider the case where the coronal portion of the loop is parallel to the solar surface.  Using the same heating functions, we re-calculate the solutions and time lags.  We find the average difference in the time lags is less than 15\%.  We then adjust the duration of the heating to 250\,s while keeping the total energy input the same.  We complete those simulations and find the average change in the time lag is less than 10\%. We conclude that these two parameters do not strongly influence the time lag and consider only perpendicular loops and 500\,s duration for the remainder of this analysis.

\subsection{Loop Length}
\label{sec:ll}

The cooling time of loops depends strongly on the loop length, with longer loops cooling more slowly. The lengths of loops in this active region range from 50 Mm to 400 Mm.  In this subsection, we investigate how the time lags depend on loop length.  We run a series of simulations for 50, 100, 200, 300, and 400\,Mm loops.  Again, we assume the loops are semi-circular  and perpendicular to the solar surface.  
For each loop length, we run several simulations with different magnitudes of impulsive heating, while maintaining the duration of heating to be 500\,s.  We choose the impulsive heating rates to sample equilibrium temperatures between $\sim 2 - 8.5$\,MK, roughly the same range of equilibrium temperatures sampled above.  We use the coronal radiative loss function.  

\begin{figure}
\begin{center}
\includegraphics[width=0.32\textwidth]{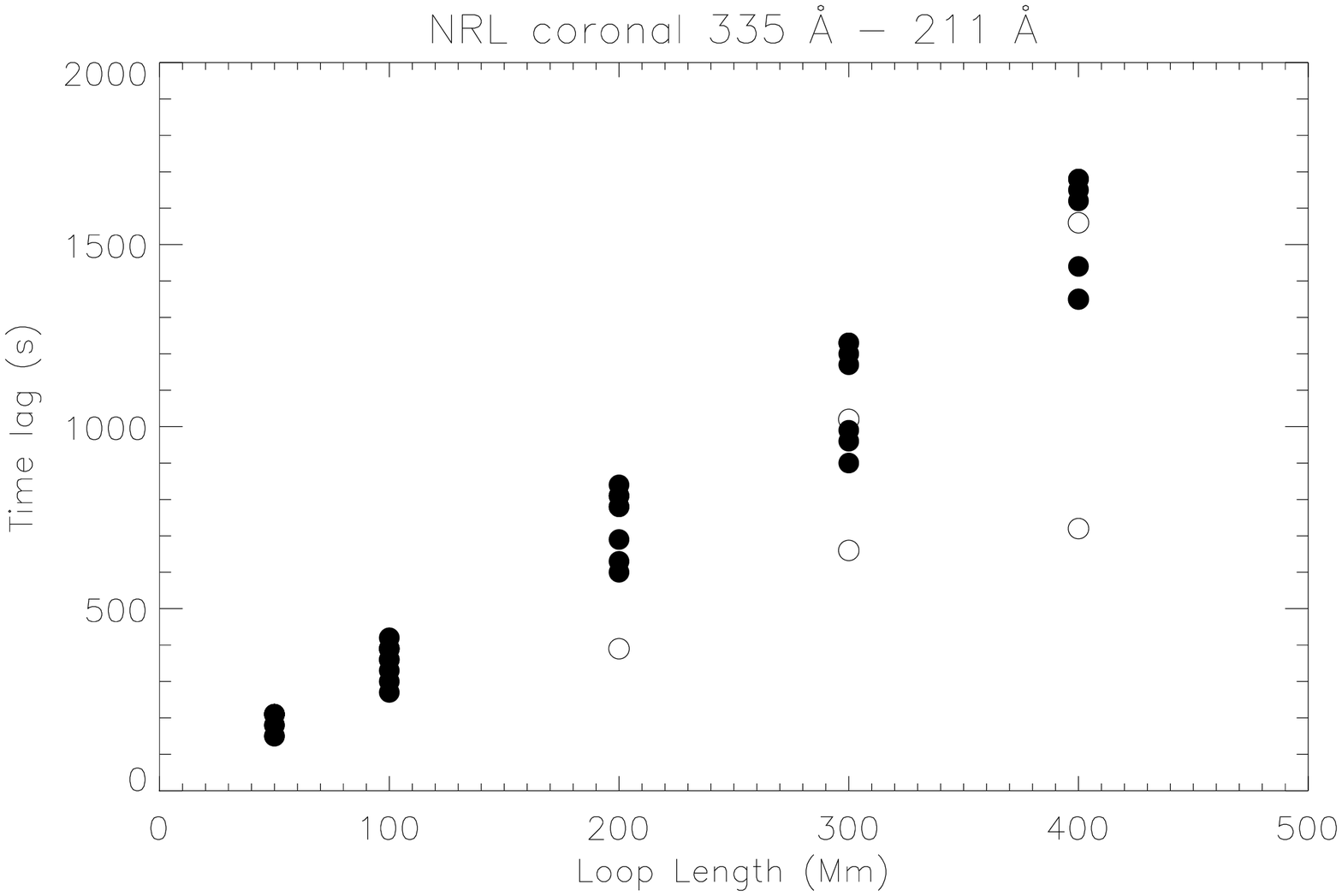}
\includegraphics[width=0.32\textwidth]{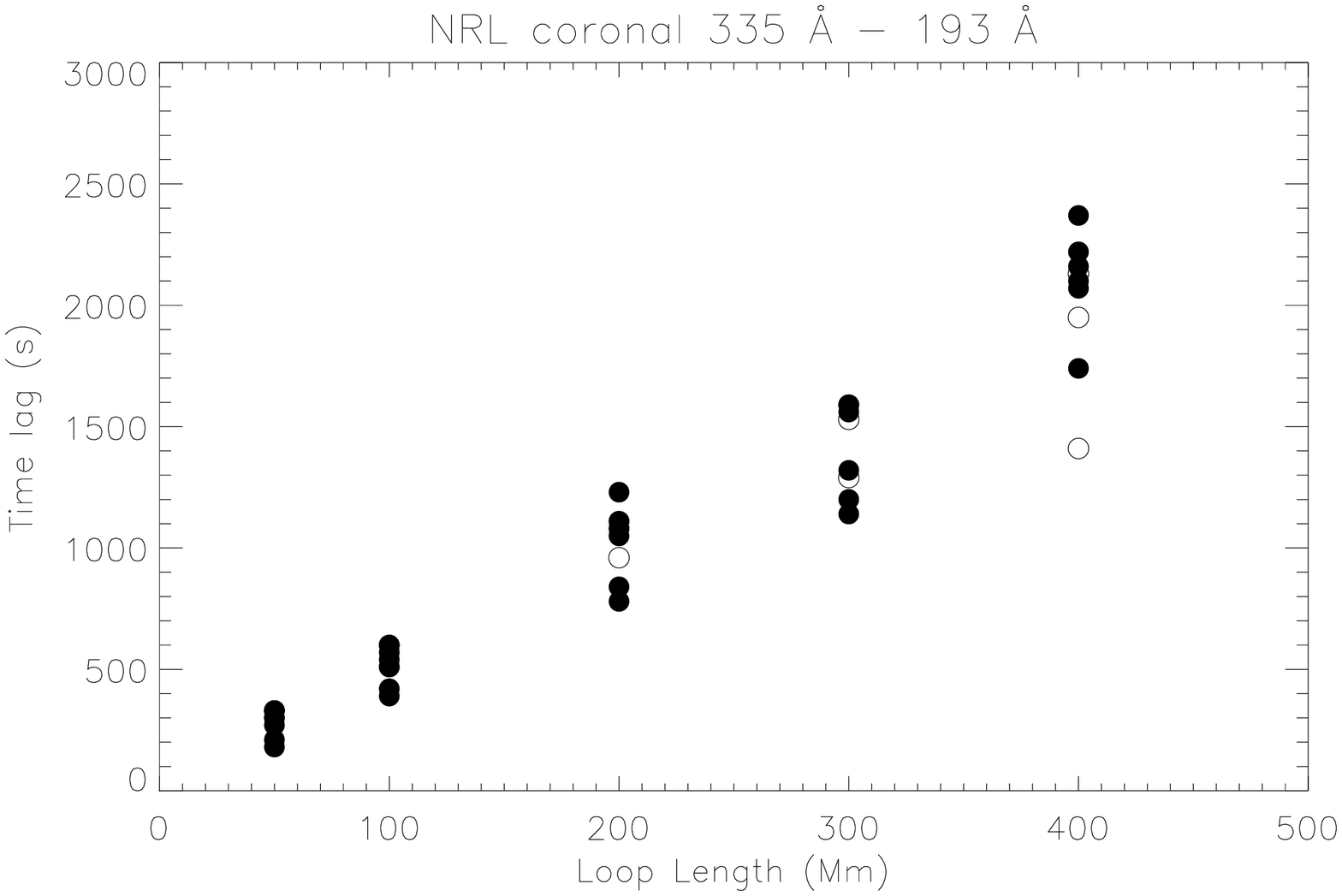}
\includegraphics[width=0.32\textwidth]{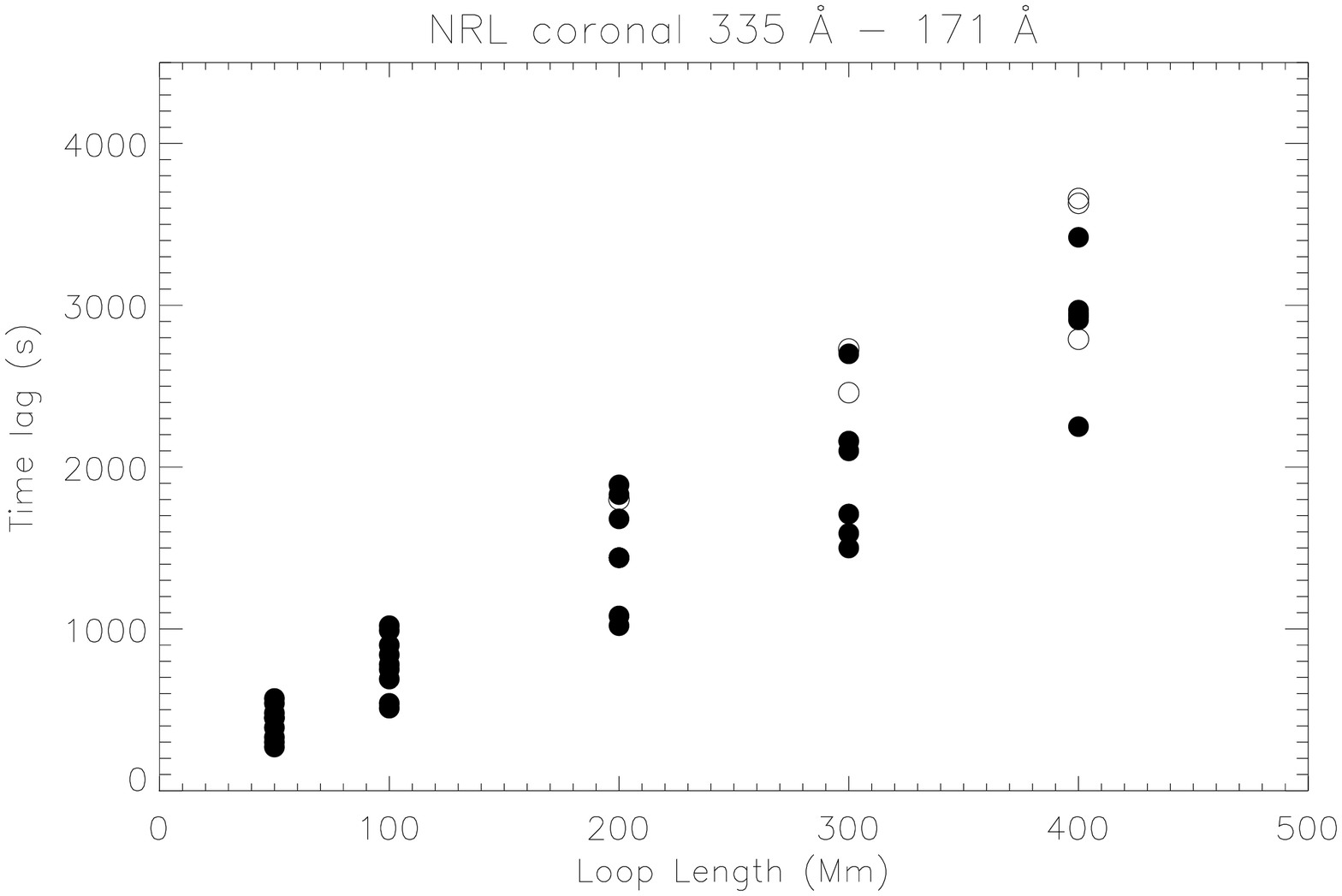}\\
\includegraphics[width=0.32\textwidth]{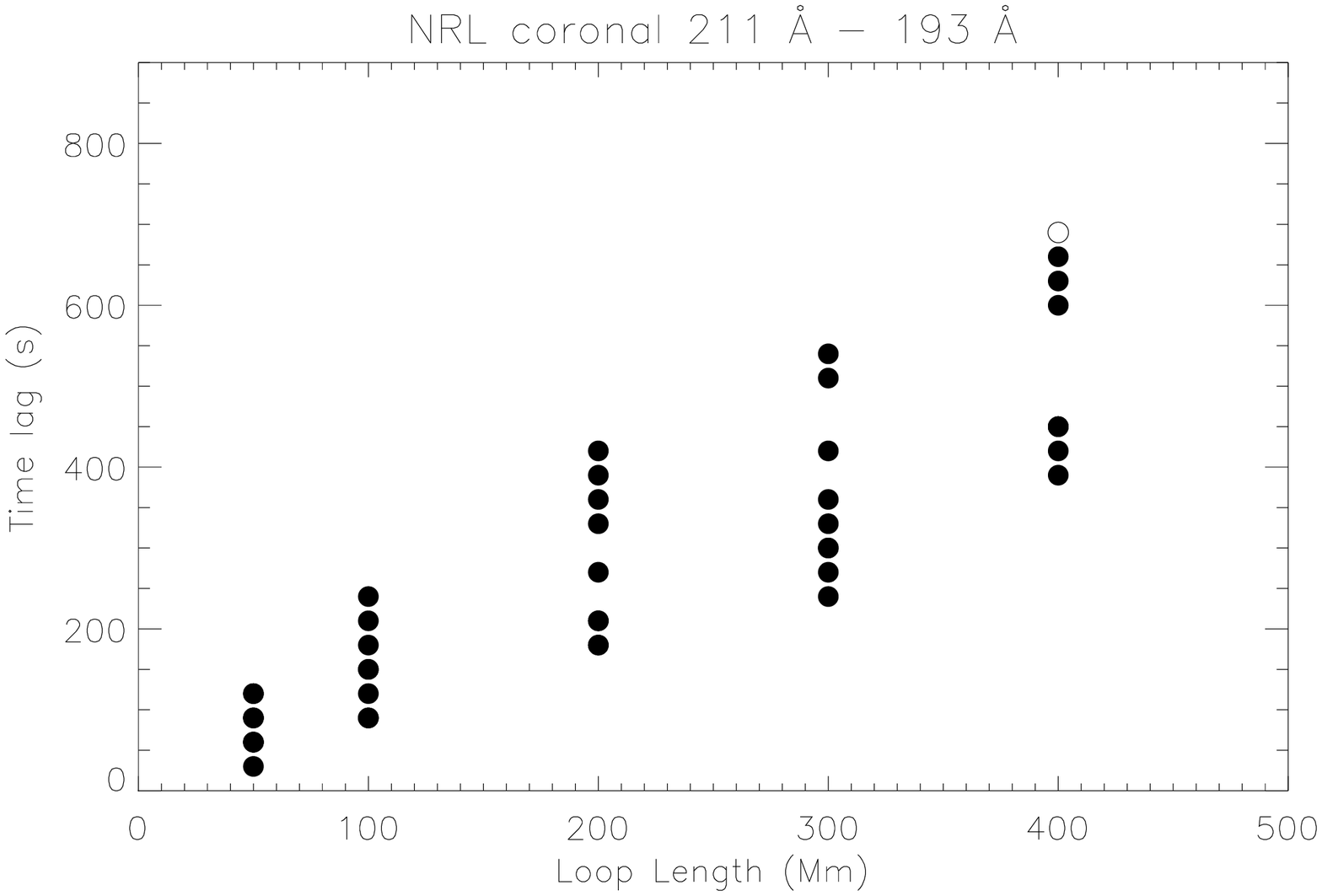}
\includegraphics[width=0.32\textwidth]{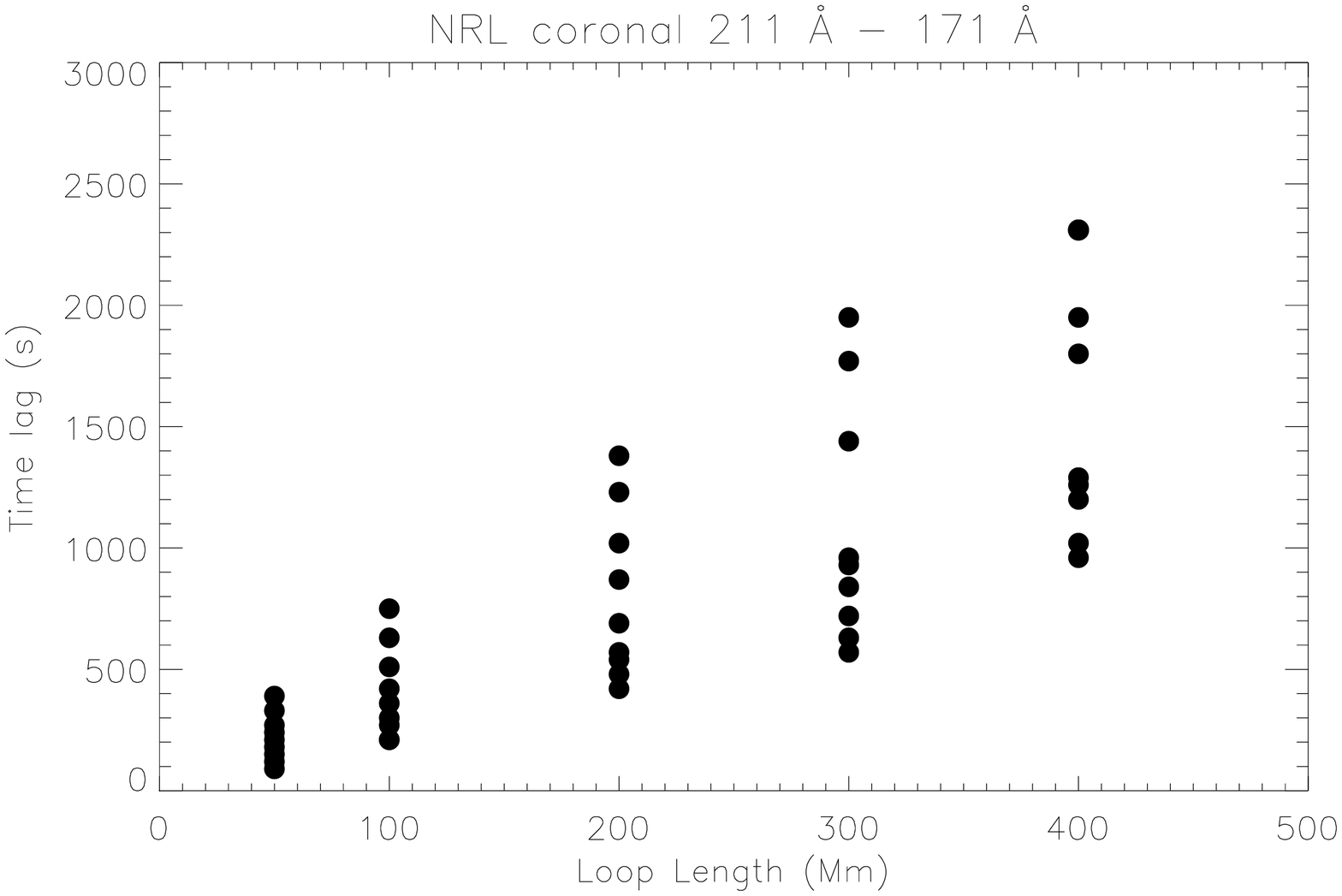}
\includegraphics[width=0.32\textwidth]{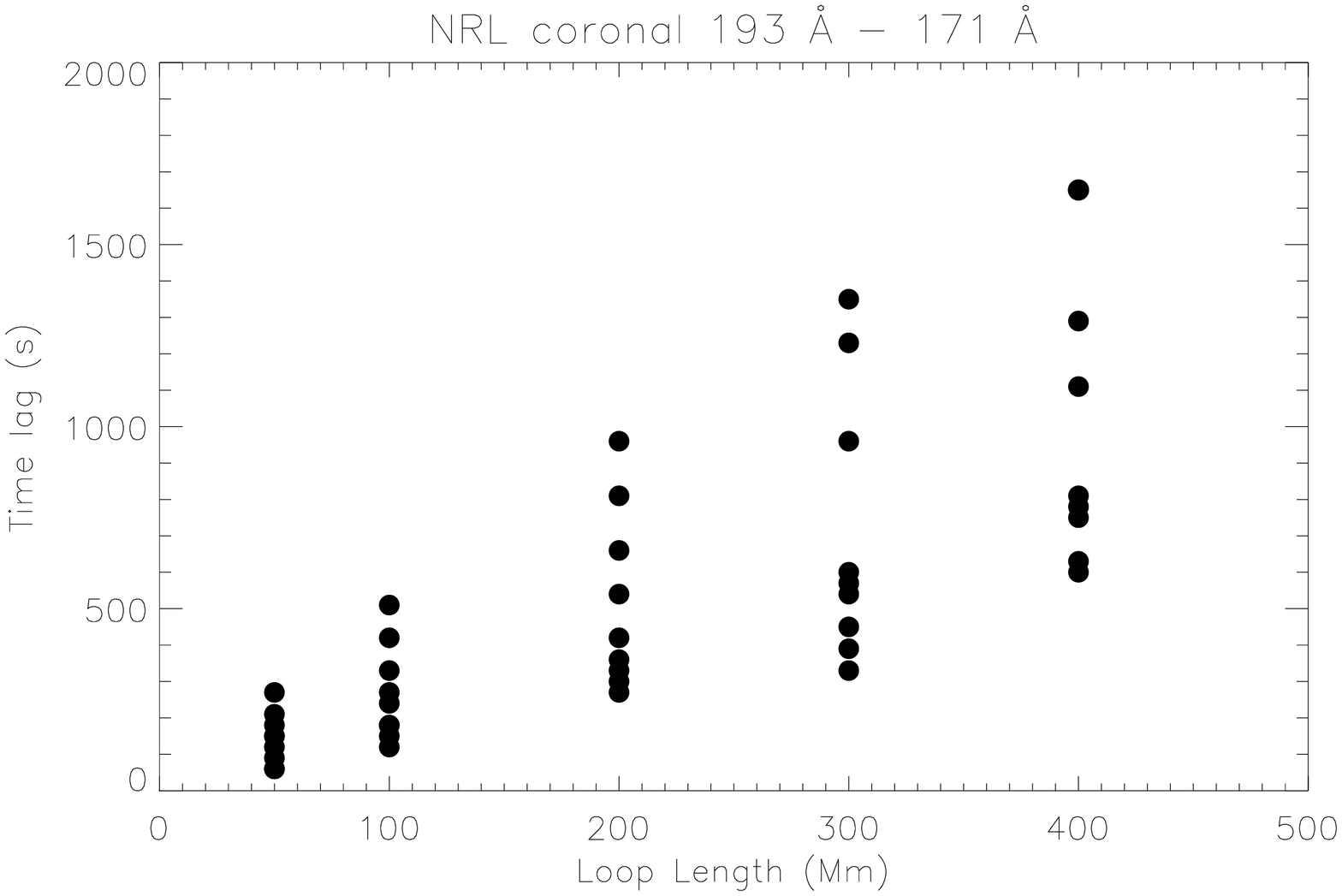}
\caption{{ The time lags calculated for different channel pairs as a function of loop length.  For each loop length, several simulations were completed with different impulsive heating rates. If the symbol is empty, it indicates the expected intensity in one or both channels was less then 0.3 DN s$^{-1}$.}} 
\label{fig:TLvsL}
\end{center} 
\end{figure}

For each of the five loop lengths and impulsive heating magnitudes, we solve for the average apex temperature and density over the upper 50\% of the loop.  We calculate the AIA 335, 211, 193, and 171\,\AA\ light curves and solve for the time lags in the channel pairs.  The resulting time lags are shown as a function of loop length in Figure~\ref{fig:TLvsL}.  For each loop length, there are several time lag measurements.  The ones shown with empty symbols imply an intensity less than 0.3 DN s$^{-1}$ in one or both of the channels.


As expected, the longest loops have the longest time lags.  The longest time lags found in this series of simulations is less than 3,800\,s in the 335-171\,\AA\ channel pair.  In all other channel pairs, the maximum time lag is less than 2,500\,s.

\subsection{Abundances}

The cooling time of loops also depends strongly on the make up of the plasma.  Typically, coronal plasma is thought to have enhanced amounts of low-first ionization potential (FIP) elements, such as iron, when compared to photospheric values.  Because these low-FIP elements account for a significant portion of the radiation at coronal temperatures, plasma with so-called ``coronal'' abundances will radiate more than that with photospheric abundances.  The radiative loss functions showing this enhancement are shown in Figure~\ref{fig_tresp_abun}.  

\begin{figure}
\begin{center}
\includegraphics[width=0.32\textwidth]{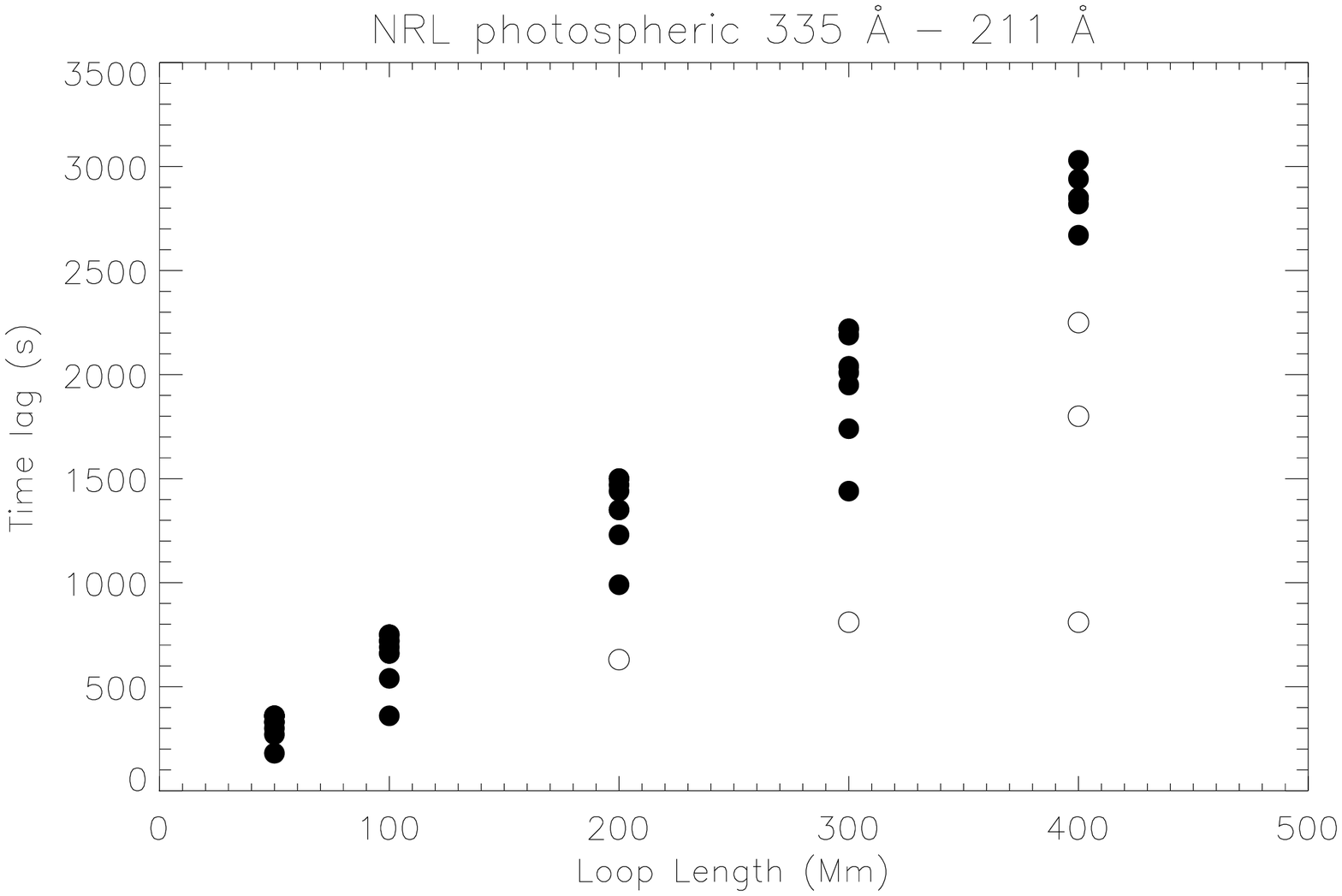}
\includegraphics[width=0.32\textwidth]{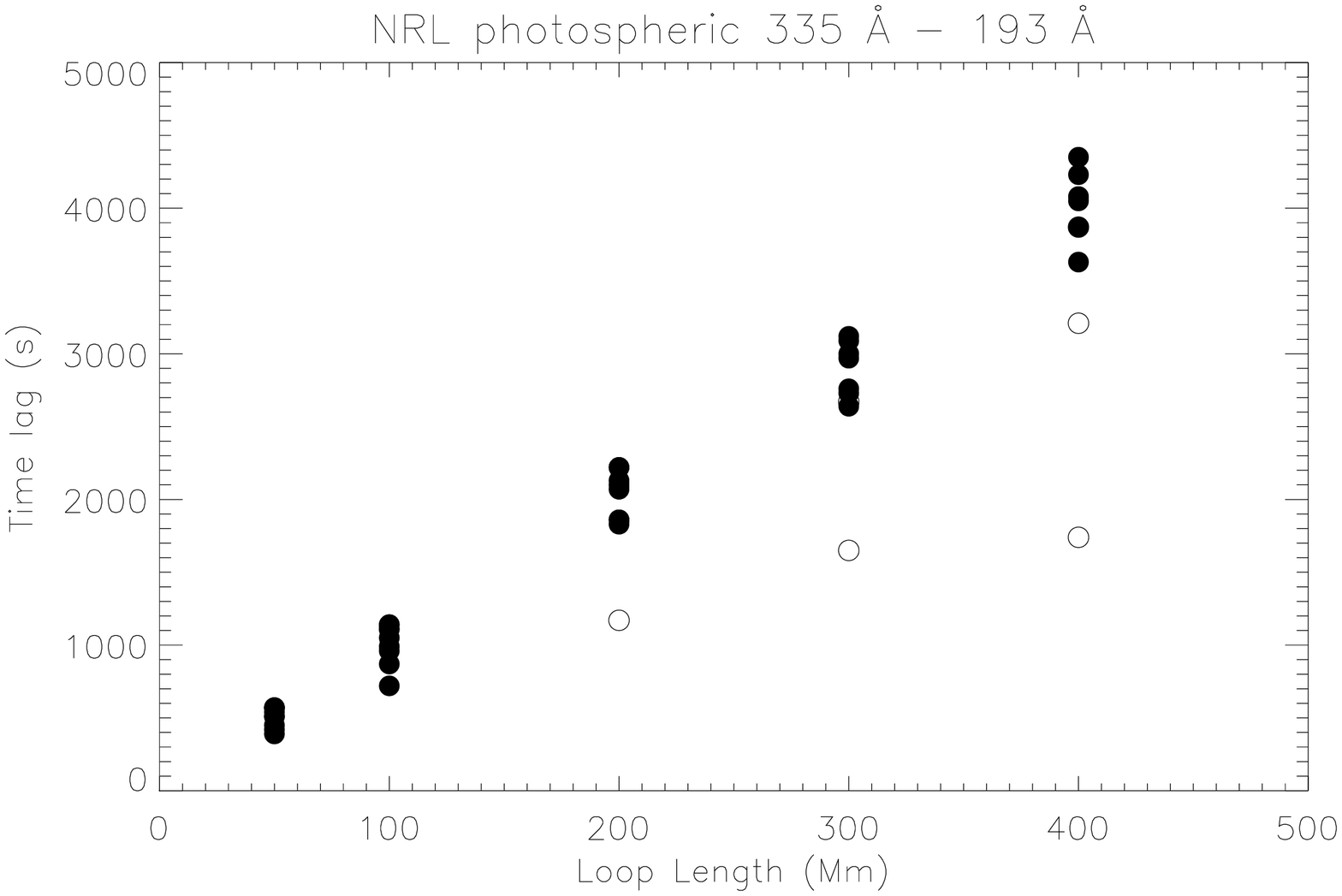}
\includegraphics[width=0.32\textwidth]{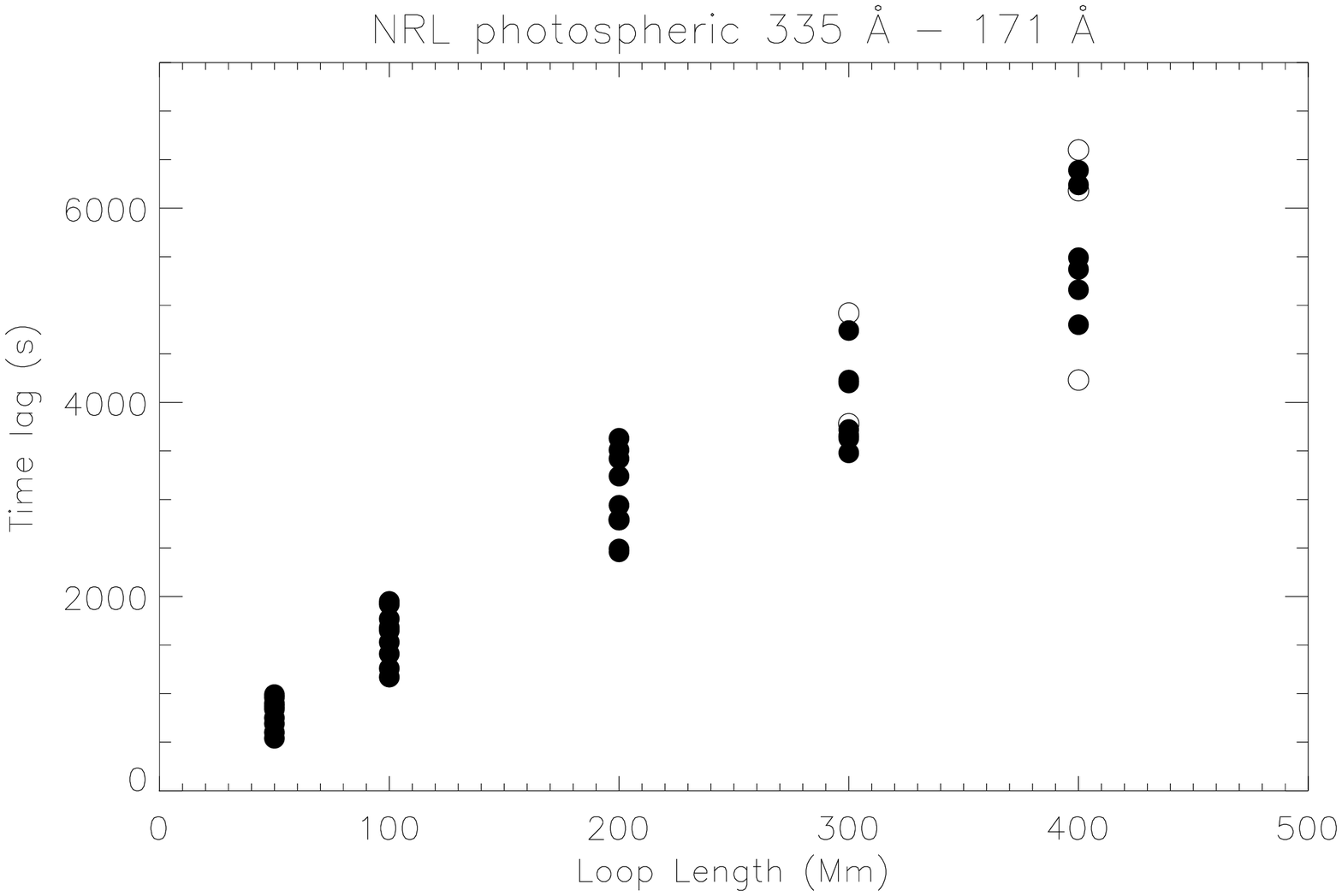}\\
\includegraphics[width=0.32\textwidth]{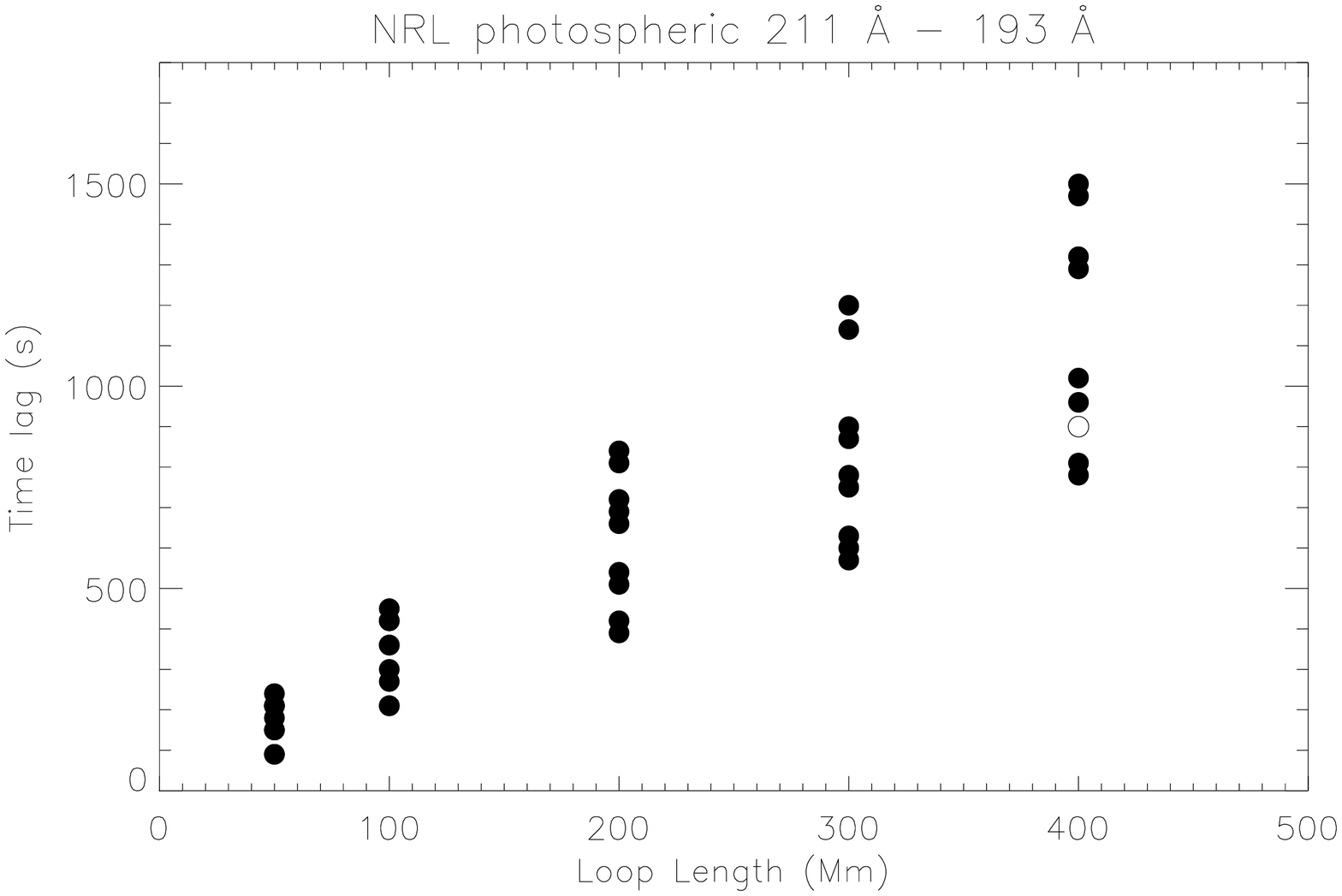}
\includegraphics[width=0.32\textwidth]{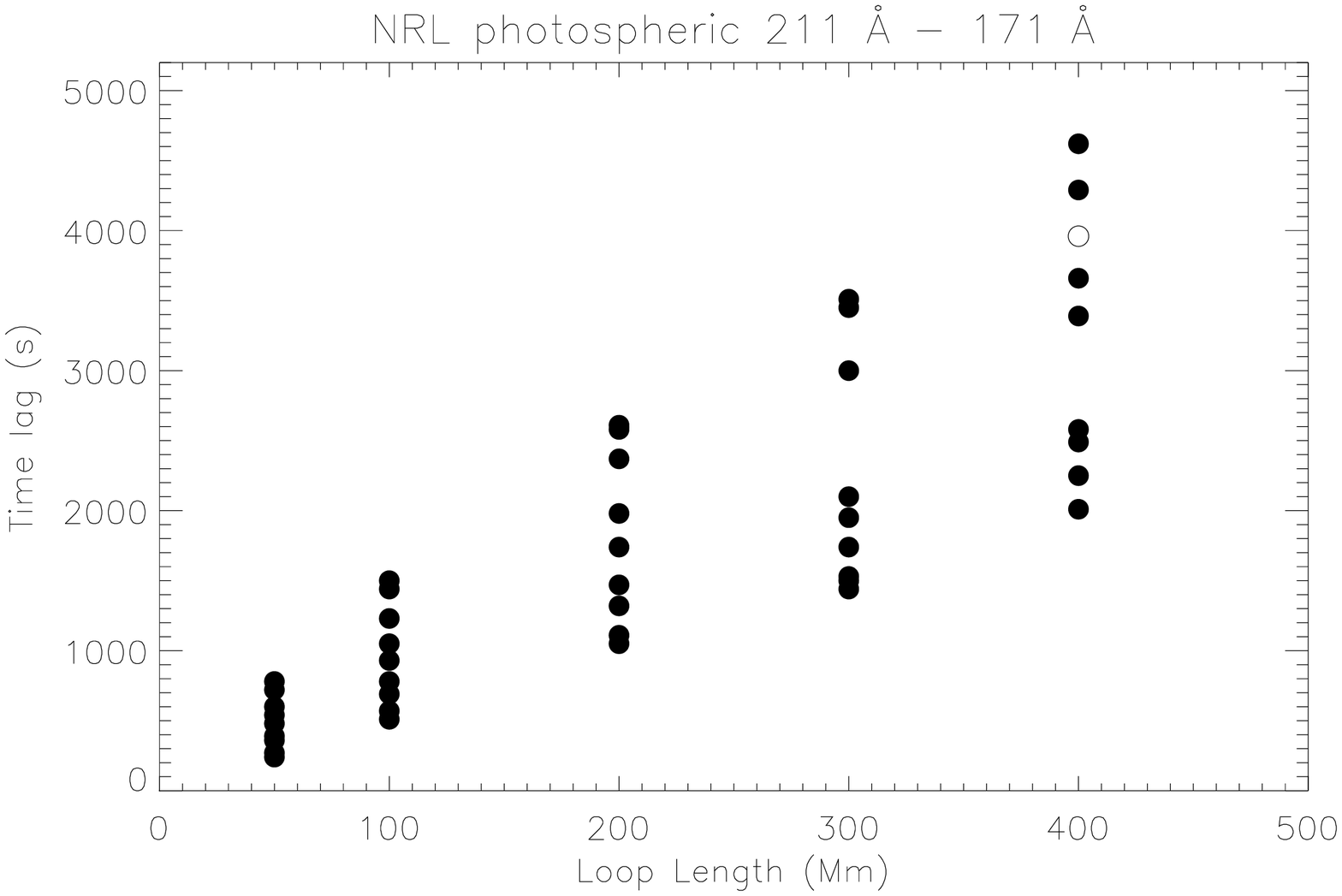}
\includegraphics[width=0.32\textwidth]{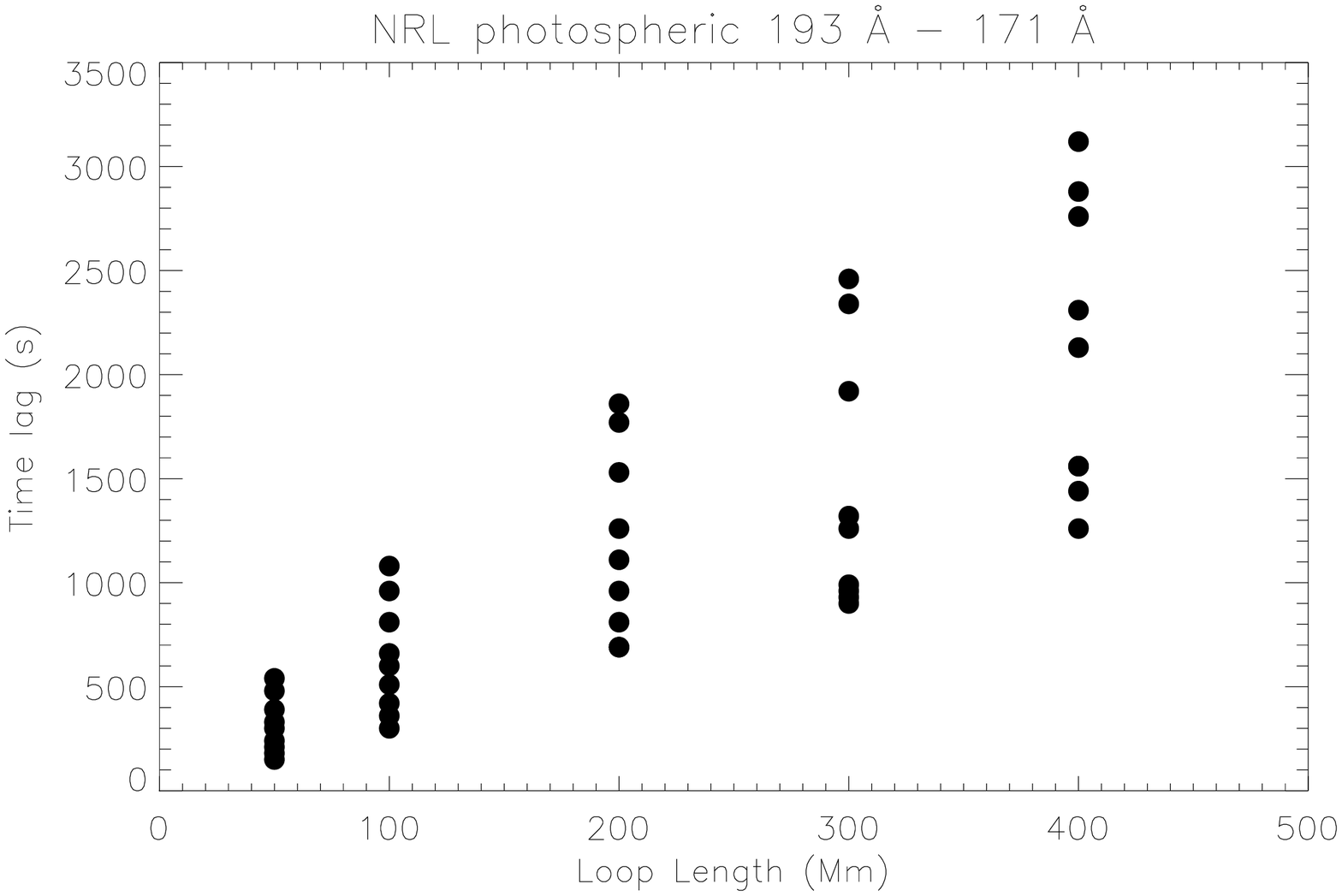}
\caption{{ The time lags expected for different channel pairs as a function of loop length using photospheric abundances.  If the symbol is empty, it indicates the expected intensity in one or both channels was less then 0.3 DN s$^{-1}$.} }
\label{fig:abundances}
\end{center} 
\end{figure}

We simulate the solutions for the photospheric abundances for the same impulsive heating rates as was calculated in Section~\ref{sec:ll}.  As above, we calculate the time lag in all channel pairs. { Figure \ref{fig:abundances} shows the same simulations as Figure~\ref{fig:TLvsL}, but with photospheric abundances. The maximum time lags in all cases have increased by a factor of approximately 2.}

\subsection{Area Expansion}

\begin{figure}
\centering
\includegraphics[width=6cm, angle=0]{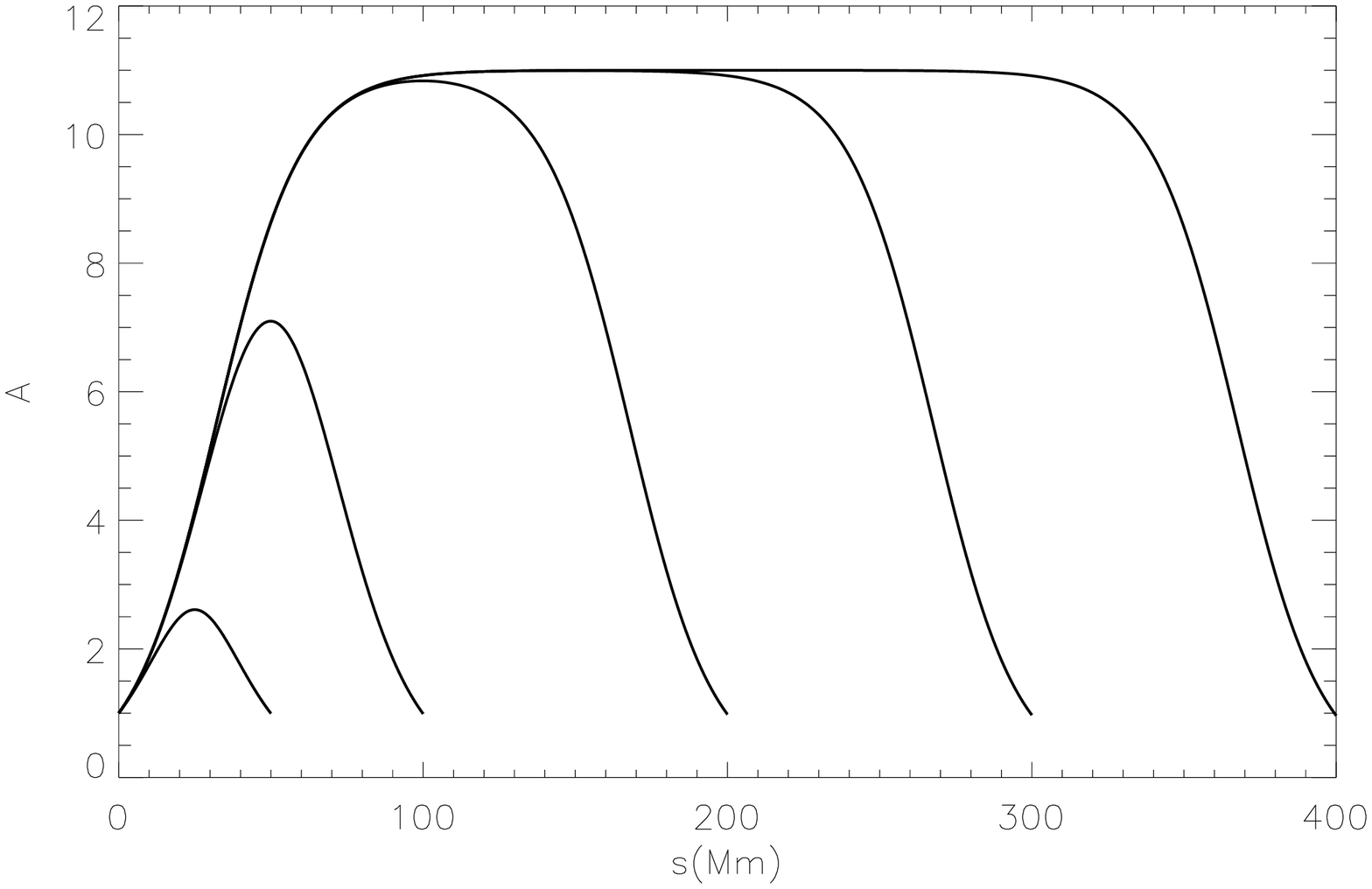}
\caption{Expansion factor $A(s)=1/B(s)$ for the different loop lengths  as a function of the distance along the loop.} 
\label{expansion}
\end{figure}

Finally, we investigate the impact of having the area of the loop expand.  To complete this, we use the PSOHM, as the standard NRLSOLFTM code does not include area expansion.  We
consider semi-circular loops with lengths: 50, 100, 200, 300 and 400\,Mm. We assume that the magnetic field strength along the loop, $B(s)$, is
exponentially decreasing with height above the solar surface and the cross-sectional area of the loop follows $1/B(s)$. The expansion factor for each loop is shown
in Figure~\ref{expansion} as a function of the length along the loop. 

\begin{figure}
\begin{center}
\includegraphics[width=0.32\textwidth]{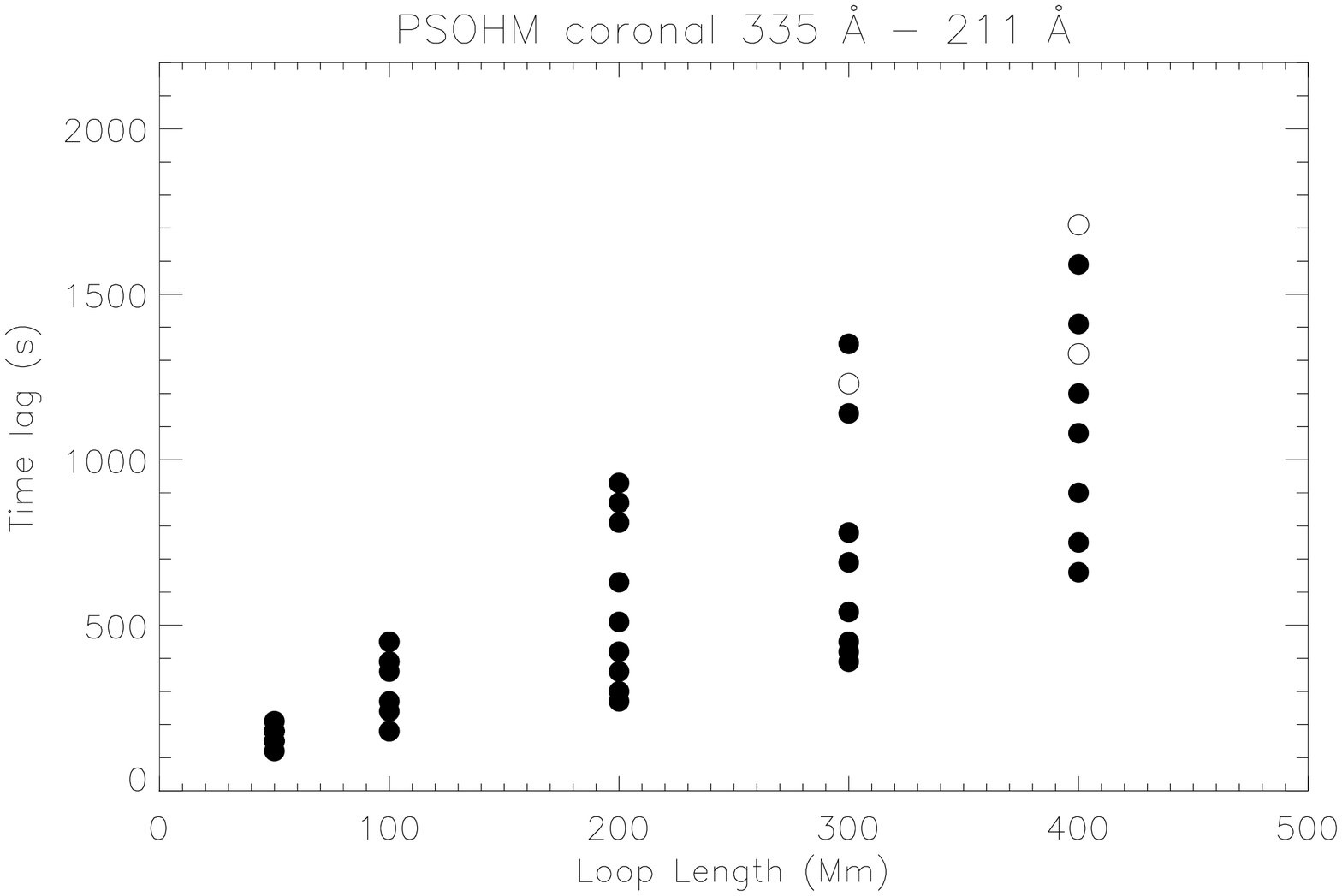}
\includegraphics[width=0.32\textwidth]{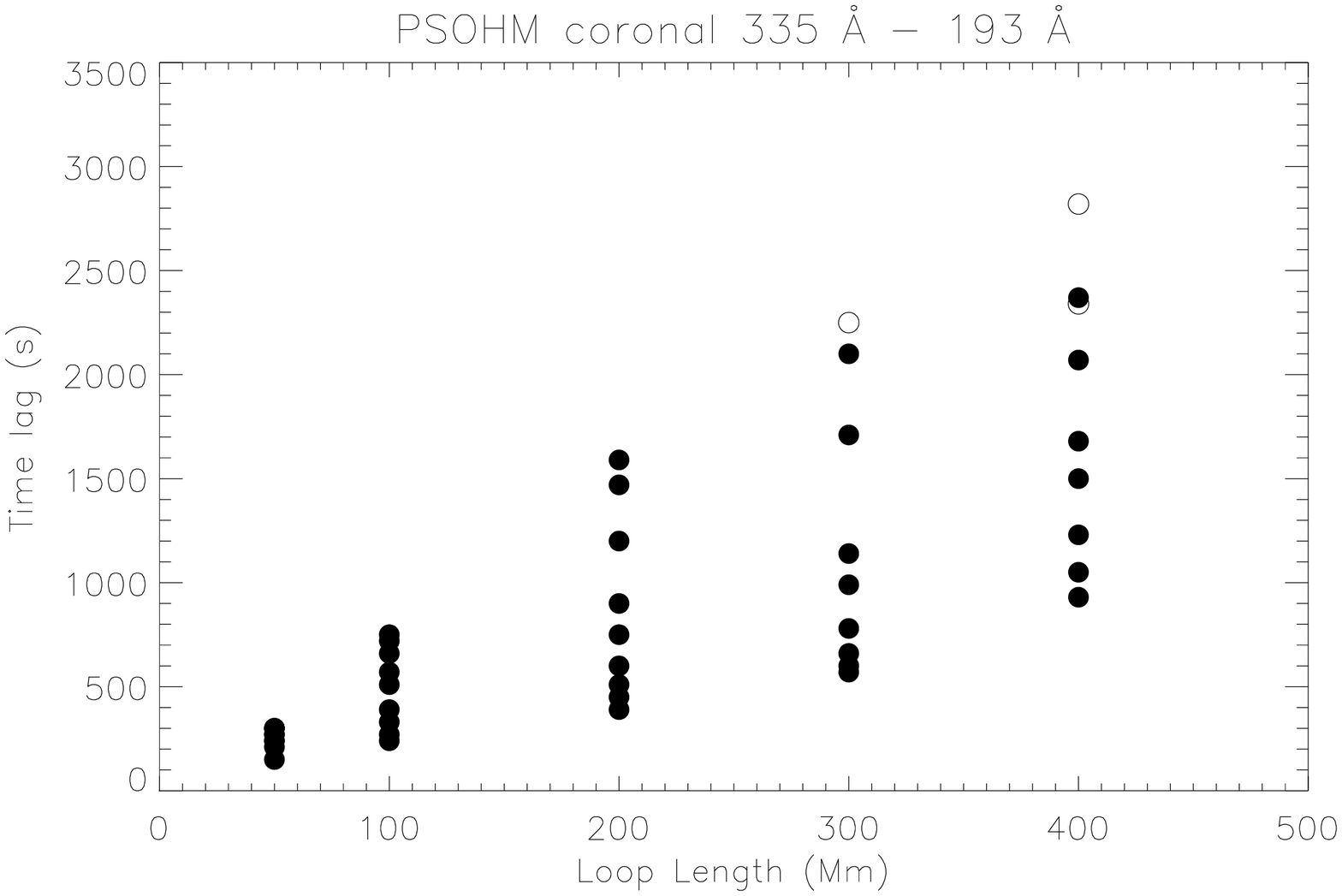}
\includegraphics[width=0.32\textwidth]{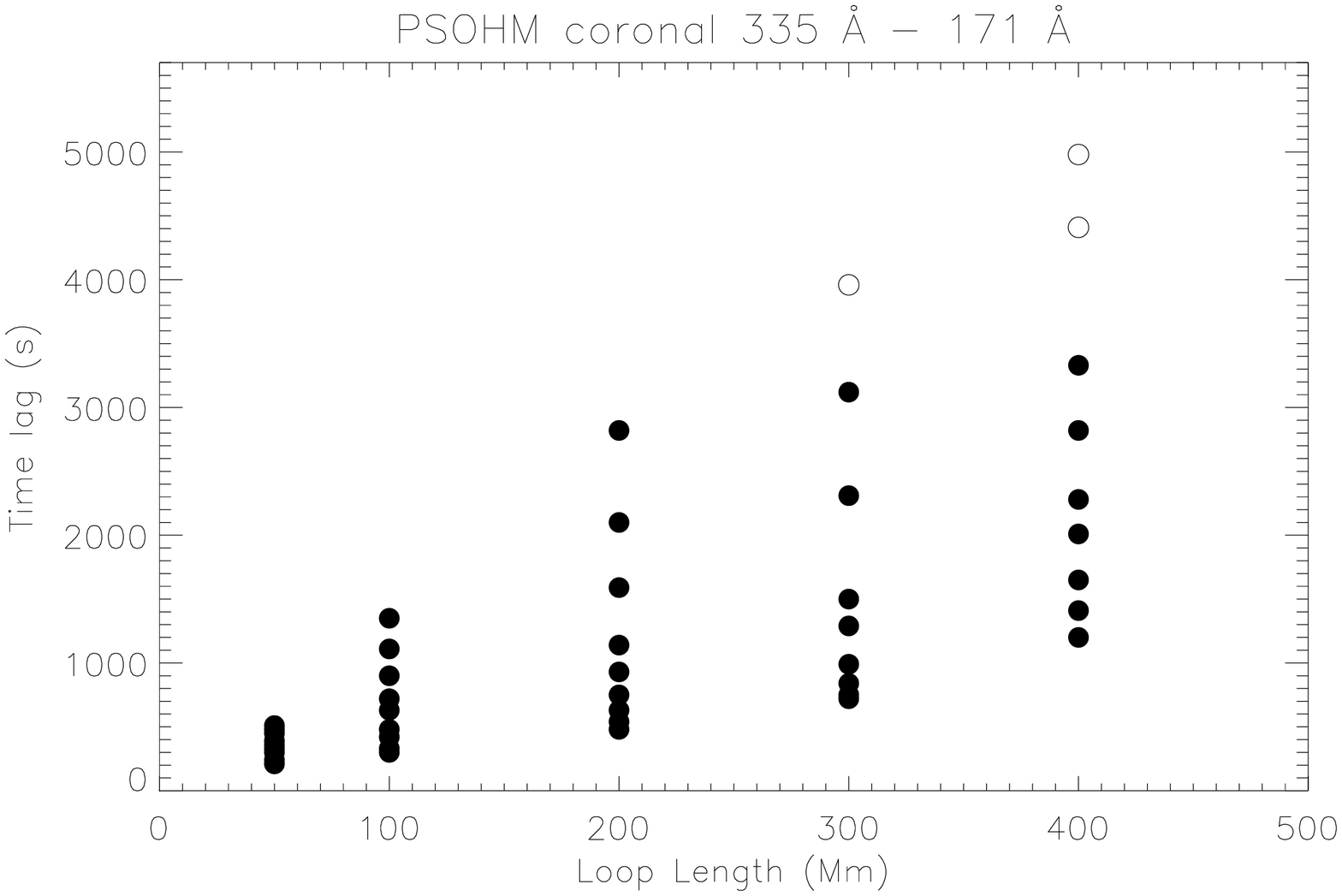}\\
\includegraphics[width=0.32\textwidth]{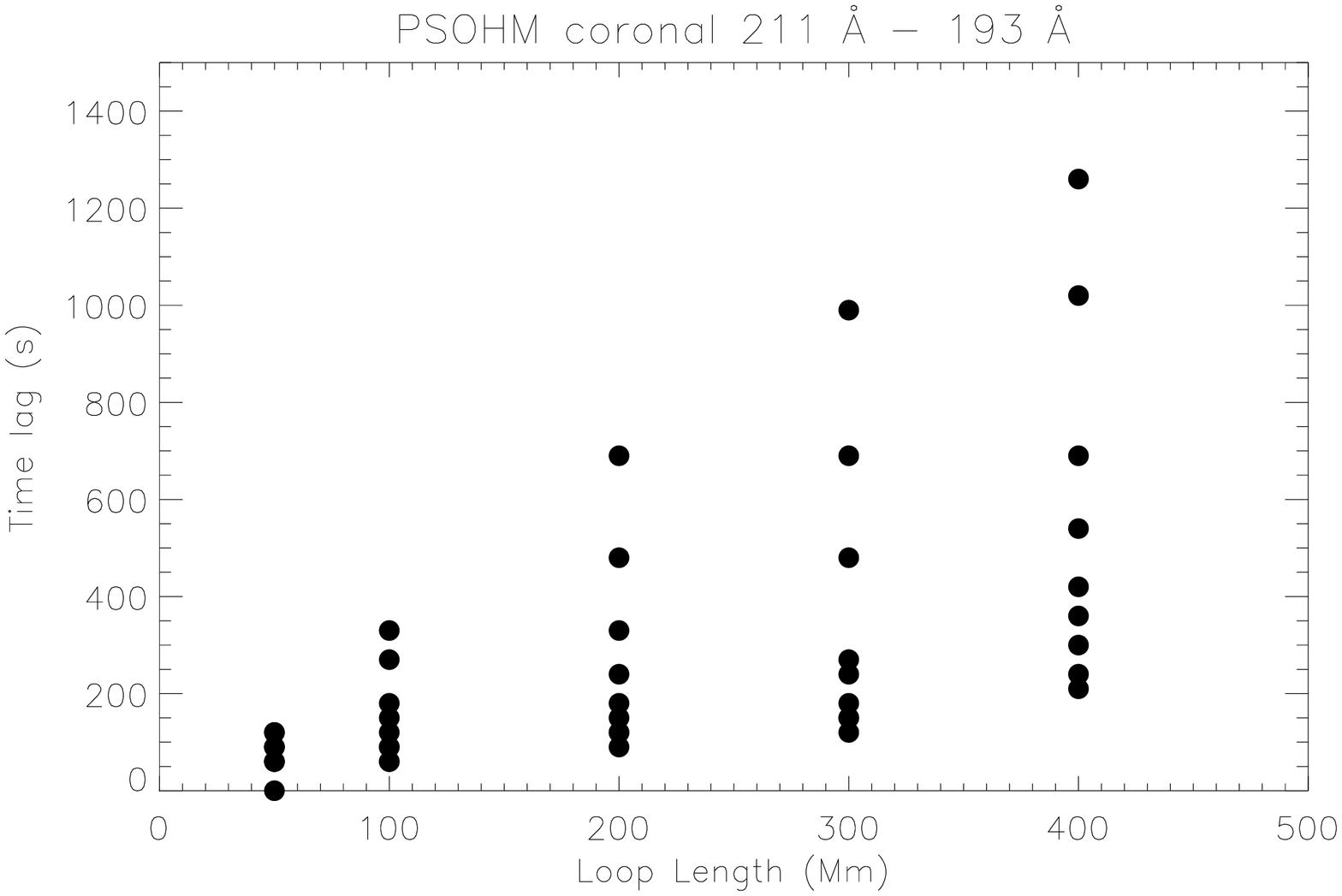}
\includegraphics[width=0.32\textwidth]{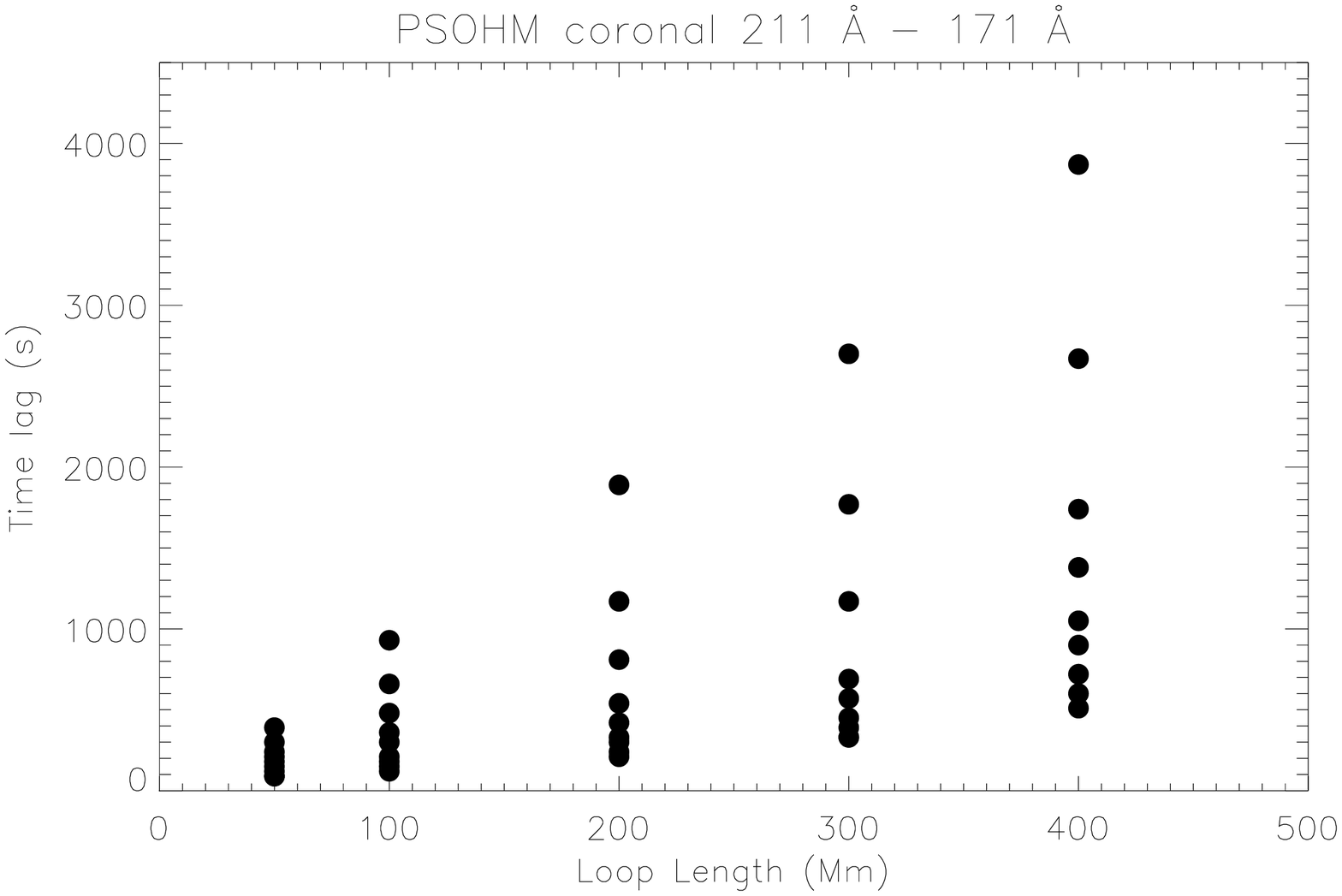}
\includegraphics[width=0.32\textwidth]{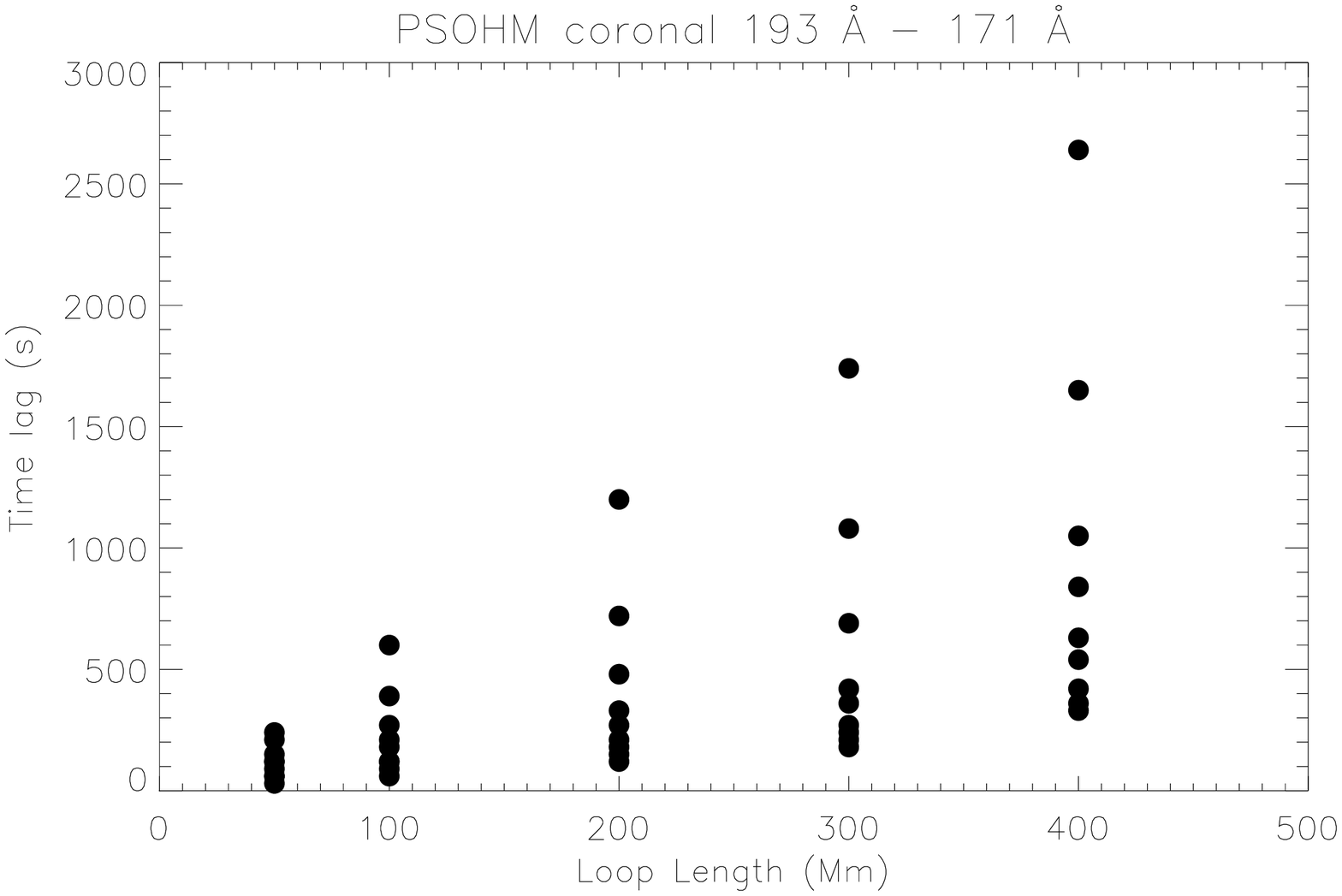}
\caption{{ Results showing the time lag for different channel pairs as a function of loop length for coronal abundances using the PSOHM code.  In these simulations, we allowed the loop to expand as a function of height.}} 
\label{area_expand} 
\end{center} 
\end{figure}

\begin{figure}
\begin{center}
\includegraphics[width=0.32\textwidth]{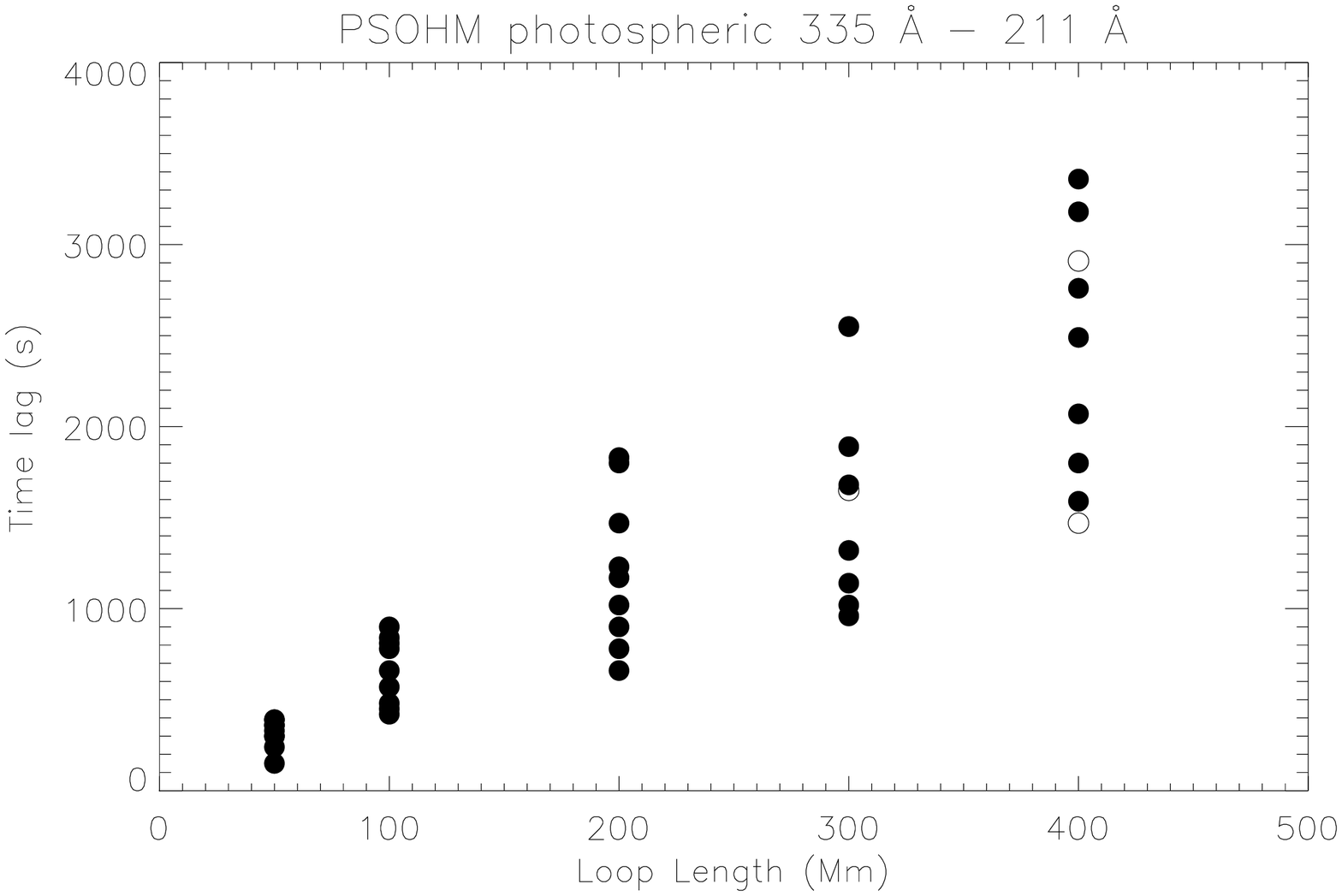}
\includegraphics[width=0.32\textwidth]{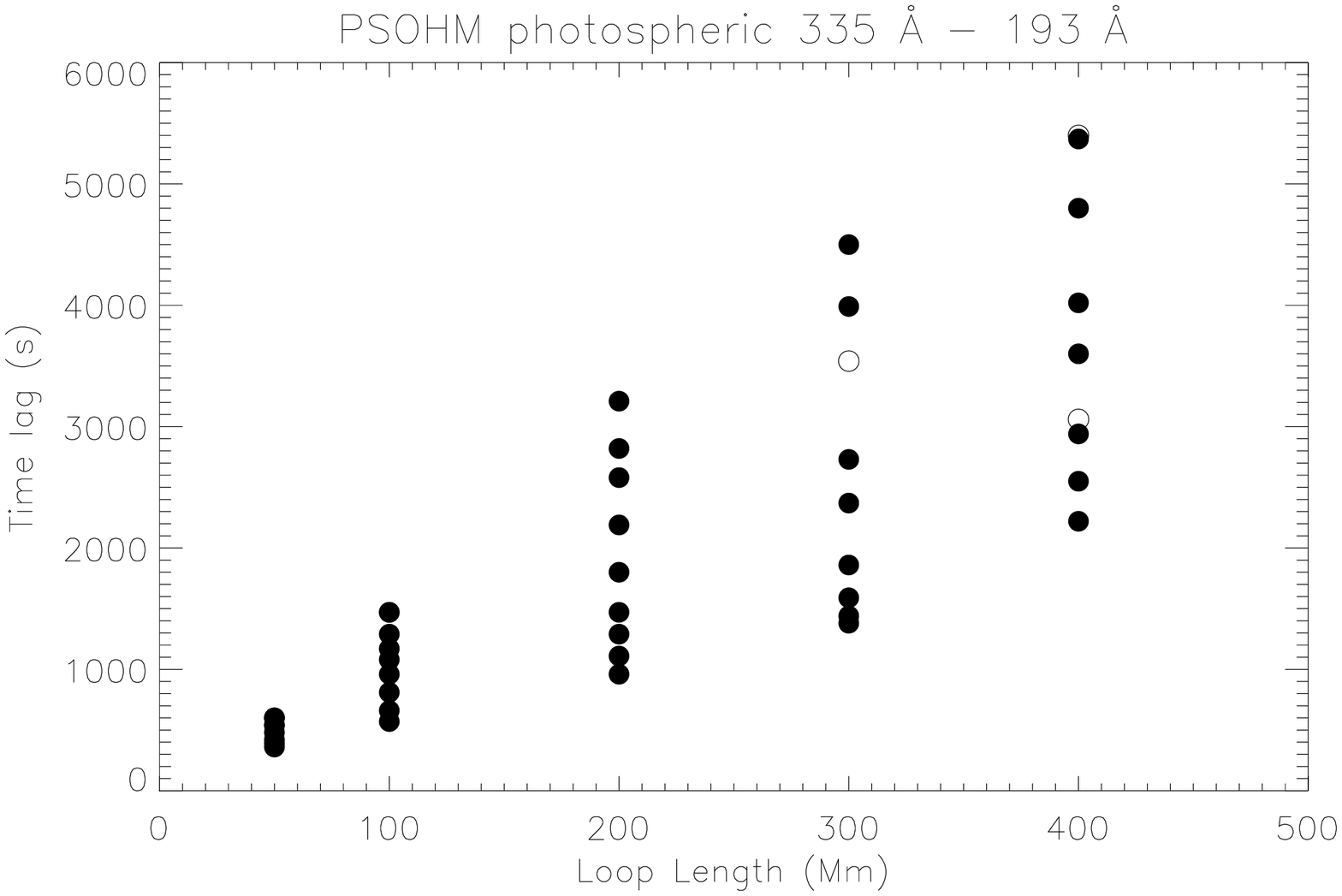}
\includegraphics[width=0.32\textwidth]{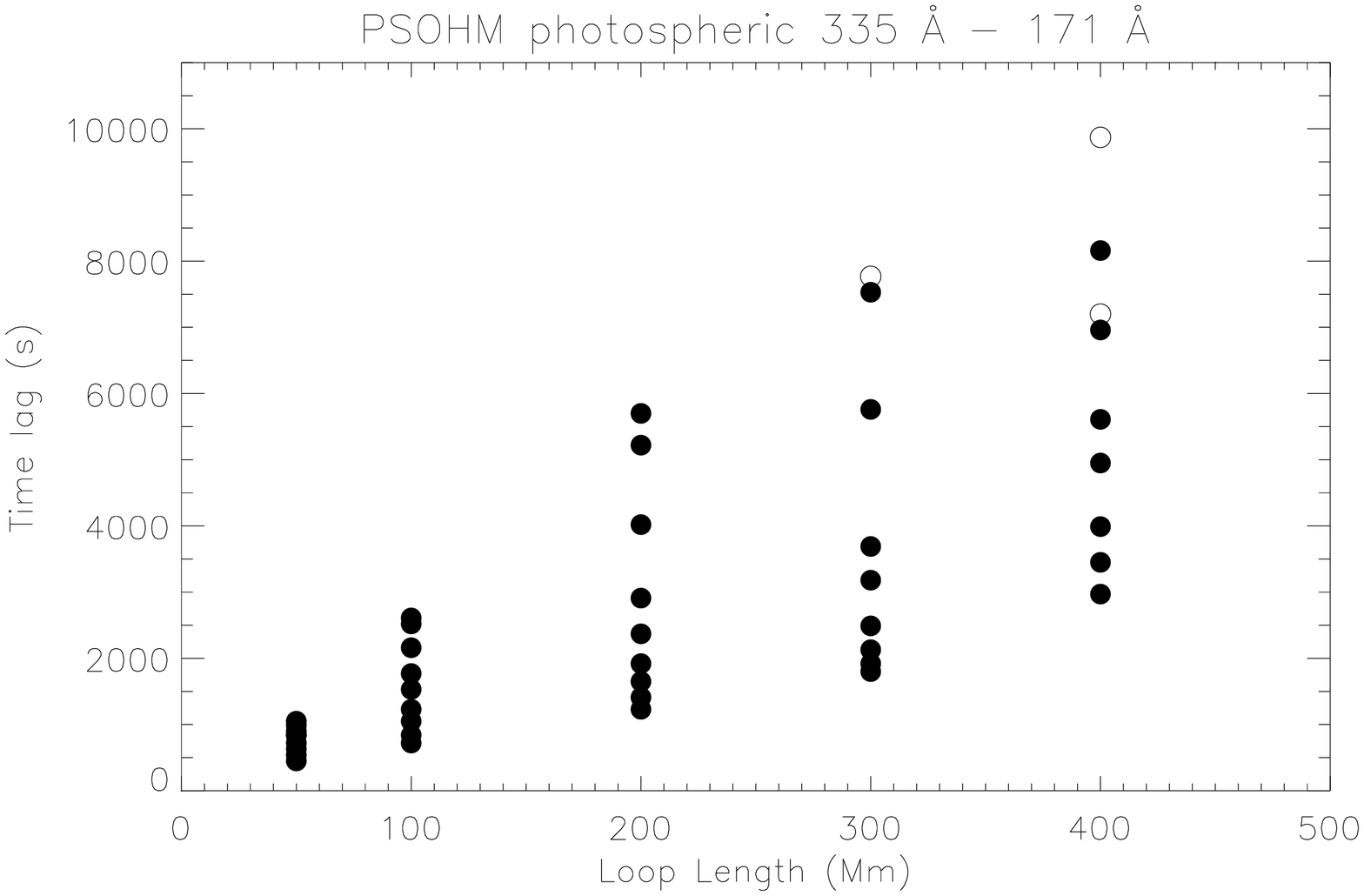}\\
\includegraphics[width=0.32\textwidth]{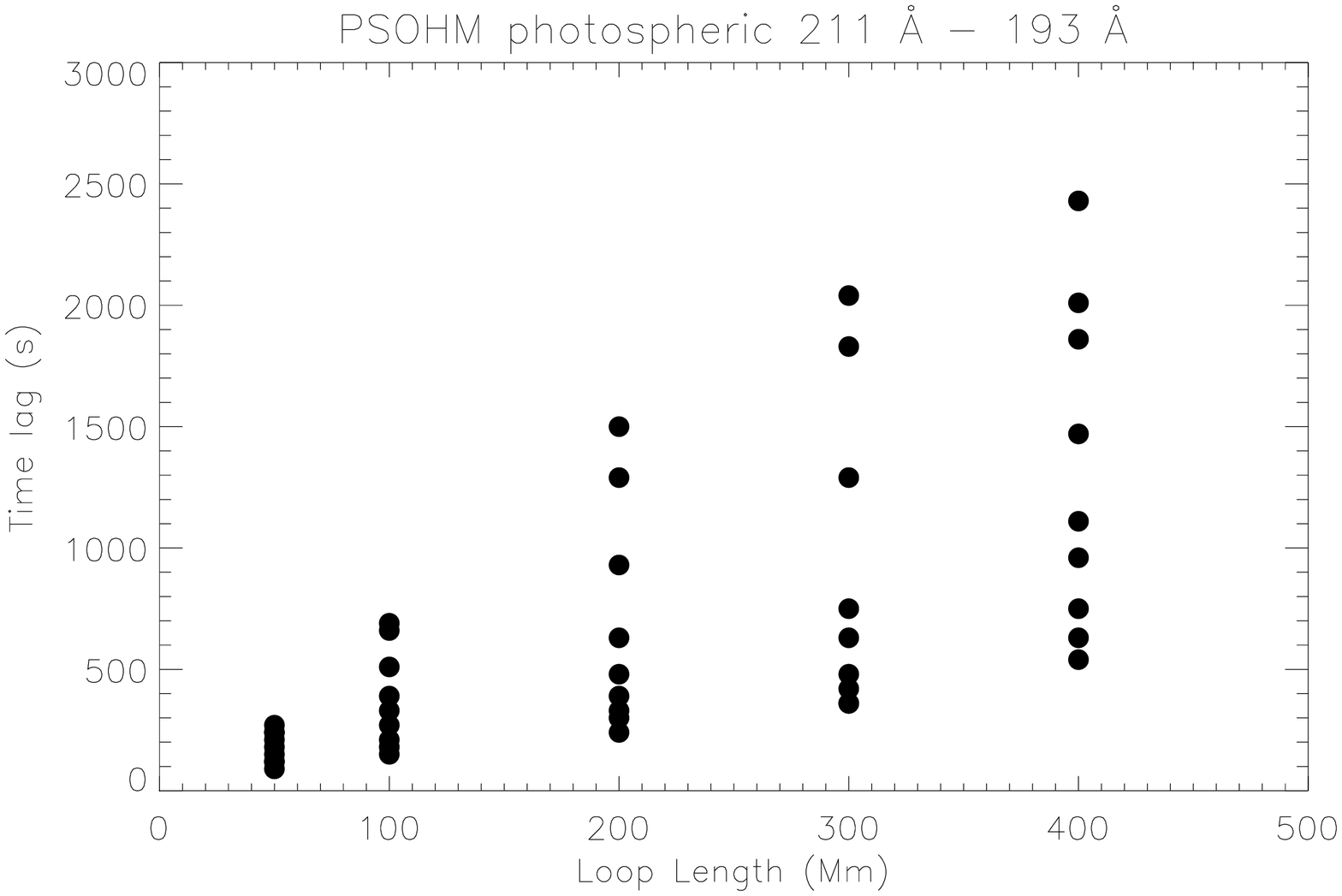}
\includegraphics[width=0.32\textwidth]{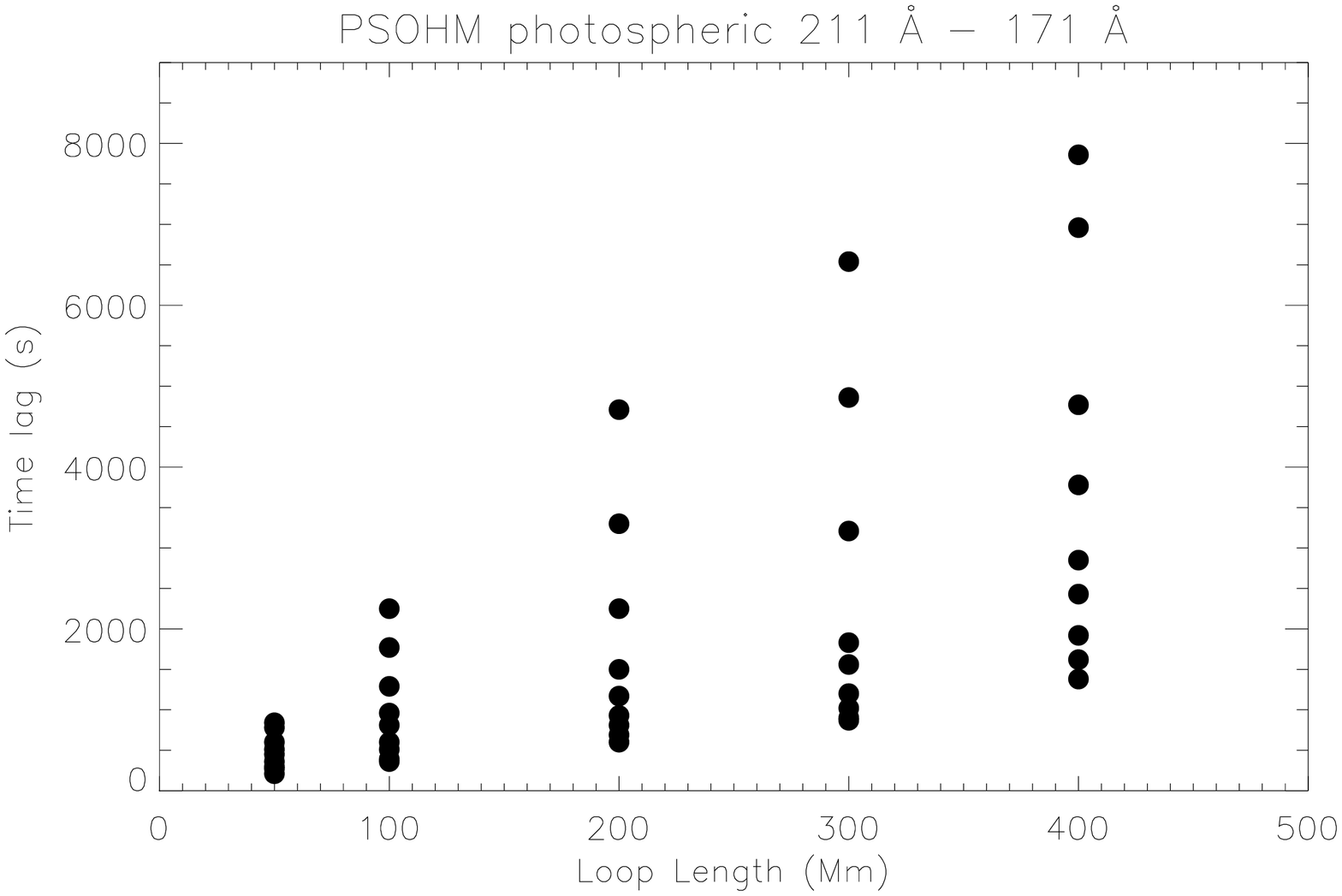}
\includegraphics[width=0.32\textwidth]{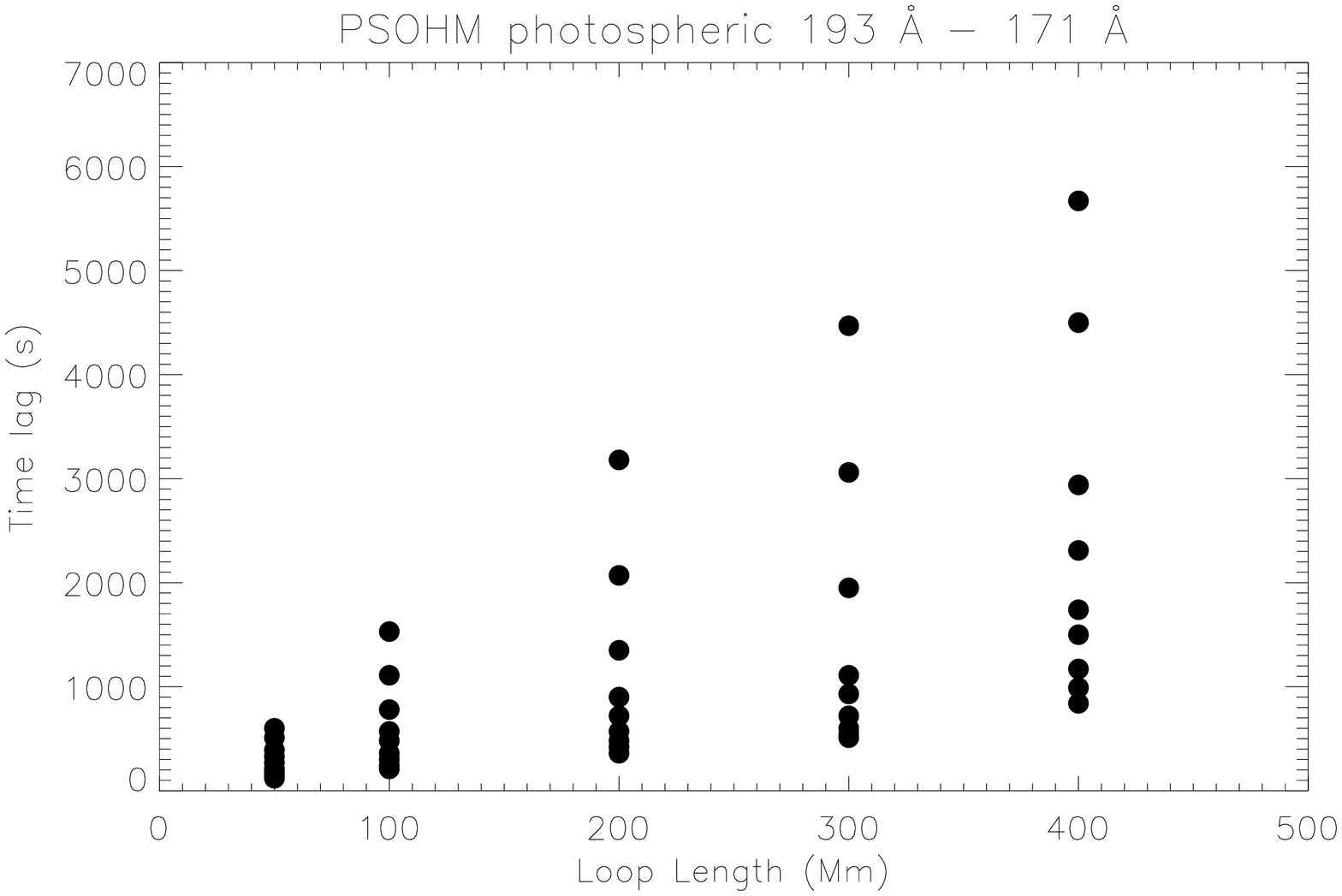}
\caption{{ Results showing the time lag for different channel pairs as a function of loop length for photospheric abundances using the PSOHM code.  In these simulations, we allowed the loop to expand as a function of height. } }
\label{fig:predphoto} 
\end{center}
\end{figure}

{ We run the same set of simulations as above (Figure \ref{fig:TLvsL}) and, in all cases, the pulse durations are 500\,s.  For each loop length, we run two sets of simulations, one with coronal abundances and one with photospheric abundances.  All solutions have an equilibrium temperature between 2 -- 8.5 MK and are given the same input energy as the NRLSOLFTM simulation runs.  The time lags as a function of loop length for the PSOHM simulations run with coronal abundances are shown in Figure~\ref{area_expand}. The results using photospheric abundances are shown in Figure~\ref{fig:predphoto}.}

{ Figures \ref{area_expand}  and \ref{fig:predphoto} show} the same trends found in the previous subsections.  The longer loops and loops with photospheric abundances have larger time lags.   When compared to the same impulsive heating rate for constant cross section, the time lags are roughly a factor of 2 larger.  In some of these channel pairs, time lags of larger than 5,000\,s are achieved.

\subsection{Summary}

In the previous subsections, we have shown the expected time lags for a variety of loops in different conditions.  We summarize the range of expected time lags in Table~\ref{tab:summary}.   For conditions that allow for delays longer than 5,000\,s, we have printed the time in bold face.

\begin{deluxetable}{ccc|c|c|c|c|c|c}
\tablecaption{Summary of results}
\tabletypesize{\scriptsize}
\tablewidth{0pt}
\tablehead{
\colhead{Loop} & \colhead{Abun-} & \colhead{Area}   & \colhead{Time lag}&  \colhead{Time lag} & \colhead{Time lag} &
\colhead{Time lag} & \colhead{Time lag} & \colhead{Time lag}\\
\colhead{Length} & \colhead{dances} & \colhead{Expan-}   &\colhead{335-211\,\AA}&  \colhead{335-193\,\AA} & \colhead{335-171\,\AA} &
\colhead{211-193\,\AA} & \colhead{211-171\,\AA} & \colhead{193-171\,\AA}\\
\colhead{[Mm]} & \colhead{} & \colhead{sion} &  \colhead{[s]} & \colhead{[s]} & \colhead{[s]} &
\colhead{[s]} & \colhead{[s]} & \colhead{[s]}
}
\startdata
50    & C & no  &150--210     & 180--330       & 270--570          &30--120           &90--390          & 60--270 \\ 
100  & C & no  & 270--420    & 390--600       &510--1020         &90--240            &210--750        & 120--510\\ 
200  & C & no &  390-840     & 780--1230     &1020--1890       &180--420         &420--1380       &270--960 \\ 
300  & C & no  &  660--1230 &1140--1590    &1500--2730       &240--540         &570--1950       &330--1350 \\ 
400  & C & no  & 720--1680  &1410--2370    &2250--3660       &390--690         & 960--2310      &600--1650 \\  \hline

50    & P & no  & 180--360    &390--570        &540--990           &90--240           &240--780         &150--540 \\ 
100  & P & no  & 360--750    &660--1140      &1170--1950        &210--450         &510--1500      &300--1080 \\ 
200  & P & no  & 630--1500  &1170--2220     &2460--3630       &390--840        &1050--2610      &690--1860 \\ 
300  & P & no  & 810--2220   & 1650--3120    & 3480--4920        &570--1200     &1440--3510      &900--2460 \\ 
400  & P & no  & 810--3030  &1740--4350    & 4230--{\bf 6600} &780--1500    &2010--4620     &1260--3120 \\ \hline

50    & C & yes   & 120--210   &150--300           &210--510              &0--120               &90--390            &30--240 \\ 
100  & C & yes   & 180--450   & 240--750          & 300--1350           & 60--330              &120--930          &60--600\\ 
200  & C & yes  & 270--930    & 390--1590        & 480--2820           & 90--690              & 210--1890       &120--1200\\ 
300  & C & yes  & 390--1350  & 570--2250        & 720--3960           &120--990               &330--2700      &180--1740 \\ 
400  & C & yes  & 660--1710  & 930--2820        & 1200--4980           &210--1260           &510--3870      & 330--2640\\ \hline

50    & P & yes   & 150--390     & 360--600           & 450--1050              & 90--270        & 210--840         &120--600 \\ 
100  & P & yes   & 420--900     & 570--1470         &720--2610             &150--690           &360--2250      &210--1530\\ 
200  & P & yes   & 660--1830   &960--3210          &1230--{\bf 5700}     &240--1500        &600--4710      &360--3180 \\ 
300  & P & yes  & 960--2550    &1380--4500        & 1800--{\bf 7770}    &360--2040        &870--{\bf 6540} &510--4470 \\ 
400  & P & yes  & 1470--3360  &2220--{\bf 5400} &2970--{\bf 9870}     &540--2430       &1380--{\bf 7860} & 840--{\bf 5670}\\ 
\enddata
\tablecomments{C $=$ Coronal, P $=$ Photospheric.}
\label{tab:summary}
\end{deluxetable}

\section{Discussion and Conclusions} 
\label{sec-conclusions}

The analysis of time lags in  Active Region 11082  observed on 19 June 2010 yielded a wide variety of time lags between AIA channel pairs  \citep{2012ApJ...753...35V}. In each of the channel pairs considered in this study, the largest time lags were larger than 5,000\,s.
 In this paper, we have done an extensive parameter search of one-dimensional models to explain the presence of these long time lags.  We have investigated how time lags depend on heating magnitude, loop length, abundances, and area expansion. We have also considered the duration of the heating event and the inclination of the loop, but found the time lag did not strongly depend on these values.   

{ When  we looked  for  evidence of cooling in AR  11082, only 30-45\% of the pixels (depending on the
 channel pair) showed positive time lags with} correlation coefficients larger than 0.2 (Table~\ref{tab:combs}).  Hence, statistically significant
evidence of cooling is not as widespread as
 \citet{2012ApJ...753...35V} suggested. 
Furthermore,
our results indicate that the models do not predict time lags larger than 
5,000\,s between the { 335-211\,\AA\ and 211-193\,\AA\ channel pairs}. However, the large time lags in the 335-171\,\AA\ channel pair may be explained by long loops with or without area expansion and photospheric abundances.  Additionally, the large time lags in the { 211-171\,\AA, 335-193\,\AA, and 193-171\,\AA\ channel pairs} may be explained by long loops with area expansion and photospheric abundances.  Though our simulations show that it is possible to recreate long time lags in these channel pairs, it is difficult to know if the conditions required to generate them are occurring in the observed active region structures. First,  not all the loops showing a $>$ 5,000\,s delay (the yellow loops seen Figure~\ref{fig:tl_map}) are associated with the longest loop lengths (color-coded in purple or black dashed in Figure~\ref{fig:mag}).  Second, it has previously been found that structures in the corona have coronal, not photospheric abundances \citep[e.g.,][]{2012ApJ...755...33S}, { and none of our simulations with coronal abundances exhibited $>$ 5,000\,s delays}.  However, recent work has indicated that coronal abundances may not be as prevalent as previously thought \citep{2015ApJ...802..104B}. 

{ 

When undertaking this analysis, we focused on time lags larger than 5,000\,s and we calculated the percentage of pixels that had trustworthy (i.e., cross correlation coefficients larger than 0.2), positive time lags larger than 5,000\,s.  These percentages are given in Table 1.  However, for the 335-211\,\AA\ and 211-193\,\AA\ channel pairs, we found that the largest time lag predicted for impulsive heating was much smaller than 5,000\,s.   The maximum allowed time lag for impulsive heating for the 335-211\,\AA\ channel pair was 3,360\,s.  The percentage of pixels that have a positive, trustworthy time lag larger than this value in this channel pair is 7.1\%.  The maximum allowed time lag for the 211-193\,\AA\ channel pair was 2,430\,s.  The percentage of pixels that have a positive trustworthy time lag larger than this value is 18.8\%.  

A large fraction of the maps in Fig.~\ref{fig:tl_map}  
is occupied by pixels with zero time-lag. Although zero time-lag 
in the moss/footpoint areas
is compatible with impulsive heating 
 \citep{2012ApJ...753...35V}, 
this does not provide conclusive evidence
in favor of any particular model.
In fact, it has not been shown that only one model can account for 
no delay.
On the contrary, 
zero time-lag is at least compatible with
equilibrium solutions and, although a study has not been completed yet, 
cannot be excluded in thermal non-equilibrium
models \citep[e.g., in][the loop solutions appear to vary mostly at the center]{2013ApJ...773...94M}. } 

\begin{figure}
\includegraphics[width=.45\textwidth]{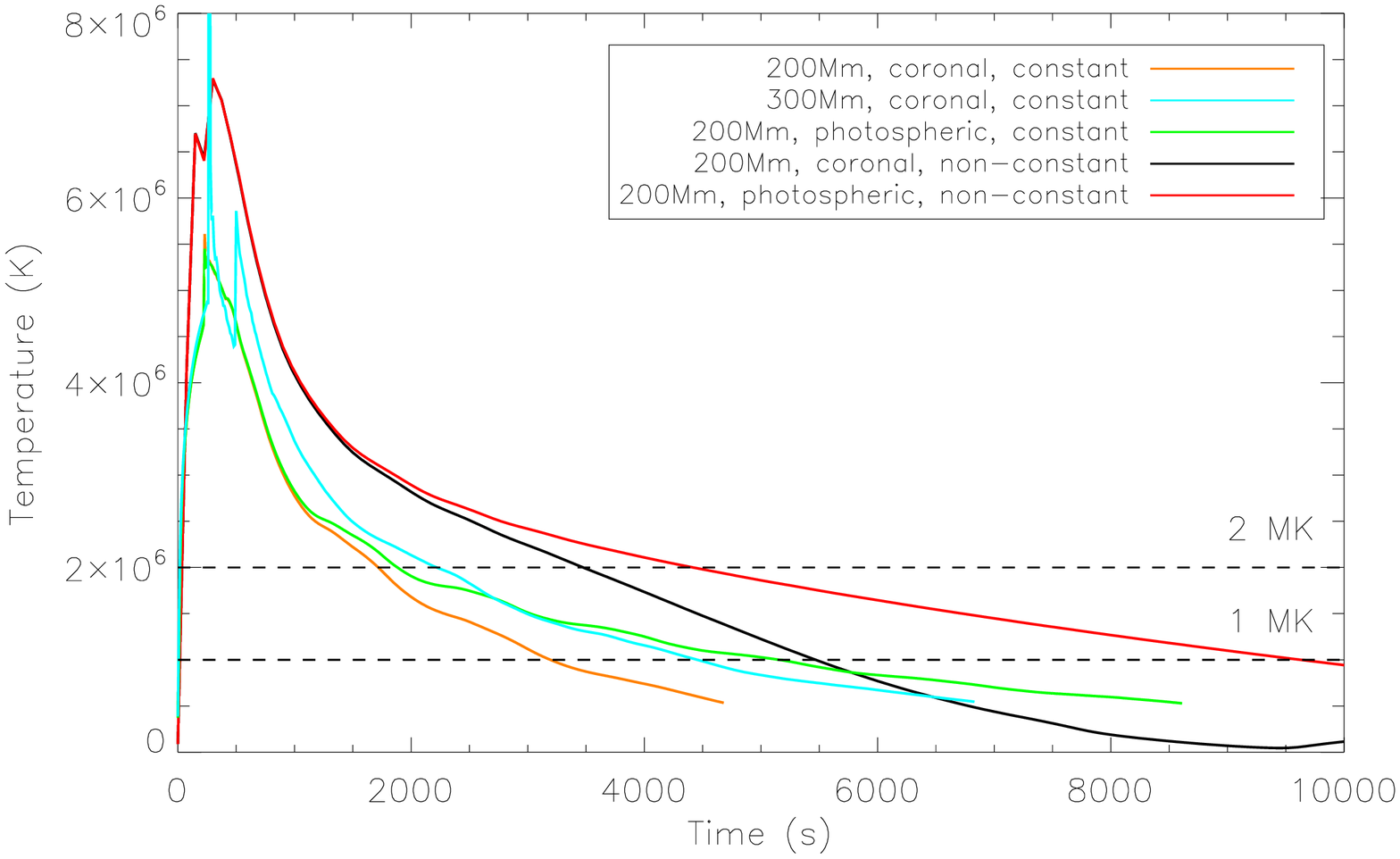}
\includegraphics[width=.45\textwidth]{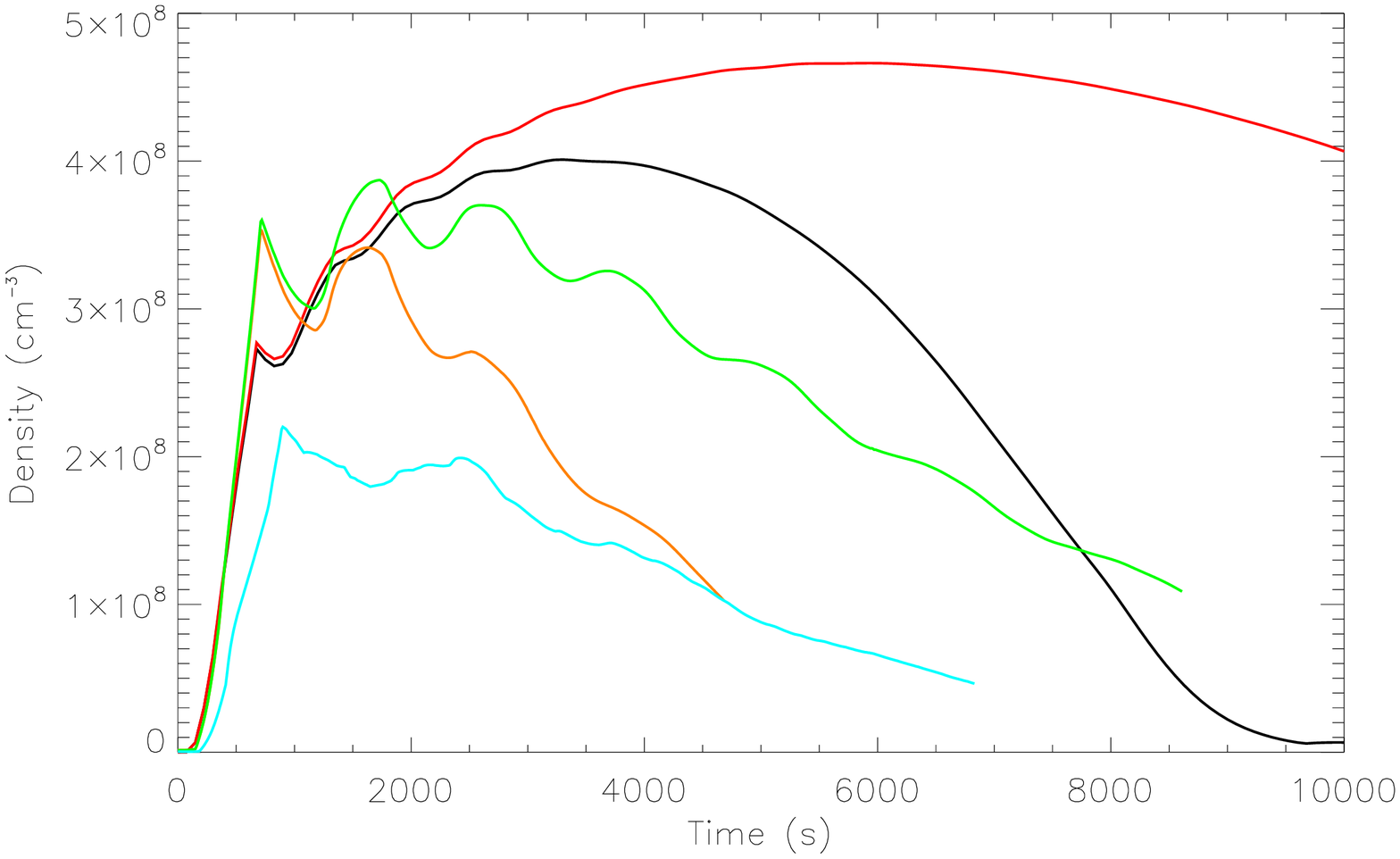}
\caption{The temperature (left) and density evolution from a sample set of simulations.  Each simulation has $\sim$ 2 MK equilibrium temperature.  Using only the temperature evolution between 2 MK and 1 MK, we calculate the cooling time of the plasma, these values are given in Table~\ref{tab:sample}.}
\label{fig:sample} 
\end{figure}

To further demonstrate the impact of the various parameters on the time lag, we plot the density and temperature evolution of a sample set of simulations in Figure~\ref{fig:sample}.  We provide information on these simulations in Table~\ref{tab:sample}.  All simulations had a comparable, $\sim$ 2 MK equilibrium temperature.  Using only the temperature evolution data from 2 MK to 1 MK, we calculate a cooling time assuming the temperature is exponentially decreasing.  These values are also provided in Table~\ref{tab:sample}.  The 200 Mm loop with coronal abundances and constant cross section has the fastest cooling time.  Increasing the length of the loop to 300\,Mm increases the cooling time by roughly 1.5.  Using photospheric abundances instead of coronal doubles the cooling time.  For an expanding area loop with coronal abundances, the cooling time increases by 1.5, while an expanding area loop with photospheric abundances more than triples the cooling time. Hence, the range of cooling times one can expect by varying the parameters from the coronal, constant cross section case is roughly a factor of three. 

\begin{deluxetable}{ccccc}
\tablecaption{Cooling times calculated for a sample set of solutions}
\tabletypesize{\scriptsize}
\tablewidth{0pt}
\tablehead{
\colhead{Code} & \colhead{Loop Length} & \colhead{Abundances} & \colhead{Cross Sectional Area} & \colhead{Cooling Time}}
\startdata
NRLSOLFTM &     200  & coronal         & constant & 5,440\,s \\
NRLSOLFTM &     300 & coronal 	    & constant  &  7,690 \\
NRLSOLFTM &     200 &  photospheric & constant  & 10,850\,s\\
PSOHM          &     200 & coronal           & non-constant & 7,950\,s\\
PSOHM          &     200 & photospheric          & non-constant & 17,430\,s
\enddata
\label{tab:sample}
\end{deluxetable}

One possible explanation for the inability of the impulsive heating shown here to reproduce these long time lags is that the time lags retrieved by the cross correlation process are not representative of actual loop evolution.  Individual loops may be cooling through the channels at a faster rate, but for some reason, the cross correlation process detects a slower period.

Our results encourage us to explore different processes (or combinations thereof) that may be at work in coronal loops and that may yield the observed time delays. Thermal non-equilibrium, which was recently investigated in a series of articles \citep{2013ApJ...773...94M,2013ApJ...773..134L,2014ApJ...795..138W}, could provide the key to the understanding of this phenomenon. Although the authors of these articles did not investigate time delays in particular, they did include some interesting details.  For example, Loop A in \citet{2014ApJ...795..138W} evolved over 
 a period of about 4.4 hours, during which the 195-171\,\AA\ time lag was approximately 1,400 s.  The length of the loop was not stated, but \citet{2013ApJ...773...94M} found a similar period for a relatively short loop of 150 Mm (Model 1; Case 4). However, the period almost
 doubled for another loop having different heating parameters but the 
same length (Model 2; Case 6).
Clearly, a loop 
with  a longer  period of evolution must also exhibit longer time lags between
emission channels.
 We expect longer loops to show even longer
periods and, consequently,   more extended time delays. We are  currently
investigating 
 whether thermal non-equilibrium can effectively 
provide emission time lags  in better agreement with observations  (Winebarger et al., in preparation). Recently, pulsations in coronal features with long time delays were detected in EIT and AIA observations \citep{2014A&A...563A...8A,2015ApJ...807..158F}.

\acknowledgements
The authors are grateful to the referee for many helpful comments.  The authors thank Drs.\ Ron Moore and Alphonse Sterling for providing many comments and discussions on the early text.  RL thanks Dr.\ Ronald Caplan for helpful elucidations.
CEA is supported by appointments to the NASA Postdoctoral Program at the NASA/MSFC, administered by ORAU through a contract with NASA.
ARW is supported by a grant from NASA SR\&T program. This work was supported by the NASA Heliophysics Theory and Living With a Star programs.

\bibliography{mybib}

\end{document}